\documentclass[11pt,a4paper]{article}

\usepackage{amsmath}
\usepackage{amsfonts}
\usepackage{amssymb}
\usepackage{amsthm}
\usepackage{graphicx}
\usepackage[left=1.2in, right=1.2in, top=1.2in, bottom=1.2in]{geometry}
\usepackage{times}
\usepackage{bm}
\usepackage{natbib}
\usepackage{multirow}
\usepackage[plain,noend]{algorithm2e}
\usepackage{xurl}
\usepackage{latexsym}
\usepackage{graphicx}
\usepackage{xcolor}
\usepackage{array}
\usepackage{multirow}
\usepackage{enumitem}
\usepackage{float}
\usepackage{tikz}
\usepackage{pgfplots}
\usepackage{booktabs}
\def\cA{\mathcal{A}}

\newcommand{\cC}{\mathcal{C}}
\newcommand{\cD}{\mathcal{D}}

\newcommand{\cG}{\mathcal{G}}

\newcommand{\cM}{\mathcal{M}}
\newcommand{\cN}{\mathcal{N}}

\newcommand{\cP}{\mathcal{P}}

\newcommand{\R}{\mathtt{R}}

\newcommand{\cU}{\mathcal{U}}

\newcommand{\Np}{{\mathtt{N}}^{+}}
\newcommand{\Nn}{{\mathtt{N}}^{-}}

\def\ba{\begin{array}}
	\def\bc{\begin{center}}
		\def\bd{\begin{description}}
			\def\be{\begin{enumerate}}
				\def\ea{\end{array}}
			\def\ec{\end{center}}
		\def\ed{\end{description}}
	\def\edt{\end{document}}
\def\ee{\end{enumerate}}
\def\ben{\begin{equation}}
\def\benn{\begin{equation*}}
\def\een{\end{equation}}
\def\eenn{\end{equation*}}
\def\benr{\begin{eqnarray}}
\def\eenr{\end{eqnarray}}
\def\benrr{\begin{eqnarray*}}
\def\eenrr{\end{eqnarray*}}

\def\al{\alpha}

\def\edt{\end{document}}
\def\ep{\epsilon}
\def\g{\gamma}

\def\h{\hat}
\def\ka{\kappa}

\def\e{\mathrm e}

\def\iny{\infty}
\def\ka{\kappa}

\def\la{\lambda}

\def\noi{\noindent}

\def\nn{\nonumber}

\def\xim{\underline\xi}
\def\xma{\overline\xi}

\def\lm{\underline \ell}

\def\si{\sigma}

\def\Si{\Sigma}

\def\vep{\varepsilon}

\def\vs{\vskip}

\def\Z{{\mathtt Z}}
\def\z{\zeta}

\DeclareMathOperator*{\argmin}{arg\,min}
\DeclareMathOperator*{\argmax}{arg\,max}

\allowdisplaybreaks
\numberwithin{equation}{section}

\newtheorem{theorem}{Theorem}[section]
\newtheorem{lemma}[theorem]{Lemma}
\newtheorem{corollary}{Corollary}[section]
\newtheorem{remark}{Remark}
\newtheorem{definition}[theorem]{Definition}

\begin{document}

	\bc
	{\Large {\bf Inference for Change Points in High Dimensional Mean Shift Models}}\\[.5cm]
	Abhishek Kaul$^{a,}$\footnote{Email: abhishek.kaul@wsu.edu.} and George Michailidis$^b$\\[0.25cm]

	$^a$Department of Mathematics and Statistics,\\ 
	Washington State University, Pullman, WA 99164, USA.\\[0.25cm]
	
	$^b$Department of Statistics and the Informatics Institute,\\University of Florida, Gainsville, FL 32611-8545, USA.
	
	\ec
	\vs .1in
	{\renewcommand{\baselinestretch}{1}
		\begin{abstract}
		We consider the problem of constructing confidence intervals for the locations of change points in a high-dimensional mean shift model. To that end, we develop a locally refitted least squares estimator and obtain component-wise and simultaneous rates of estimation of the underlying change points. The simultaneous rate is the sharpest available in the literature by at least a factor of $\log p,$ while the component-wise one is optimal. These results enable existence of limiting distributions. Component-wise distributions are characterized under both vanishing and non-vanishing jump size regimes, while joint distributions for any finite subset of change point estimates are characterized under the latter regime, which also yields asymptotic independence of these estimates. The combined results are used to construct asymptotically valid component-wise and simultaneous confidence intervals for the change point parameters. The results are established under a high dimensional scaling, allowing for diminishing jump sizes, in the presence of diverging number of change points and under subexponential errors. They are illustrated on synthetic data and on sensor measurements from smartphones for activity recognition.
	\end{abstract} }
	\noi {\it Keywords: Multiple change points
		Inference,
		Rate of convergence,
		Limiting distributions,
		High dimensions.}

\section{Introduction}\label{sec:intro}

Detection of change points constitutes a canonical statistical problem due to numerous applications in diverse areas, including economics and finance (\cite{basseville1993detection}, \cite{frisen2008financial}), quality process control (\cite{qiu2013introduction}), functional genomics and neuroscience (\cite{koepcke2016single}). The \textit{offline} version of the problem, wherein one examines the data retrospectively and aims to detect the presence and/or location of change points has been studied extensively for a variety of statistical models, including signal plus noise, regression, graphical, random graph, factor and time series models and various algorithms have been developed to accomplish this task -dynamic programming, regularized cost functions, binary segmentation, multiscale methods, etc., see, e.g. the review article \cite{niu2016multiple}. In the presence of multiple change points, consistency of the estimated location of the change points under certain regularity assumptions on the temporal spacing between change points and on the magnitude of the changes in the underlying model parameters have been established, see, e.g. \cite{fryzlewicz2014wild}, \cite{frick2014multiscale} and \cite{wang2018high} amongst several others, here the former two are under a fixed $p$ framework and the latter under a high dimensional framework. Further, when a single change point has been assumed, the asymptotic distribution of the change point estimator has been established for various statistical models, see, e.g., \citep{bai1994, bai1997estimation}, \cite{csorgo1997limit}, under fixed $p$ setting, and \citep{bhattacharjee2017common, bhattacharjee2019change}, \citep{kaul2020inference, kaul2021graphical}, under diverging dimensionality, where the last two articles allow potential high dimensionality.

However, under multiple change points the literature on their asymptotic distributions is very limited, even for models involving univariate data. One potential roadblock is the absence of consistency results exhibiting an optimal rate, which constitute an intermediate building step for establishing the existence of limiting distributions for the underlying change points. To that end, this paper addresses the problem of inference for multiple change points for high dimensional mean shift models. Specifically, we consider the following data generation mechanism:
\benr\label{model:rvmcp}
y_t&=&\begin{cases}\theta_{(1)}^*+\vep_t, & t=1,...,\tau_1^0\\
	\theta_{(2)}^*+\vep_t, & t=\tau_1^0+1,...,\tau_2^0\\
	\vdots\\
	\theta_{(N+1)}^*+\vep_t, & t=\tau_N^0+1,...,T.
\end{cases}\nn\\
&=&\sum_{j=1}^{N+1}\theta_{(j)}^* {\bf 1}[\tau^0_{j-1}< t\le \tau_{j}^0]+\vep_t,\quad\textrm{ for}\,\,t=1,...,T,
\eenr
wherein $y_t=(y_{t1},y_{t2},...,y_{tp})^T\in\R^{p}$ denotes the response, and the noise $\vep_t=(\vep_{t1},...,\vep_{tp})^T\in\R^p$ are zero mean, subexponential random variables. The model parameters that need to be estimated from the available data are the number of change points $N\in\Np=\{1,2,...\}$, their locations  $\tau^0=(\tau^0_1,\tau^0_2,...,\tau^0_N)^T\subseteq\{1,...,T\}^N,$ with all $N$ components assumed to be distinct and in ascending order, and finally the mean vectors $\theta^*_{(j)}\in\R^p,$ $j=1,...,(N+1).$ The location parameters $\tau^0_j,$ $j=1,..,N$ are of prime interest, while $\tau^0_0=0$ and $\tau^0_{N+1}=T$ are defined for notational convenience. Finally, the dimension $p$ (number of data streams under consideration), as well as the number of change points $N$ can diverge with the sample size $T$, with the former diverging potentially exponentially, as specified in the sequel.

To proceed further we require more notation. Define the jump vector's and the jump size's associated with model (\ref{model:rvmcp}). These quantities are known to be fundamentally related to statistical properties of any change point estimator, see, for e.g. \cite{liu2019minimax}. For $j=1,...,N,$ let,

\benr\label{def:jumpsize}
\eta^*_{(j)}=(\theta^*_{(j)}-\theta^*_{(j+1)}),\qquad\xi_j=\|\eta^*_j\|_2,\qquad \xim=\min_{1\le j \le N}\xi_j\quad{\rm and}\quad \xma=\max_{1\le j \le N}\xi_j
\eenr

The key objective is to obtain limit distributions for $\tau^0_j, j=1,...,N$. To that end, we first obtain an \textit{optimal component-wise} estimation rate for the location of the change points, for which we proposed a refitted least squares estimation procedure. Subsequently, the distributional behavior of the proposed estimates is characterized under the following regimes: (i) vanishing ($\xim\to 0$), and (ii) non-vanishing ($\xim\to\xi_{\iny}, 0<\xi_{\iny}<\iny$ jump sizes. For both regimes, the component-wise limiting distributions is obtained, together with the joint distribution under the non-vanishing jump size regime. These in turn enable construction of asymptotically valid confidence intervals, for any finite subset of the potentially diverging number of change points. To the best of our knowledge, the ability to perform inference on locations of change point parameters is unavailable in the current literature under high dimensionality (or even diverging dimensionality which is slower than $T$). The only result available in the literature is that of \cite{lu2018intelligent} that considers a univariate mean shift model and establishes the joint asymptotic distribution of the location of a finite number of change points, assuming Gaussian noise.

The remainder of this article is organized as follows. Section \ref{sec:tech.p} describes the proposed estimation procedure and a brief summary of the main results developed in this article. Section \ref{sec:main} provides a rigorous description of the estimation and inference results together with the sufficient assumptions made for their validity. This section proceeds under an assumption of available preliminary mean and change point estimates that are slower than optimal that are necessary for the construction of the proposed optimal estimation procedure. Thus, the results of this section remain only theoretical and unimplementable without the availability of these preliminary estimates. Section \ref{sec:feasible} develops feasible and implementable in practice algorithms by aggregating results developed in Section \ref{sec:main} together with estimators and results from the literature. Section \ref{sec:numerical} provides numerical support to our methodology and results via monte-carlo simulations. Section \ref{sec:application} provides an application of the proposed methodology to smartphone based human activity recognition. We conclude this section with a short note on the notation used throughout.

\vspace{1.5mm}
\noi{\it Notation}: $\R$ represents the real line. For any vector $\delta\in\R^p,$ $\|\delta\|_1,$ $\|\delta\|_2,$ $\|\delta\|_{\iny}$ represent the usual 1-norm, Euclidean norm, and sup-norm respectively. For any set of indices $U\subseteq\{1,2,...,p\},$ let $\delta_U=(\delta_j)_{j\in U}$ represent the subvector of $\delta$ containing the components corresponding to the indices in $U.$ Let $|U|$ and $U^c$ represent the cardinality and complement of $U.$ We denote by $a\wedge b=\min\{a,b\},$ and $a\vee b=\max\{a,b\},$ for any $a,b\in\R.$ We use a generic notation $c_u>0$ to represent universal constants that do not depend on $T$ or any other model parameter. All limits are with respect to the sampling periods $T.$ The notation $\Rightarrow$ represents convergence in distribution. For any vector $\delta,$ the notation $\delta^T$ represents its transpose (note: there is a notational overlap with the sampling period $T$, however the distinction shall be contextually clear throughout the article).

\section{Technical Preliminaries and Estimation Procedure}\label{sec:tech.p}

Given the high dimensional nature of the posited model (\ref{model:rvmcp}) (diverging dimension $p$ as a function of the sample size $T$), we further assume a sparsity condition on the jump vectors, $\|\eta^0_{(j)}\|_0\le s,$ $1\le j\le N,$ where $1\le s<<T,$ see, e.g., \cite{wang2018high} and \cite{harchaoui2010multiple}. Next, we consider the following reparameterized version of the model through global centering
($x_t=y_t-\bar y$)
\benr\label{model:center.rvmcp}
&&x_t=\sum_{j=1}^{N+1}\theta_{(j)}^0 {\bf 1}[\tau^0_{j-1}< t\le \tau_{j}^0]+\vep_t^*,\,\,\,\textrm {for}\,\,\,t=1,...,T,\quad \textrm{where,}\nn\\
&&\theta^0_{(j)}=\theta_{(j)}^*-w(\theta^*),\quad  w(\theta^*)=\frac{1}{T}\sum_{j=1}^{N+1}(\tau_j^0-\tau_{j-1}^0)\theta_{(j)}^*\quad\textrm{and}\nn\\
&&\vep_t^*=\vep_t-\bar\vep,\quad\bar\vep=\frac{1}{T}\sum_{t=1}^T\vep_t,\quad t=1,...,T.
\eenr
that transfers the $s$-sparsity of the jump vectors to $Ns$-sparsity of individual mean vectors. Despite the simplicity of this transformation, doing so allows us to exploit the assumption of sparsity quite differently than \cite{wang2018high} and \cite{harchaoui2010multiple}, without any loss of generality in the assumptions being made. A further algebraic manipulation of $\theta_{(j)}^0$ yields a more insightful expression for these reparameterized means:
\benr\label{eq:trans.mean}
\theta^0_{(j)}=\frac{1}{T}\Big[-\sum_{k=1}^{j-1}\tau_k^0\eta_{(k)}^*+\sum_{k=j}^N(T-\tau_k^0)\eta_{(k)}^*\Big],\quad j=1,...,N+1\footnotemark
\eenr
\footnotetext{Here $\sum_{j=1}^{k-1},$ and $\sum_{j=k}^{N}$ are defined to be zero at $k=1$ and $k=N+1,$ respectively.}

Note that in (\ref{eq:trans.mean}) the $\theta_{(j)}^{0},$ $j=1,...,N+1,$ are expressed as a linear combination of $s$-sparse jump vectors $\eta^*_{(j)},$ $j=1,...,N$. This centering is in the spirit of the same operation in high dimensional linear regression that removes the intercept term and is known not to impact estimation rates. The only consequence is a slight alteration to the re-defined noise term $\vep_t^*$ of (\ref{model:center.rvmcp}), which we shall show has no statistical impact on the results to follow.

\begin{remark}(On the jump sizes of the reparameterized model (\ref{model:center.rvmcp}))\label{rem:jump.equivalence} {\rm Note that the mean parameters $\theta^*_{(j)}$ and $\theta^0_{(j)},$ $j=1,...,N+1$ of models (\ref{model:rvmcp}) and (\ref{model:center.rvmcp}) are distinct. However, the jump vectors and jump sizes that control properties of the change point remain identical, since
		\benr
		\big(\theta_{(j)}^*-\theta_{(j+1)}^*\big)=\eta^*_{(j)}=\big(\theta_{(j)}^0-\theta_{(j+1)}^0\big)=\eta^0_{(j)},\quad j=1,...,N.\nn
		\eenr
		Consequently, the jump size parameters $\xi_j,$ $j=1,...,N,$ $\xim,$ and $\xma$ defined in (\ref{def:jumpsize}) remain identical for the two models. Thus, in the remainder we do not distinguish between jump vector and jump size parameters of models (\ref{model:rvmcp}) and (\ref{model:center.rvmcp}) that are denoted by $\eta^0_{(j)},$ $\xi_j,$ $j=1,...,N,$ and $\xim,$ and $\xma,$ irrespective of the underlying model.}
\end{remark}

Next, we provide a roadmap for obtaining the main results. The methods and results presented below assume the availability of preliminary near optimal estimates that are necessary for implementing the proposed methodology. Such near optimal estimates are available in the literature, and this aggregation to obtain feasible implementations are presented in Section \ref{sec:feasible}.

Consider any $T\ge N+1,$ $p\ge 1$ (such that $\log(p\vee T)\ge 1$) and for any $j=1,...,N,$ let $\tau_{-j}=(\tau_1,...,\tau_{j-1},\tau_{j+1},...,\tau_N)^T\in\{1,...,T-1\}^{N-1}$ be any vector with the $j^{th}$ component removed, additionally satisfying $\tau_{j-1}<\tau_{j+1}.$ Consider any  $\theta_{(j)}\in\R^p,$ $j=1,...N+1$ and let $\theta$ represent the concatenation of all $\theta_{(j)}'$s. Next, define the following squared loss function evaluated at any point $\tau_j\in\{\tau_{j-1},...,\tau_{j+1}\}$ w.r.t. realizations $x_t,$ $t=1,...,T$ of model (\ref{model:center.rvmcp}),
\benr\label{def:Q}
Q_j\big(\tau_j,\tau_{-j},\theta\big)=\sum_{t=\tau_{j-1}+1}^{\tau_j} \|x_t-\theta_{(j)}\|_2^2 +\sum_{t=\tau_j+1}^{\tau_{j+1}} \|x_t-\theta_{(j+1)}\|_2^2.
\eenr

Assume for the time being the availability of preliminary estimates $\h\tau=(\h\tau_1,...,\h\tau_{\h N})^T\in\R^{\h N}$ of change points, and  $\h\theta_{j}\in\R^{p},$ $j=1,...,\h N+1$ of mean parameters of model (\ref{model:center.rvmcp}). Then, for each $j=1,...,\h N,$ define a locally refitted plug in least squares estimator utilizing these estimates,
\benr\label{est:optimalcp}
\tilde\tau_j:=\tilde\tau_j\big(\h\tau_{-j},\h\theta\big)=\argmin_{\h\tau_{j-1}<\tau_j<\h\tau_{j+1}} Q_j\big(\tau_j,\h\tau_{-j},\h\theta\big),\quad j=1,...,\h N
\eenr

This local refitting based on \textit{slower than optimal} preliminary nuisance parameter estimates leads to an improved estimate of the $j^{th}$ change point parameter that is \textit{optimal}. This in turn provides sufficient regularity for limiting distributions of these updated estimates to exist, in the presence of potentially diverging number of change points and high dimensionality of means. The preliminary estimates should satisfy,
\benr\label{eq:40}
\max_{1\le j\le N+1}\|\h\theta_{(j)}-\theta_{(j)}^0\|_2\le c_u\si\Big\{\frac{N s\log (p\vee T)}{T\lm}\Big\}^{\frac{1}{2}},\quad\textrm {and}\nn\\
\h N=N,\quad \max_{j=1,...,N} |\h\tau_j-\tau^0_j|\le c_{u1}T\lm,\hspace{0.5in}
\eenr
with probability at least $1-o(1),$ under the following rate condition on the relationship between jump size and dimensionality of the problem
\benr
\Big(\frac{\si}{\xim}\Big)\Big\{\frac{Ns\log^{3/2} (p\vee T)}{\surd (T\lm)}\Big\}\le c_{u1}\label{eq:41},
\eenr
wherein $c_{u1}>0$ is a suitably chosen small enough constant. The parameter $\lm$ is the least separation sequence (spacing between change points) satisfying $\min_{1\le j\le N+1} \big(\tau^0_{j}-\tau_{j-1}^0\big)\ge T\lm\ge 1,$ (Cond. B), and $\si^2$ is a variance proxy parameter of the data generating process (Cond. A).

Then we shall show that the locally refitted estimator (\ref{est:optimalcp}) exhibits the following component-wise and simultaneous estimation rates, 
\benr\label{eq:loc.est.optimal}
&(i)& |\tilde\tau_j-\tau_j^0|=O_p\big(\xi_j^{-2}\big),\quad\text{for any given}\,\, j=1,...,N,\quad\textrm {and,} \nn\\
&(ii)& \max_{1\le j\le N} |\tilde\tau_j-\tau_j^0|=O\big(\xim^{-2} \log^2 T\big),\quad{\text{w.p. at least}}\,\,1-o(1).
\eenr
Note that (\ref{eq:loc.est.optimal}ii) is the sharpest simultaneous estimation rate in the literature, for the posited mean shift model by at least a factor of $\log p$. Under high dimensionality, this yields a polynomial in $T$ improvement. More significantly, the component-wise rate of estimation of (\ref{eq:loc.est.optimal}i) is minimax optimal for the setting under consideration allowing both the dimension $p$ and the number of change points $N$ to be diverging. Finally, the posited results hold for subexponential distributions of the noise term. vis-a-vis the more commonly assumed Gaussian case in the literature.

The most important consequence of of result (\ref{eq:loc.est.optimal}) is the existence of limiting distributions for both vanishing and non-vanishing jump size regimes. Specifically, under the vanishing jump regime, i.e., for any given $j=1,...,N,$ let $\xi_j\to 0,$ then we establish
\benr\label{eq:limiting.d.vanishing}
\xi_j^2\si^{-2}_{(\iny,j)}(\tilde\tau_j-\tau^0_j)\Rightarrow \argmax_{\z\in\R} \big(2W_j(\z)-|\z|\big),
\eenr
wherein $\si^2_{(\iny,j)}=\lim_{T\to\iny}(\eta^{0T}_{(j)}\Sigma\eta^0_{(j)}\big)/\xi_j^2$,  with $\Sigma=E\vep_t\vep_t^T$, for each $j=1,...,N,$ and $W_j(\cdot)$ denotes a two-sided standard Brownian motion\footnote{A two-sided Brownian motion $W(\z)$ is defined as $W(0) = 0,$ $W(\z) = W_1(\z),$ $\z > 0$ and $W(\z) = W_2(-\z),$ $\z < 0,$ where $W_1(\z)$ and $W_2(\z)$ are two independent Brownian motions defined on the non-negative half real line} in $\R.$ This limiting distribution is the same to the one that would be obtained, if the nuisance estimates $\tau^0_{-j}$ and $\theta^0$ were known. The distribution of $\argmax_{\z\in\R}\big(2W(\z)-|\z|\big)$ is well studied in the literature and its cdf and thus its quantiles are readily available  (\cite{yao1987approximating}). The result also highlights the fundamental property  of adaptation as defined in \cite{bickel1982adaptive}; i.e., the estimator (\ref{est:optimalcp}) statistically behaves as if these nuisance parameters were known.

Next, consider the non-vanishing jump size regime of $\xi_j\to\xi_{(\iny,j)},$ with $0<\xi_{(\iny,j)}<\iny,$ for any given $j=1,...,N.$ To describe the limiting distributions in this case we first need to define the following negative drift two sided random walk initialized at the origin,
\benr\label{eq:neg.drift.rw}
\cC_{\iny}(\z,\xi,\si^2)=
\begin{cases}\sum_{t=1}^{\z} z_t, & \z\in \Np=\{1,2,3,...\} \\
	0,                                                              &          \z=0 \\
	\sum_{t=1}^{-\z} z_t^*,                            &          \z\in \Nn=\{-1,-2,-3,...\},
\end{cases}
\eenr
wherein $z_t,z_t^*$ are independent copies of $\cP\big(-\xi^2,4\xi^2\si^2\big),$ which are also independent over all $t,$ for a distribution law $\cP$\footnote{If one assumes $\vep_{t}\sim^{i.i.d} \cN(0,\Si),$ then $\cP$ shall also be a normal distribution.} that is determined by the underlying distribution of the noise term in model (\ref{model:rvmcp}) (Condition A$'$). The notation in the arguments of $\cP(\cdotp,\cdotp)$ is representative of the mean and variance of this distribution. Finally, let,
\benr\label{eq:42}
\cC_{(\iny,j)}(\z)=\cC_{\iny}\big(\z,\xi_{(\iny,j)},\si_{(\iny,j)}^2\big)\qquad j=1,...,N,
\eenr
wherein $\si_{(\iny,j)}^2,$ $j=1,...,N,$ are asymptotic variance parameters as defined earlier in the context of the vanishing regime. Then, for any given $j=1,...,N,$ we establish
\benr\label{eq:limiting.d.non.vanishing}
(\tilde\tau_j-\tau^0_j)\Rightarrow \argmax_{\z\in \Z}\cC_{(\iny,j)}(\z),
\eenr
with $\Z$ denoting the set of integers. Quantiles of this distribution can be obtained numerically by simulating sample paths of the limit process.

\begin{remark}{\textnormal{
	The results (\ref{eq:loc.est.optimal}i), (\ref{eq:limiting.d.vanishing}) and (\ref{eq:limiting.d.non.vanishing}) share similarities to the inference procedures for debiased lasso (\cite{van2014asymptotically}) and othogonalized moment estimators (\cite{belloni2011inference}) for high dimensional regression parameters. Recall that these estimators also involve a refitting step that allows target estimates to achieve an optimal rate of estimation; however, the key distinction is that in those methods one needs to perform an additional debiasing step while refitting, in order to eliminate the statistical disturbance caused due to the interaction between the target and the potentially high dimensional nuisance estimates in the respective model. In contrast, for the problem at hand, a squared loss based refitting directly enables this property without the need for any debiasing step; the reason being that the disturbance caused by the interaction between a target change point estimate $(\tilde\tau_j)$ and the nuisance estimates ($\h\tau_{-j}$ and $\h\theta$) is of smaller order than the rate of estimation of $\tilde\tau_j$ itself.}}
\end{remark}

The limiting results (\ref{eq:limiting.d.vanishing}) and (\ref{eq:limiting.d.non.vanishing}) allow construction of asymptotically valid confidence intervals with any desired coverage $(1-\al)$ as,
\benr\label{def:comp.interval}
&\big[(\tilde\tau_j-ME_j^{\al}),\, (\tilde\tau_j+ME_j^{\al})\big],\quad\text{where,}\nn\\
& ME_j^{\al}=q_{\alpha}^v\si^2_{(\iny,j)}/\xi_j^2\quad\text{or}\quad ME_j^{\al}=q_{(\al,j)}^{nv},
\eenr
in the vanishing and non-vanishing regimes, respectively. The values $q_{\alpha}^v$ and $q_{(\al,j)}^{nv}$ represent quantiles at $(1-\al)$ coverage of the distributions of (\ref{eq:limiting.d.vanishing}) and (\ref{eq:limiting.d.non.vanishing}). These intervals shall guarantee a componentwise nominal coverage asymptotically at $(1-\al)$ for any given $j=1,...,N.$

The next result extends the component-wise coverage of intervals (\ref{def:comp.interval}) to simultaneous coverage over any finite subset $H$ of indices of change points, under the non-vanishing jump size regime. This is established by first obtaining the joint limiting distribution of the sub-vector $\tilde\tau_{H},$ whose consequence is that the components of $\tilde\tau_H$ are \textit{asymptotically pairwise independent}. Specifically,
consider any $H\subseteq\{1,...,N\}$ such that $|H|\le c_u.$ Then, under the non-vanishing regime $\xi_j\to \xi_{(\iny,j)},$ $0<\xi_{(\iny,j)}<\iny,$ $\forall\, j \in H,$ the following holds
\benr\label{eq:limiting.d.simultaneous.non.vanish}
(\tilde\tau_{H}-\tau^0_{H})\Rightarrow \Pi_{j\in H}\argmax_{\z_j\in\Z} \cC_{(\iny,j)}(\z_j).
\eenr
A few clarifications on notation used in (\ref{eq:limiting.d.simultaneous.non.vanish}) are in order. Here $\tilde\tau_H$ and $\tau^0_H$ represent subvectors of $\tilde\tau,$ $\tau^0,$ with entries corresponding to indices in $H.$  The product notation $\Pi_{j\in H}$ in the r.h.s of  (\ref{eq:limiting.d.simultaneous.non.vanish}) represents a joint distribution of  dimension $|H|,$ where the marginal distributions of components $j\in H$ are those of the multiplicands. Additionally the components $j\in H$ are pairwise independent. More generally, the two sided random walks $\cC_{(\iny,j)}(\cdotp)'$s of (\ref{eq:limiting.d.simultaneous.non.vanish}), are independent as stochastic processes, over components $j\in H.$  These results shall allow evaluation of asymptotic simultaneous coverage of confidence intervals under the considered jump size regime as,
\benr\label{eq:siml.intervals}
pr\Big(\tilde\tau_j-q_{(\al,j)}^{nv}\le \tau_j^0\le \tilde\tau_j+q_{(\al,j)}^{nv},\,\,\forall j\in H\Big)\to (1-\al)^{|H|},
\eenr
where $q_{(\al,j)}^{nv},$ $j\in H$ is defined in (\ref{def:comp.interval}). One may adjust componentwise significance level to $\al'=\big(1-(1-\al)^{1/|H|}\big),$ in order to obtain simultaneous coverage at any desired level $(1-\al).$

The above discussion concludes the main results of this article. We remind the reader that utilizing these results require preliminary estimates satisfying (\ref{eq:40}), a discussion on this together with the necessary required theoretical guarantees is provided in Section \ref{sec:feasible}.

\section{Theoretical results}\label{sec:main}\hfill

We provide sufficient conditions for the estimation and inference results for the locally refitted estimator $\tilde\tau$ of (\ref{est:optimalcp}). All assumptions are made on the first level model (\ref{model:rvmcp}), even though the estimator $\tilde\tau$ is based on the centered data obtained from the reparametrized model (\ref{model:center.rvmcp}). Additional technical issues caused by this transformation are also addressed.

\vspace{1.5mm}
{\it {{\noi{\bf Condition A (on distributions):}} The vectors $\vep_t=(\vep_{t1},...,\vep_{tp})^T,$ $t=1,..,T,$ are independent and identically distributed subexponential random vectors with variance proxy $\si^2<\iny$ (see, Definition \ref{def:sube} and \ref{def:submult} in Appendix \ref{app:auxiliary})}}

\vspace{1.5mm}
The posited class contains distributions exhibiting heavier tails than the Gaussian, and also includes discrete ones. Examples in the class include the Laplace, mean centered Exponential, mean centered Chi-square,  mean centered Bernoulli, mean centered Poisson distribution (for further details, see \cite{vershynin2019high}).

\vspace{1.5mm}
{\it {{\noi{\bf Condition B (on parameters):}} \\
		(i)\,{\bf (covariance)}\,\, The matrix $\Sigma:=E\vep_{t}\vep_{t}^T$ has bounded eigenvalues, i.e., $0<\ka^2\le\rm{min eigen}(\Si) <\rm{max eigen}(\Si)\le\phi^2<\iny,$ for constants $\ka^2,$ $\phi^2.$ \\~
		(ii)\,{\bf (existence and separation of jumps)}\,\, Assume there exists at least one change point $(N\ge 1)$; further, all $N$ change points are distinct and sufficiently separated, i.e., for $(\tau^0_{j}-\tau^0_{j-1})=T\ell_j,$ $j=1,...,N+1,$ we have  $\min_{1\le j\le N+1\}}T\ell_j\ge T\lm\ge 1,$ for a positive sequence $\lm\to 0,$ such that $\log T=o\big(\surd(T\lm)\big).$}\\~
	(iii)\,{\bf (sparsity of jumps)}\,\, Let $\eta^0_{(j)},$ $j=1,...,N$ be jump vectors as in (\ref{def:jumpsize}), so that $\max_{1\le j\le N}\|\eta^0_{(j)}\|_0\le s,$ wherein $s\ge 1$ is a positive sequence of integers.\\~
	(iv)\,{\bf (relative order of jump sizes)}\,\, For $\xim,$ $\xma$ as in (\ref{def:jumpsize}), let $\xma\le c_u\xim,$ for some  $c_u\ge 1.$}

\vspace{1.5mm}
All parts of Condition B are fairly standard in the literature.
The upper bound of Condition B(i) ensures finiteness of the asymptotic variances of the limiting processes defined in (\ref{eq:limiting.d.vanishing}) and (\ref{eq:limiting.d.non.vanishing}), while the lower bound plays a role in ensuring existence of the same distributions.  Condition B(ii) assumes existence of at least one change point and separation of all $N$ change points. In practice, existence of at least one change is usually established via boundary tests, such as that in (\cite{jirak2015uniform}); however, focusing on our objective of post-estimation inference we assume this existence apriori. Further conditions on the rate of the least separation sequence $\lm$ shall be placed later in the article. Condition B(iii) assumes sparsity of the jump vectors and as discussed in Section \ref{sec:tech.p}, we exploit this sparsity via the reparametrized model (\ref{model:center.rvmcp}) which transfers the assumed $s$-sparsity of Condition B(iii) to an $Ns$-sparsity on the individual means $\theta^0_{(j)}.$

Condition B(iv) assumes that all jump sizes of model (\ref{model:rvmcp}) are of the same order. Stated conversely, no particular jump dominates the others in the order of its magnitude. We require this assumption to ensure that for any fixed $j,$ the neighbors $\tau^0_{j-1}$ and $\tau_{j+1}^0$ do not interfere in the estimation of $\tau^0_j.$ Stronger versions of this assumption are also common in the literature and have been assumed in other ways such as by assuming jump sizes to be bounded below and above, e.g. \cite{fryzlewicz2014wild} and \cite{lu2018intelligent}.

Next, define sets of non-zero components associated with the mean parameters $\theta^0_{(j)}$ of (\ref{model:center.rvmcp}),
\benr\label{def:setS}
S_j=\big\{k\in\{1,...,p\};\,\,\theta^0_{(j)k}\ne 0\big\},\quad j=1,...,N+1.
\eenr
and let $S_j^c,$ $j=1,...N+1$ denote the complement sets. The earlier discussion in context of model (\ref{model:center.rvmcp}), yields,  $\max_{j}|S_j|\le Ns.$ Our analysis is agnostic on the choice of the estimators used to obtain the preliminary estimates $\h\tau$ and $\h\theta.$ Instead, we shall rely on the following assumption describing the sufficient conditions required for the validity of our results.


\vspace{1.5mm}
{\it {{\noi{\bf Condition C (for preliminary estimates $\h\tau,$ and $\h\theta$):}} Let $\pi_T\to 0$ be a positive sequence and assume that (i) and (ii) below hold with probability at least $1-\pi_T.$ \\~
		(i) ({\bf Preliminary change point estimate $\h\tau=(\h\tau_1,...,\h\tau_{\h N})$} of $\tau^0$): For an appropriately chosen small enough constant $c_{u1}>0,$ we assume that,
		\benr
		\h N=N,\qquad \max_{1\le j\le N}|\h\tau_j-\tau_j^0|\le c_{u1}T\lm, \nn
		\eenr
		wherein $\lm$ is the separation sequence defined in Condition B(ii).  \\~
		(ii)\,\, ({\bf Preliminary mean estimates of $\theta^0$ of (\ref{eq:trans.mean})}): Assume that the following two properties hold.\\~  	
		(a)\,\, The estimates $\h\theta_{(j)},$ $j=1,...,N+1,$ satisfy $\|(\h\theta_{(j)})_{S_j^c}\|_1\le 3\|(\h\theta_{(j)}-\theta_{(j)}^0)_{S_j}\|_1,$  with $S_j,$ $j=1,...,N+1$ being sets of non-zero components defined in (\ref{def:setS}). \\~
		(b)\,\, Assume there exists a sequence $r_T\ge 0,$ such that,
		\benr
		\max_{1\le j\le N+1}\|\h\theta_{(j)}-\theta_{(j)}^0\|_2\le r_T= \frac{c_{u1}\xim}{(Ns)^{1/2}\log (p\vee T)},\nn
		\eenr
		for a suitably small constant $c_{u1}>0,$ wherein $\xim$ is the least jump size defined (\ref{def:jumpsize}).\\~}}

\vspace{1.5mm}
Condition C is carefully constructed with the following two considerations in mind. First, it is stated in the weakest form that is sufficient for optimality of $\tilde\tau,$ and second that it is feasible. Specifically, Condition C(i) ensures that for any given $j=1,...,N,$ $\tau^0_j$ lies between the neighboring preliminary estimates $\h\tau_{j-1}$ and $\h\tau_{j+1},$ w.p. $1-o(1).$ Additionally, it ensures that the interval $(\h\tau_{j-1},\h\tau_{j+1})$ contains at most three change points $\tau^0_{j-1},\tau^0_{j},$ $\tau^0_{j+1}$ and no other ones, w.p. $1-o(1).$ These consequences are observable from the assumed $\ell_{\iny}$-rate in Condition C(i), together with Condition B(ii) which defines $T\lm$ as the least change point separation. A stronger version of Condition C(i) is met by existing estimation methods, see, e.g. \cite{wang2018high} under high dimensionality and \cite{harchaoui2010multiple} under fixed $p.$  further, we show in Section \ref{sec:feasible} that Condition C(ii) on mean estimates is satisfied as a consequence of this stronger version of Condition C(i) via regularized sample means. Thus, this condition shall effectively only require slower than optimal change points estimates.

\subsection{Rates of convergence}

Next, both a component-wise rate, as well as an $\ell_{\iny}$ rate of estimation for the change point locations are provided below. The consequences of these results become apparent in the sequel.

\begin{theorem}\label{thm:cpoptimal}(component-wise rate of estimation) Assume that Conditions A, B and C hold. Then, for any given $j=1,...,N,$ and any $0<a<1$ with $c_{a}\ge \surd{(1/a)},$ the following holds
	\benr
	|\tilde\tau_j-\tau_j^0|\le c_uc_a^2\si^2\xi_j^{-2}\nn
	\eenr
	with probability at least $1-2a-o(1)-\pi_T.$ Equivalently, $\si^{-2}\xi_j^{2}(\tilde\tau_j-\tau_j^0)=O_p(1),$ for any given $j=1,...,N.$
\end{theorem}

Theorem \ref{thm:cpoptimal} provides componentwise rates of convergence of $\tilde\tau_j,$ $j=1,...,N,$ that are optimal (see, e.g., Proposition 3 of Supplement of \cite{wang2018high}). This is the key result that allows the feasibility of performing inference on the change point parameters. It is the same rate, to the one obtained if perfect knowledge about the nuisance parameters $\theta^0$ and $\tau_{-j}^0$ (all other change points) was available. It is an instance of the adaptation property, as described in \cite{bickel1982adaptive}, in the presence of a diverging number of change points and underlying high dimensionality. The next result establishes an $\ell_{\iny}$ rate of convergence of proposed refitted estimates.

\begin{theorem}\label{thm:cpoptimal.simul}($\ell_{\iny}$ rate of estimation) Assume that Conditions A, B and C hold. Then, the following holds
	\benr
	\max_{1\le j\le N}|\tilde\tau_j-\tau_j^0|\le c_u\si^2\xim^{-2}\log^2 T,\nn
	\eenr
	with probability at least $1-o(1)-\pi_T.$
\end{theorem}

Note that Theorem \ref{thm:cpoptimal} is crucial for inference purposes, while Theorem \ref{thm:cpoptimal.simul} is most relevant for characterizing the behavior of all $\tilde\tau_j,$ $j=1,...,N,$ simultaneously. We also note that the local refitting undertaken in (\ref{est:optimalcp}) does not alter the number of change points $\h N$ of the preliminary estimate $\h\tau,$ thus, Condition C(i) ensures $\tilde N=\h N= N,$ w.p. $\to 1.$

\begin{remark}\label{rem:cost.subE} {\textnormal{(The cost of subexponential errors)
	The $\log^2T$ term in the $\ell_{\iny}$ bound is a consequence of that assumption.
	In case of subgaussian errors, the sharper rate $\max_{1\le j\le N}|\tilde\tau_j-\tau_j^0|\le c_u\si^2\xim^{-2}\log T,$ w.p. $1-o(1),$ would be obtained, due to the availability of sharper tail bounds on residual error terms (\cite{kaul2020inference}). Nevertheless, this rate is the sharpest available in the literature under high dimensionality. For example, it is at least $\log p/\log T$ faster than that obtained in \cite{wang2018high} for their corresponding estimator. Further, heuristic comparisons to $\ell_\iny$ rates under other high dimensional models exhibiting change points -linear regression or covariance- the same observation holds ; see, e.g, \cite{rinaldo2021localizing, wang2021optimal}.} }
\end{remark}

\subsection{Limiting distributions}

Next, we obtain component-wise and joint limiting distributions for the refitted change point estimators. We start by positing a few additional assumptions.

\vspace{1.5mm}
{\it {{\noi{\bf Condition D (stability of asymptotic variances):}} For jump sizes $\xi_j,$ $j=1,...,N$ as defined in (\ref{def:jumpsize}) and covariance $\Si$ as in Condition B(i), assume the following limits exists,
		\benr
		\xi_{j}^{-2}\big(\eta^{0T}_{(j)}\Si\eta^0_{(j)}\big)\to \si^2_{(\iny,j)},\quad 0<\si^2_{(\iny,j)}<\iny,\quad \text{for each given}  \ \ j=1,...,N.
		\eenr
}}

\vspace{1.5mm}
Recall that all limits in this work are with respect to the observation period $T.$ The limits in Condition D are acting in $T$ via the dimension $p$ and the jump sizes $\xi_j,$ $j=1,...,N.$ The quantities $\si^2_{(\iny,j)}$ $j=1,...,N,$ serve as variance parameters of the limiting processes described in (\ref{eq:limiting.d.vanishing}) and (\ref{eq:limiting.d.non.vanishing}), thus the need for their stability.
Note that finiteness of these limits is already guaranteed by Condition B(i), while the current condition only assumes their stability. To see this, observe that the assumed convergence is on a sequence that is guaranteed to be bounded below and above, i.e.,
\benr
0<\ka^2\le \min_{1\le j\le N}\xi_{j}^{-2}\big(\eta^{0T}_{(j)}\Si\eta^0_{(j)}\big)<\max_{1\le j\le N}\xi_{j}^{-2}\big(\eta^{0T}_{(j)}\Si\eta^0_{(j)}\big)\le \phi^2<\iny.
\eenr
The above inequalities follow from the bounded eigenvalues assumption on $\Si$ \big(Condition B(i)\big). An easier to interpret, but stronger sufficient condition for the finiteness of these limits is by assuming absolute summability of each row or column of $\Si.$ This condition is satisfied by large classes of covariances such as banded and toeplitz type matrices. We refer to Condition D of \cite{kaul2021graphical} for further details on this argument.

\vspace{1.5mm}
{\it {{\noi{\bf Condition E (on rate of convergence of jump size):}} Let $\xim$ and $\lm$ be as defined in (\ref{def:jumpsize}) and Condition B(ii), respectively. Then, we assume that $\xim^{-1}\log T=o\big(\surd{(T\lm)}\big).$
}}

\vspace{1.5mm}
Recall from Section \ref{sec:tech.p} that our results allow potentially diminishing jump sizes. Condition E is the first requirement imposed on the rate at which the least jump size ($\xim$) can potentially converge to zero. It requires this rate to be at least $\log T$ slower than $1\big/\surd{(T\lm)}.$ 
Note that this assumption seems fairly weak for a high dimensional problem,
given the minimax result in \cite{liu2019minimax}. However, additional stronger restrictions on $\xim$ are needed and presented in Section \ref{sec:feasible}. In the interim, the burden of these additional rate restrictions have been transferred to Condition C on the preliminary estimates in a guise that shall become apparent in Section \ref{sec:feasible}.

\begin{remark}\label{rem:degenerate}{\textnormal{(Degenerate distributions under diverging jump sizes) It may be observed that as a direct consequence of Theorem \ref{thm:cpoptimal}, for any given $j=1,..,N,$ when $\xi_j\to\iny,$ we have $pr(\tilde\tau_j=\tau^0_j)\to 1,$ i.e., the limiting distribution of the $j^{th}$ component is degenerate when the corresponding jump size is diverging. This result extends to the case of any finite subset $H\subseteq\{1,...,N\},$ i.e. $pr(\tilde\tau_j=\tau^0_j,\,\,\forall j\in H)\to 1,$ if $\xi_j\to \iny,$ $j\in H.$ Furthermore, if one assumes a faster divergence $(\xim\big/\log T)\to \iny,$ then Theorem \ref{thm:cpoptimal.simul} yields $pr(\tilde\tau_j=\tau^0_j,\,\,\forall j)\to 1.$ Thus, in the following we restrict our analysis to $\xim\le c_u,$ where limiting distributions of $\tilde\tau$ are non-trivial. This case is further subdivided into two distinct regimes as described in what follows.}}
\end{remark}

\begin{theorem}\label{thm:wc.vanishing} (component-wise distributions for the vanishing regime) Assume that Conditions A, B, D and E hold. Consider any given change point $j=1,...,N,$ and assume that the jump size $\xi_j\to 0$ is vanishing, and that $\tau^0_{-j},$ $\theta^0$ are known. Denote  $\tilde\tau^*_j=\tilde\tau_j(\tau^0_{-j},\theta^0).$ Then, 
	\benr\label{eq:wc.vanishing}
	\xi^{2}_j(\tilde\tau^*_j-\tau^0_j)\Rightarrow \argmax_{\z\in\R}\big\{2\si_{(\iny,j)}W_j(\z)-|\z|\},
	\eenr
	where $W_j(\z)$ is a two sided standard Brownian motion.  Alternatively, when $\tau^0_{-j}$ and $\theta^0$ are unknown, let $\tilde\tau_j$ be as defined in (\ref{est:optimalcp}) and assume $\h\tau_{-j}$ and $\h\theta$ satisfy Condition C. Further, assume that the sequence $r_T$ in Condition C(ii) satisfies $r_T=\{o(1)\xim\}\big/\{{(Ns)^{1/2}\log (p\vee T)}\}.$ Then, the convergence (\ref{eq:wc.vanishing}) also holds when $\tilde\tau^*_j$ is replaced with $\tilde\tau_j.$
\end{theorem}

The limiting distributions of $\tilde\tau_j$ $j=1,...,N,$ can be used to construct asymptotically valid component-wise confidence intervals for the locations of the change points under the assumed vanishing jump regime. Observe that a change of variable to $\z=\si_{(\iny,j)}^2\z',$ yields $\argmax_{\z\in\R}\big\{2\si_{(\iny,j)}W_j(\z)-|\z|\}=^d\si_{(\iny,j)}^2\argmax_{\z'\in\R}\big\{2W_j(\z')-|\z'|\},$ which in turn yields the relations in (\ref{eq:limiting.d.vanishing}) provided in Section \ref{sec:tech.p}. 

Next, we consider the non-vanishing regime for $\xi_j\to \xi_{(\iny,j)},$ $0<\xi_{(\iny,j)}<\iny.$ For this purpose, we require the following additional distributional assumption.

\vspace{1.5mm}
{\it {{\noi{\bf Condition A$'$ (additional distributional assumptions):}} Suppose Condition A, B(i) and D hold and additionally assume for any given $j=1,...,N$ and any constants $c_1,c_2\in\R,$ the r.v.'s $c_1+c_2\vep^T_t\eta^0_{(j)}\Rightarrow\cP\big(c_1,c_2^2\xi_{(\iny,j)}^2\si^2_{(\iny,j)}\big),$ for  $t=1,...,T,$ for some  distribution law $\cP,$ which is continuous and supported in $\R.$ }}

\vspace{1.5mm}
As in Condition D, the limits here are acting in $T$ via the dimension $p$ and the jump sizes $\xi_j.$  The only additional requirement in Condition A$'$ is that the random variables under consideration are continuously distributed (recall Condition A may allow discrete distributions). If one assumes an underlying Gaussian distribution, then Condition A$'$ is redundant, i.e., $\vep_t\sim\cN(0,\Si),$ then $\cP\big(c_1,c_2^2\xi_{(\iny,j)}^2\si^2_{(\iny,j)}\big)\sim \cN\big(c_1,c_2^2\xi_{(\iny,j)}^2\si^2_{(\iny,j)}\big),$ which follows directly. More generally, we note that the variance expression in $\cP\big(c_1,c_2^2\xi_{(\iny,j)}^2\si^2_{(\iny,j)}\big)$ follows from Condition D together with the jump size regime assumption of $\xi_j\to\xi_{(\iny,j)},$ the expression for its mean follows trivially. Consequently, the limiting distribution of the sequence $c_1+c_2\vep^T_t\eta^0_{(j)}$ is well defined, i.e. supported in $\R.$ Thus, Condition A$'$ simply reflects notation for the underlying distribution $\cP$ and the representation $\cP(\mu,\si^2),$ with $E\cP(\mu,\si^2)=\mu,$ and $\textrm {var}\big(\cP(\mu,\si^2)\big)=\si^2,$ is only for ease of presentation and does not imply that $\cP$ is characterized only by its mean and variance.

The two-sided random walks $\cC_{(\iny,j)}(\z)$ of (\ref{eq:42}) can now be used to characterize the limiting distributions of $\tilde\tau_j$ for $j=1,...,N,$ in the current non-vanishing regime. The only additional requirement of Condition A$',$ of continuity of the distribution $\cP$ is assumed for the regularity of the \textit{argmax} of these two sided negative drift random walks.

\begin{theorem}\label{thm:wc.non.vanishing} (componentwise distributions for the  non-vanishing regime) Suppose Conditions A, B, D and E hold. Consider any given $j=1,...,N,$ and assume that the jump size is non-vanishing, $\xi_j\to \xi_{(\iny, j)},$ $0<\xi_{(\iny, j)}<\iny,$ and that $\tau^0_{-j},$ $\theta^0$ are known. Let $\tilde\tau^*_j=\tilde\tau_j(\tau^0_{-j},\theta^0),$ then, we have,
	\benr\label{eq:wc.non.vanishing}
	(\tilde\tau^*_j-\tau^0_j)\Rightarrow \argmax_{\z\in \Z}\cC_{(\iny,j)}(\z),
	\eenr
	where $\cC_{(\iny,j)}(\z)$ is  defined in (\ref{eq:42}). Alternatively, when $\tau^0_{-j}$ and $\theta^0$ are unknown, let $\tilde\tau_j$ be as defined in (\ref{est:optimalcp}) assume $\h\tau_{-j}$ and $\h\theta$ satisfy Condition C. Additionally assume sequence $r_T$ of Condition C(ii) satisfies $r_T=\{o(1)\xim\}\big/\{{(Ns)^{1/2}\log (p\vee T)}\}.$ Then, the convergence (\ref{eq:wc.non.vanishing}) also holds when $\tilde\tau^*_j$ is replaced with $\tilde\tau_j.$
\end{theorem}

The sole distinction in assumptions of Theorem \ref{thm:wc.vanishing} and Theorem \ref{thm:wc.non.vanishing} is the change of regime from vanishing to non-vanishing jump size. Since the analytical form of $\argmax_{\z\in \Z}\cC_{(\iny,j)}(\z)$ is unavailable, one can obtain the quantiles of these distributions by simulating sample paths of the two sided random walks under consideration.

Theorem \ref{thm:wc.vanishing} and Theorem \ref{thm:wc.non.vanishing} both yield component-wise control for the asymptotic coverage of the corresponding confidence intervals in their respective regime. The following result provides the joint limiting distribution of any finite subset of the proposed change point estimates, under the non-vanishing jump size regime.

\begin{theorem}\label{thm:wc.joint} (joint distributions for the non-vanishing regime\footnotemark) Suppose Condition A$'$, B, C, D and E hold and assume $r_T$ of Condition C(ii) satisfies $r_T=\{o(1)\xim\}\big/\{{(Ns)^{1/2}\log (p\vee T)}\}.$ Let $H\subseteq\{1,...,N\}$ be any finite subset of change point indices and $\tilde\tau_H=\tilde\tau_H(\h\tau_{-H},\h\theta)$ be a subvector of change point estimates as defined in (\ref{est:optimalcp}). Additionally assume the jump size regime is non-vanishing, i.e., $\xi_j\to\xi_{(\iny,j)},$ $0<\xi_{(\iny,j)}<\iny,$ $\forall j\in H,$ then, we have,
	\benr\label{eq:27}
	(\tilde\tau_H-\tau^0_H)&\Rightarrow& \argmax_{\z\in\Z^{|H|}}\sum_{j\in H}\cC_{(\iny,j)}(\z_j),
	\eenr
	where increments $z_{tj}$ and $z_{tj}^*$ of $\cC_{(\iny,j)}(\z_j),$ are pairwise independent for all combinations of each other, over $t$ as well as over $j\in H.$ Moreover, the convergence (\ref{eq:27}) is equivalent to,
	\benr\label{eq:28}
	(\tilde\tau_H-\tau^0_H)&\Rightarrow& \Pi_{j\in H}\argmax_{\z_j\in\Z}\cC_{(\iny,j)}(\z_j),
	\eenr
	Consequently, $\tilde\tau_j$ are also asymptotically independent over $j\in H.$
\end{theorem}
\footnotetext{see, discussion after (\ref{eq:limiting.d.simultaneous.non.vanish}) for clarifications on notations used in this theorem}

Theorem \ref{thm:wc.joint} provides two equivalent ways of constructing simultaneously valid confidence intervals over the set $H.$ The convergence in (\ref{eq:27}) to the $|H|$ dimensional maximizer of the  $\Z^{|H|}\to \R$ random field $\sum_{j\in H}\cC_{(\iny,j)}(\z_j),$ provides the first such approach. However, doing so requires obtaining Monte Carlo approximations of the quantiles from this distribution. This can be fairly computationally intensive depending on the cardinality $|H|.$ From a practical perspective, the more important finding of this result is that the increments $z_{tj}$ and $z_{tj}^*$ of $\cC_{(\iny,j)}(\z_j),$ are independent over $j\in H.$ This allows the equivalent representation of (\ref{eq:28}), which in turn yields asymptotic independence of $\tilde\tau_j$ over $j\in H.$ The latter justifies computing component-wise intervals with an adjusted component-wise coverage that maintains the  simultaneous nominal coverage as described in (\ref{eq:siml.intervals}).

The next Section resolves the following two issues: (i) the availability of preliminary estimates $\h\tau$ and $\h\theta$ satisfying Condition C, and (ii) positing explicit restrictions on the rate of divergence of the model dimensions ($s,p$).

\section{Construction of feasible change point estimators}\label{sec:feasible}

We start by constructing estimates for the mean parameters. For any $\tau=(\tau_1,...,\tau_N)^T\in\{1,...,(T-1)\}^N$ satisfying $\tau_{j-1}<\tau_{j},$ $j=1,...,N+1,$ let,
\benr\label{def:empmeans}
\bar x_{(j)}(\tau)=\frac{1}{(\tau_{j}-\tau_{j-1})}\sum_{t=\tau_{j-1}+1}^{\tau_j}x_t,\quad j=1,....,N+1,
\eenr
be piece-wise means evaluated on the partitioning of $\{1,...,T\}$ induced by $\tau.$ Next, consider the soft-thresholding operator, $k_{\la}(x)={\rm sign}(x)(|x|-\la)_{+},$ $\la>0,$ $x\in\R^p,$ wherein ${\rm sign}(\cdotp),$ $|\cdotp|,$ and $(\cdotp)_{+}$\footnote{For $x\in \R,$ $(x)_{+}=x,$ if $x\ge 0,$ and $x=0$ if $x<0.$} are applied component-wise. Then, for any $\la_j>0,$ define $\ell_1$ regularized mean estimates,
\benr\label{est:softthresh}
\h\theta_{(j)}(\tau)=k_{\la_j}\big(x_{(j)}(\tau)\big),\quad j=1,...,N+1.
\eenr
It is well known (\cite{donoho1995noising}, \cite{donoho1995wavelet}) that the soft-thresholding operator (\ref{est:softthresh}) is equivalent to the $\ell_1$ regularization,
\benr\label{est:softL1construction}
\h\theta_{(j)}(\tau)&=&\argmin_{\theta\in\R^p}\big\|\bar x_{(j)}(\tau)-\theta\big\|^2_2+\la_j\|\theta\|_1,\quad \la_j>0,\quad j=1,...,N+1.
\eenr
Note again that these soft-thresholded means are evaluated on the transformed $x_t,$ $t=1,...,T$ of (\ref{model:center.rvmcp}) and recall that the associated mean parameters of that model are $Ns$-sparse. Next, we establish that $\h\theta(\tau)$ evaluated with a plug-in change point estimate whose rate is slower than the optimal, still satisfies all requirements in Condition C(ii) being satisfied, provided some additional rate conditions on model parameters hold, and stated next.

\vspace{1.5mm}
{\it {{\noi{\bf Condition E$'$ (additional condition on rates of model parameters):}} Assume one of the following three conditions written sequentially in order of strength.
		\benr
		&(i)&\,\, \Big(\frac{\si}{\xim}\Big)\Big\{\frac{Ns\log^2(p\vee T)}{T\lm}\Big\}^{\frac{1}{2}}\le c_{u1},\quad  (ii)\,\, \Big(\frac{\si}{\xim}\Big)\Big\{\frac{Ns\log^{3/2}(p\vee T)}{\surd(T\lm)}\Big\}\le c_{u1},\nn\\
		&(iii)&\,\, \Big(\frac{\si}{\xim}\Big)\Big\{\frac{Ns\log^{3/2}(p\vee T)}{\surd(T\lm)}\Big\}=o(1),
		\eenr
		wherein $c_{u1}>0$ is a suitably chosen small constant.}}

\begin{remark} {\textnormal{
	The rate restrictions in Condition E$'$ are progressively stronger and viewed together with the following theorem provide important insights on the parametric rate requirements needed for both estimation and inference for the change point parameters. This discussion is provided immediately following the next result.}}
\end{remark}

\begin{theorem}\label{thm:mean.rate} Suppose Condition A, B and E$'$(i) hold and let $\h\tau=(\h\tau_1,...,\h\tau_{\h N})^T$ be a preliminary change point estimate satisfying,
	\benr\label{eq:29}
	\h N= N\quad{\rm and}\quad\max_{1\le j\le N}|\h\tau_j-\tau^0_j|\le  c_{u}\si^2\xim^{-2}Ns\log^2(p\vee T),
	\eenr
	for some $c_u>0,$ with probability at least $1-\pi_T.$ Then, this preliminary estimate $\h\tau$ satisfies Condition C(i). Let $\psi=\max_j\|\eta^0_{(j)}\|_{\iny}$ and assume $\psi/\xim=O(1).$ Further,  assume the tighter rate restriction of Condition E$'$(ii) and that $T\lm\ge \log (p\vee T).$ Then, upon choosing $\la_j=\la=c_u\si\big\{\log(p\vee T)\big/T\big\}^{1/2},$ $j=1,...,N+1,$ the mean estimates $\h\theta_{(j)}(\h\tau)$ satisfy Condition C(iia) and the bound,
	\benr\label{eq:32}
	\max_{1\le j\le N+1}\|\h\theta_{(j)}-\theta_{(j)}^0\|_2\le c_u\si\Big\{\frac{Ns\log (p\vee T)}{T\lm}\Big\}^{\frac{1}{2}},
	\eenr
	with probability at least $1-o(1)-\pi_T.$ Consequently, $\h\theta_{(j)}(\h\tau),$ $j=1,...,N+1$ satisfy all requirements of Condition C(ii).
\end{theorem}

From a practical perspective, Theorem \ref{thm:mean.rate} shows that the only requirement for the main results of Section \ref{sec:main} to hold is solely the availability of preliminary near optimal estimates $\h\tau$ satisfying (\ref{eq:29}), which in turn yield estimates $\h\tau$ and $\h\theta$ that satisfy all requirements of Condition C. An example from the literature that can be used to obtain $\h\tau$ is provided in Remark \ref{rem:method.example}.

Theorem \ref{thm:mean.rate} together with Condition E$'$ provide some interesting and possibly surprising insights. First, restrictions similar to E$'$(i) are commonly thought of in the literature as necessary for change point estimation. However, the results of Section \ref{sec:main} and Theorem \ref{thm:mean.rate} illustrate a further subtlety. Such rate conditions arise instead from the need to estimate nuisance parameters and not the estimation of a target change point itself, for e.g., Theorem \ref{thm:cpoptimal} yields that if $\tau^0_{-j}$ and $\theta^0$ are known, then one can estimate $\tau_j^0$ at an optimal rate, even if $p$ is diverging arbitrarily fast w.r.t. $T.$ 

The second and more consequential observation is regarding E$'$(ii).
Existing literature typically assumes restrictions similar to Condition E$'$(i), which leads to only near optimal rates. The tighter Condition E$'$(ii) aids in obtaining an optimal estimation rate for the change points.

It is worth noting that heuristic comparisons to double machine learning methods, such as the debiased lasso or orthogonalized moment estimators for inference in high dimensional regression settings can also be made here.  Condition E$'$(ii) is analogous to the super-sparse assumption imposed in these methods that yields optimal rates of estimation and then inference, with the distinction that in the present setting it also involves other parameters $N, \xim,$ and $\lm$ that arise by necessity of the change point model under consideration. Finally, note that
Condition E$'$ is sufficient, while its necessity, in accordance with equivalent conditions required in  double machine learning methods, remains unknown.

Condition E$'$(iii) is used in Corollary \ref{cor:alg1.validity} below and the additional rate tightening from $O(1)$ to $o(1)$ in comparison to E$'$(ii) ensures existence of limiting distributions. The requirement of this condition can also be observed directly in Theorem's \ref{thm:wc.vanishing},  \ref{thm:wc.non.vanishing} and \ref{thm:wc.joint}, since they require sequence $r_T$ of Condition C to satisfy $r_T=\{o(1)\xim\}\big/\{{(Ns)^{1/2}\log (p\vee T)}\},$ as opposed to $r_T=\{c_{u1}\xim\}\big/\{{(Ns)^{1/2}\log (p\vee T)}\}$ for Theorem's \ref{thm:cpoptimal} and \ref{thm:cpoptimal.simul}. This slight tightening to obtain limiting distributions is also in accordance with classical results in a fixed $p$ and single change point setting, wherein the conditions reduce to a relationship between $\xim,\lm$ and $T,$ see, e.g.,  \cite{bai1994}.

Finally, as noted earlier, an additional price is paid for subexponential distributions of the noise terms. In particular, due to sharper available tail bounds for sub-Gaussian distribution, one can obtain analogous results by requiring $\log (p\vee T)$ in E(i), E(ii) and E(iii), respectively. 

Algorithm 1 presents all necessary steps for a feasible implementation of the proposed methodology, while Corollary \ref{cor:alg1.validity} summarizes estimation and inferential properties of the resulting change point estimate.

\begin{figure}[H]
	\noi\rule{\textwidth}{0.5pt}
	
	\vspace{-2mm}
	\flushleft {\bf Algorithm 1:} Locally refitted estimation of $\tau^0=(\tau^0_1,...,\tau^0_N)^T.$
	
	\vspace{-1.25mm}
	\noi\rule{\textwidth}{0.5pt}
	
	\vspace{-1.25mm}
	\flushleft{\bf Step 1:} Implement any estimator $\h\tau=\big(\h\tau_1,....,\h\tau_{\h N}\big)^T$ from the literature that satisfies the near optimal bounds (\ref{eq:29}), with probability $1-o(1).$
	
	\vspace{-1.25mm}
	\flushleft{\bf Step 2:} Compute mean estimates $\h\theta_{(j)}(\h\tau),$ $j=1,...,\h N+1,$ and obtain locally refitted change point estimates,
	
	\vspace{-3.25mm}
	\benr
	\tilde\tau_j=\argmin_{\h\tau_{j-1}<\tau_j<\h\tau_{j+1}}Q_j\big(\tau_j,\h\tau_{-j},\h\theta\big),\qquad j=1,...,\h N,\nn
	\eenr
	
	\vspace{-3.25mm}
	\flushleft{\bf (Output):}\,\, $\tilde\tau=\big(\tilde\tau_1,....,\tilde\tau_{\h N}\big)^T.$
	
	\vspace{-1.25mm}
	\noi\rule{\textwidth}{0.5pt}
\end{figure}

\begin{corollary}\label{cor:alg1.validity} Assume that Conditions A, B and E$'$(ii) hold, together with $T\lm\ge \log (p\vee T)$ and that $\psi/\xim=O(1).$  Then, $\tilde\tau$ of Algorithm 1 satisfies the component-wise and $\ell_{\iny}$ bounds of Theorems \ref{thm:cpoptimal} and \ref{thm:cpoptimal.simul}, respectively. If in addition 
	Conditions D and E$'$(iii) hold, $\tilde\tau$ satisfies the component-wise limiting distribution in Theorem \ref{thm:wc.vanishing}, under a vanishing jump size, while if Condition A$'$ holds,  $\tilde\tau$ satisfies the component-wise limiting distribution in Theorem \ref{thm:wc.non.vanishing}, under a non-vanishing jump size. Finally, it satisfies the joint limiting distributions of Theorem \ref{thm:wc.joint} under the same non-vanishing regime and for any finite subset $H$ of the change point indices.
\end{corollary}

The additional assumption made in Theorem \ref{thm:mean.rate} and Corollary \ref{cor:alg1.validity} that has thus far not been discussed is $\psi/\xim=O(1).$ This assumption is very similar to Condition B(iv). It is also restricting the relative order of the maximum and minimum jump size, with the distinction that the former is evaluated in the sup-norm instead of the $\ell_2$ norm. The reasoning is identical as before, i.e., if a target jump is `too small w.r.t a neighboring jump, then mean estimation across the target jump may not be precise. Further insight to relationship with Condition B(iv) can be obtained by the inequality $(\xma/\xim)\le \surd s(\psi/\xim),$ thus if one assumes a further stronger condition of $\surd s (\psi/\xim)=O(1),$ then both conditions of interest are satisfied.

We conclude with the following important remarks. The first provides an example of a method that provides the preliminary estimate $\h\tau,$ and the associated theoretical considerations.

\begin{remark}\label{rem:method.example} {\textnormal{To our knowledge, for the high dimensional mean shift model, the projected CUSUM estimator of \cite{wang2018high} provides the thus far sharpest $\ell_{\iny}$ rate of estimation available in the literature. Under similar assumptions as made here, Theorem 2 in that paper establishes
	\benr\label{eq:39}
	\h N= N\quad{\rm and}\quad\max_{1\le j\le N}|\h\tau_j-\tau^0_j|\le  c_{u}\si^2\xim^{-2}\lm^{-4}\log(p\vee T),
	\eenr
	w.p. at least $1-o(1),$ for their proposed estimator. Under a fixed number of change points, $N\le c_u$ and $\lm\ge c_u,$ this estimator satisfies the requirement (\ref{eq:29}) of the preliminary estimate for our methodology. When $N$ is diverging, (\ref{eq:29}) holds under the relation $\ell^{-4}\le Ns.$ Consequently, this estimator can serve as a theoretically valid preliminary estimate for Algorithm 1. We note that in that work, a Gaussian assumption on the noise terms is required, which is not needed for our results. Development of a near optimal estimator that is able to yield (\ref{eq:29}) under the weaker distributional assumptions made in this article remains a further question left to future work. The article \cite{kaul2020inference} provides an estimation method under these weaker conditions, but is limited to a single change point.}}
\end{remark}

\begin{remark}\label{rem:subseq.change} {\textnormal{(Relaxing (\ref{eq:29}) and Condition C for preliminary change point estimates) From a practical perspective, the weakest link holding up our inference results is the selection consistency $\h N=N,$ w.p. $\to 1,$ required of the preliminary change point estimates. While this consistency is typically theoretically guaranteed by all near optimal procedures proposed in the literature (in both finite and high dimensional frameworks), it is often observed to be violated due to a variety of possible reasons; for example, data adaptive tuning parameter choices may not align with the required theoretical choices which are infeasible to implement in practice. This observation has also recently been made in \cite{romano2021detecting} under additional model relaxations such as dependence amongst errors.
	In view of this observation, we provide the following relaxation of (\ref{eq:29}) which in turn provides relaxations to the assumed Condition C on preliminary estimates. In place of (\ref{eq:29}) one may instead assume,
	\benr\label{eq:45}
	&&\h N\ge N,\quad{\textrm{and there exists a subsequence}}\,\,\{j_1,...,j_N\}\subseteq\{1,...,\h N\}\,\,\textrm{such that,}\nn\\
	&(i)& \max_{1\le m\le N} |\h\tau_{j_m}-\tau^0_m|\le c_{u}\si^2\xi^{-2}Ns\log^2(p\vee T),\quad{\rm and}\nn\\
	&(ii)&\min_{1\le j\le (\h N+1)}|\h\tau_j-\h\tau_{j-1}|\ge c_u T\lm,\quad{\textrm{for some}\,\,c_u>0},
	\eenr
	with probability at least $1-o(1).$	Then, using the same structure of arguments developed in the Supplement, it can be shown that all results of Section \ref{sec:main} remain valid on the subsequence $\{j_1,...,j_N\},$ i.e., validity of component-wise limiting distributions as well as joint distributions over any finite subset of this subsequence. This relaxation to Condition C allows the possibility of spurious change points detected by the preliminary estimation method in addition to the true set of change points. It is clearly infeasible for (\ref{eq:45}ii) to hold consistently (w.p. $1-o(1)$) since it requires a separation between potentially spurious change points in the estimated vector. Instead, such conditions on minimum separation of estimated change points are typically forced directly into optimization problems, which can alternatively supply validity of (\ref{eq:45}ii). Even with this relaxation, the end result is
	that $N$ of the $\h N$ intervals obtained are asymptotically valid, but does not specify which are valid and which are spurious. This would be an important problem to address in future work, since it would significantly enhance the applicability of the inference results established in this work. However, this topic 
	is disconnected from the inferential objectives of this article and is thus outside its scope.}}
\end{remark}

\section{Numerical experiments}\label{sec:numerical}

Next, we illustrate Algorithm 1 and the results developed in Section \ref{sec:main} and summarized in Corollary \ref{cor:alg1.validity}. For numerical experiments, data are generated as per model (\ref{model:rvmcp}) and as required by the proposed methodology, estimation and inference procedures are carried out after a centering operation leading to model (\ref{model:center.rvmcp}). 
The mean vectors are set as $\theta_{(1)}^*=\big(1_{s\times 1}^T,0...,0\big)^T_{p\times 1},$ $\theta_{(2)}^*=\big(0_{s\times1}^T,1_{s\times 1}^T,0...,0\big)^T_{p\times 1},$ and  $\theta_{(3)}^*=\big(0_{2s\times1}^T,1_{s\times 1}^T,0...,0\big)^T_{p\times 1}.$ These vectors are repeated iteratively depending on $N,$ i.e., $\theta_{(j)}^*=\theta^*_{(1)},$ $\theta_{(j)}^*=\theta^*_{(2)},$ or $\theta_{(j)}^*=\theta^*_{(3)},$ if $j \equiv 1 \pmod{3},$ $j \equiv 2 \pmod{3},$ or $j \equiv 0 \pmod{3},$ respectively. The matrix $\Si$ is chosen to be Toeplitz type $\Si_{ij}=\rho^{|i-j|},$ $i,j=1,...,p$ with $\rho=0.5.$ We consider all combinations of  $T\in\{450,600,750\},$ $p\in\{50,200,350,500\},$ and $N\in\{2,4\}.$ The locations of change points $\tau^0_j,$ $j=1,...,N,$ are set to $N$ evenly spaced values, e.g., when $T=300,$ $N=2,$ we have $\tau^0=(100,200)^T.$  We consider both subgaussian and subexponential noise, specifically, for Scenario A and B below $\vep_{t}\in\R^p,$ are generated as i.i.d. zero mean Gaussian r.v.'s, i.e., we set $\vep_{t}\sim^{i.i.d} \cN(0,\Si),$ $t=1,...,T.$ For Scenario A$'$ and B$'$ we generate noise as $\vep_t=\Si^{\frac{1}{2}}w_t,$ $t=1,...,T,$ where $w_t=(w_{t1},....,w_{tp})^T,$ and each component $w_{tj}^*\sim^{i.i.d} {\rm Laplace}(0,1),$ $j=1,...,p,$ with zero mean and unit variance. This yields i.i.d random variables $\vep_t,$ $t=1,..,T$ which are subexponential random vectors with a covariance $\Si$ amongst components. The remaining specifications are provided in the following.

\vspace{1.5mm}
\noi{\bf Scenario A and A$'$} reflects an idealized setting. Algorithm 1 is implemented with the true  $\tau^0=(\tau^0_1,...,\tau^0_N)^T$ supplied to its Step 1. Then, Step 2 carries out mean estimation and local refitting as required by the method. Although not useful in practice, it nevertheless serves the following two purposes. It provides evidence towards a numerical proof of principle of inference results supporting Algorithm 1 and also serves as a benchmark for the realistic Scenario B.

\vspace{1.5mm}
\noi{\bf Scenario B and B$'$}, wherein all parameters need to be estimated. Step 1 of Algorithm 1 is carried out by one of two methods. First, the projected CUSUM estimator of \cite{wang2018high} (referred to as WS below) which is implemented in the \texttt{R}-package \texttt{InspectChangepoint}. 
The underlying tuning parameters of this preliminary estimator are chosen as the default values implemented in the package. We refer to this as WS+LR, wherein LR refers to local refitting.

We also implement Step 1 with a second method and some heuristics. The $\ell_0$ regularized near optimal estimator of \cite{kaul2020inference} (Remark 4.2 in that paper, referred to as KFJS below) designed for a single change point is extended to multiple change points via binary segmentation, i.e., recursive application of the method, repeated until no further change points are detected. This is described as Algorithm 3 in Appendix \ref{app:numerical.supplement}. Tuning parameters are chosen via a BIC-type criterion at each segmenting recursion, as recommended in that paper. The overall procedure is referred to as KFJS+BS+LR, wherein BS refers to binary segmentation.

In each of the two preliminary estimation methods, two additional filtering steps are included. First, a minimum separation criterion is enforced by sequentially removing any change point $\h\tau_j$ estimated within $10$ indices of a prior change point $\h\tau_{j-1}.$ Second, any component $\h\tau_j$ is removed from the preliminary estimated vector $\h\tau_j,$ if the corresponding jump vector (and jump size) are estimated as identically zero, i.e, $\h\eta_{(j)}=0_{p\times 1},$ $\h\xi_j=0.$

For all settings considered, we construct component-wise intervals as
\benr\label{eq:43}
CI_j= \big[(\tilde\tau_j-ME_j),\, (\tilde\tau_j+ME_j)\big],\quad j=1,...,N,
\eenr
with $\tilde\tau$ being the output of Algorithm 1. The margin of error ($ME_j$) is computed as described in (\ref{def:comp.interval}). In all cases, we set $\al=0.05$ (coverage $(1-\al)=0.95$). The critical value $q_{\al}^v$ under the vanishing regime is evaluated as $q_{\al}^v=11.03$ by using its distribution function provided in \cite{yao1987approximating}. The quantile $q_{(\al,j)}^{nv}$ of the argmax of the two sided random walk is computed by simulating $3000$ sample paths from it. The distribution $\cP$ in Condition A$'$ is Gaussian for Scenario A and B, and is assumed to be Laplace for Scenario A$'$ and B$'$. We utilize plugin estimates of $\si^2_{(\iny,j)}$ and $\xi^2_j,$ whose computational details are provided in Appendix \ref{app:numerical.supplement}. We assess the validity of the joint distribution in Theorem \ref{thm:wc.joint} under the non-vanishing regime by computing the simultaneous coverage yielded by these component-wise constructed intervals. Since by construction we have a finite number of change points, recall that owing to the \textit{asymptotic independence} of $\tilde\tau_j,$ Theorem \ref{thm:wc.joint} dictates that simultaneous coverage should satisfy 
\benr
pr\big(\tau^0_j\in \,\,CI_j;\,\,\forall 1\le j\le N\big)\to (1-\al)^N,\nn
\eenr
wherein $CI_j,$ $j=1,...,N$ are as in (\ref{eq:43}). We set $(1-\al)^N=0.902,0.814,$ for $N=2,4.$

\textit{Selection of tuning parameters for mean estimates in Step 2}: The regularizers $\la_j,$ $j=1,...,N+1$ used to obtain soft thresholded mean estimates in Step 2 of Algorithm 1 are tuned via a BIC type criterion. Specifically, we set $\la_j=\la,$ $j=1,...,N,$ and evaluate $\h\theta_{(j)}^{\la}(\h\tau)$ for each value of $\la$ in an equally spaced grid of twenty five values in the interval $(0,0.5).$ Upon letting $\h S=\Big\{k\in\{1,...,p\};\,\,\cup_{j=1}^{N+1}\h\theta_{(j)k}(\h\tau)\ne 0\Big\}$ we evaluate the criterion
\benr\label{eq:bic}
BIC(\la)= \sum_{j=1}^{N+1}\sum_{t=\h\tau_{j-1}+1}^{\h\tau_j}\big\|x_t-\h\theta_{(j)}^{\la}(\h\tau)\big\|_2^2+ |\h S|\log T.
\eenr
Then, we set $\la$ as the minimizer of $BIC(\la).$

The following metrics are employed to summarize the simulation results: (1) Hausdorff distance (haus. d.): average over replications of $d_H(\h\tau,\tau^0),$ where
\benr
d_H(\h\tau,\tau^0)= \max\Big\{\max_{1\le j\le \h N}d(\tilde\tau_j,\tau^0),\,\,\max_{1\le j\le N}d(\tilde\tau,\tau^0_j)\Big\},
\eenr
and $d(\cdotp,\cdotp)$ denoting the absolute difference. (2) Standard deviation over replications of the Hausdorff distance (sd). (3) $N$-match: relative frequency of number of times $\h N=N.$ Note that by design in Scenario A and A$'$, we have $N$-match $=1,$ thus this metric is only reported for Scenario B and B$'$. Recall that while our methodology does not concern estimation of $N,$ nevertheless, the quality of this estimate plays a critical role in the simulation results. To measure inference performance, for Scenario A and A$'$ we report, (4) Component-wise coverage for the first change point $\tau^0_1$ (Comp. coverage): relative frequency of the number of times $\tau^0_1$ lies in its confidence interval, these are obtained by both the vanishing and non-vanishing regime results. (5) Average margin of error (av. ME) for $\tau^0_1$: average over replications of the margin of errors of each confidence interval of the first change point $\tau^0_1$. (6) Simultaneous coverage over all change point parameters (Simul. coverage): relative frequency of the number of times $\tau^0_j$ lies in corresponding confidence interval for all $j=1,...,N,$ obtained under the non-vanishing jump size result.

Note that metrics (4), (5) and (6) are not meaningful for replicates, wherein $\h N\ne N.$ Consequently, we report instead for Scenario B and B$'$ conditional versions of these metrics, i.e., (4)' Component-wise coverage for the first change point $\tau^0_1$ conditioned on $\h N=N$ (Comp. coverage \big|$\h N=N $): relative frequency over those intervals where $\h N=N$ of the number of times $\tau^0_1$ lies in its confidence interval. Analogous versions for (5) and (6) are reported as (5)' Average margin of error conditioned on $\h N=N$ (av. ME \big|$\h N=N $). (6)' Simultaneous coverage over all change point parameters conditioned on $\h N=N$ (Simul. coverage \big|$\h N=N $).

All results are based on $500$ replicates. Partial results are reported in Tables \ref{tab:simA.N2} and \ref{tab:simB.N2} below (Scenario A and B with $N=2$). Results of the remaining cases are reported in Tables \ref{tab:simA.N4} and \ref{tab:simB.N4} (Scenario A and B with $N=4$), Tables \ref{tab:simA'.N2} and \ref{tab:simB'.N2} (Scenario A$'$ and B$'$ with $N=2$) and Tables \ref{tab:simA'.N4} and \ref{tab:simB'.N4} (Scenario A$'$ and B$'$ with $N=4$), all included in Appendix \ref{app:numerical.supplement} of the Supplement.

\begin{table}[]
	\centering
	\resizebox{0.6\textwidth}{!}{
		\begin{tabular}{cccccc}
			\toprule
			\multicolumn{2}{c}{\begin{tabular}[c]{@{}c@{}}$N=2,$\\ $s=4$\end{tabular}} & \multirow{2}{*}{haus.d (sd)} & \multicolumn{2}{c}{\begin{tabular}[c]{@{}c@{}}Comp. coverage (av. ME)\\ $(1-\alpha)=0.95$\end{tabular}} & \multirow{2}{*}{\begin{tabular}[c]{@{}c@{}}Simul. \\ Coverage\\ $(1-\alpha)^N=0.902$\end{tabular}} \\ \cmidrule{1-2} \cmidrule{4-5}
			$T$                                  & $p$                                 &                              & Vanishing                                          & Non-Vanishing                                      &                                                                                                    \\ \midrule
			450                                  & 50                                  & 0.77 (1.09)                  & 0.924 (2.15)                                       & 0.948 (2.04)                                       & 0.884                                                                                              \\
			450                                  & 200                                 & 0.80 (1.14)                   & 0.942 (2.14)                                       & 0.962 (2.04)                                       & 0.870                                                                                               \\
			450                                  & 350                                 & 0.71 (0.93)                  & 0.958 (2.13)                                       & 0.982 (2.03)                                       & 0.902                                                                                              \\
			450                                  & 500                                 & 0.74 (1.02)                  & 0.954 (2.11)                                       & 0.964 (2.03)                                       & 0.886                                                                                              \\ \midrule
			600                                  & 50                                  & 0.70 (0.95)                   & 0.966 (2.17)                                       & 0.976 (2.05)                                       & 0.898                                                                                              \\
			600                                  & 200                                 & 0.72 (1.11)                  & 0.962 (2.17)                                       & 0.968 (2.05)                                       & 0.892                                                                                              \\
			600                                  & 350                                 & 0.77 (1.06)                  & 0.962 (2.12)                                       & 0.968 (2.04)                                       & 0.898                                                                                              \\
			600                                  & 500                                 & 0.72 (0.92)                  & 0.962 (2.14)                                       & 0.974 (2.02)                                       & 0.898                                                                                              \\ \midrule
			750                                  & 50                                  & 0.82 (1.18)                  & 0.958 (2.19)                                       & 0.970 (2.03)                                        & 0.876                                                                                              \\
			750                                  & 200                                 & 0.86 (1.14)                  & 0.952 (2.17)                                       & 0.968 (2.04)                                       & 0.872                                                                                              \\
			750                                  & 350                                 & 0.82 (1.15)                  & 0.962 (2.18)                                       & 0.970 (2.03)                                        & 0.850                                                                                               \\
			750                                  & 500                                 & 0.81 (1.07)                  & 0.954 (2.17)                                       & 0.972 (2.04)                                       & 0.872                                                                                              \\ \bottomrule
	\end{tabular}}
	\caption{Results of Scenario A with $N=2$ based on 500 monte-carlo replications. Coverage metrics rounded to three decimals, all other metrics rounded to two decimals.}
	\label{tab:simA.N2}
\end{table}

\begin{table}[]
	\centering
	\resizebox{0.8\textwidth}{!}{
		\begin{tabular}{cccccccc}
			\toprule
			\multirow{2}{*}{Method}                                                    & \multicolumn{2}{c}{\begin{tabular}[c]{@{}c@{}}$N=2,$\\ $s=4$\end{tabular}} & \multirow{2}{*}{haus.d (sd)} & \multirow{2}{*}{N-match} & \multicolumn{2}{c}{\begin{tabular}[c]{@{}c@{}}Comp. cov. (av. ME) $\big|\h N=N$\\ $(1-\alpha)=0.95$\end{tabular}} & \multirow{2}{*}{\begin{tabular}[c]{@{}c@{}}Simul. \\ cov. $\big| \h N= N$\\ $(1-\alpha)^N=0.902$\end{tabular}} \\ \cmidrule{2-3} \cmidrule{6-7}
			& $T$                                  & $p$                                 &                              &                          & Vanishing                                               & Non-Vanishing                                           &                                                                                                                \\ \midrule
			\multirow{12}{*}{\begin{tabular}[c]{@{}c@{}}KFJS+\\ BS+\\ LR\end{tabular}} & 450                                  & 50                                  & 15.81   (23.74)              & 0.68                     & 0.947 (2.16)                                            & 0.956 (2.04)                                            & 0.856                                                                                                          \\
			& 450                                  & 200                                 & 17.59 (26.64)                & 0.69                     & 0.945 (2.14)                                            & 0.965 (2.05)                                            & 0.908                                                                                                          \\
			& 450                                  & 350                                 & 16.93 (26.19)                & 0.70                     & 0.948 (2.18)                                            & 0.963 (2.08)                                            & 0.862                                                                                                          \\
			& 450                                  & 500                                 & 17.57 (26.74)                & 0.69                     & 0.948 (2.24)                                            & 0.965 (2.13)                                            & 0.851                                                                                                          \\ \cmidrule{2-8}
			& 600                                  & 50                                  & 23.76 (33.48)                & 0.64                     & 0.944 (2.15)                                            & 0.953 (2.04)                                            & 0.850                                                                                                          \\
			& 600                                  & 200                                 & 25.37 (35.22)                & 0.65                     & 0.933 (2.17)                                            & 0.951 (2.05)                                            & 0.831                                                                                                          \\
			& 600                                  & 350                                 & 27.44 (36.50)                & 0.63                     & 0.956 (2.18)                                            & 0.975 (2.05)                                            & 0.892                                                                                                          \\
			& 600                                  & 500                                 & 23.21 (34.14)                & 0.67                     & 0.931 (2.18)                                            & 0.955 (2.07)                                            & 0.871                                                                                                          \\ \cmidrule{2-8}
			& 750                                  & 50                                  & 28.09 (41.55)                & 0.65                     & 0.951 (2.18)                                            & 0.963 (2.05)                                            & 0.884                                                                                                          \\
			& 750                                  & 200                                 & 34.06 (44.16)                & 0.61                     & 0.970 (2.18)                                            & 0.977 (2.05)                                            & 0.882                                                                                                          \\
			& 750                                  & 350                                 & 34.11 (44.26)                & 0.61                     & 0.957 (2.19)                                            & 0.964 (2.06)                                            & 0.868                                                                                                          \\
			& 750                                  & 500                                 & 29.51 (42.71)                & 0.66                     & 0.933 (2.19)                                            & 0.945 (2.04)                                            & 0.861                                                                                                          \\ \midrule
			\multirow{12}{*}{\begin{tabular}[c]{@{}c@{}}WS\\ +LR\end{tabular}}         & 450                                  & 50                                  & 17.52 (30.92)                & 0.68                     & 0.92 (2.16)                                             & 0.938 (2.07)                                            & 0.834                                                                                                          \\
			& 450                                  & 200                                 & 15.62 (29.65)                & 0.72                     & 0.967 (2.12)                                            & 0.989 (2.03)                                            & 0.953                                                                                                          \\
			& 450                                  & 350                                 & 16.82 (31.40)                & 0.73                     & 0.937 (2.12)                                            & 0.959 (2.02)                                            & 0.837                                                                                                          \\
			& 450                                  & 500                                 & 16.98 (32.13)                & 0.73                     & 0.946 (2.14)                                            & 0.956 (2.05)                                            & 0.861                                                                                                          \\ \cmidrule{2-8}
			& 600                                  & 50                                  & 19.68 (38.88)                & 0.69                     & 0.948 (2.15)                                            & 0.956 (2.06)                                            & 0.854                                                                                                          \\
			& 600                                  & 200                                 & 21.78 (41.97)                & 0.71                     & 0.941 (2.14)                                            & 0.958 (2.03)                                            & 0.843                                                                                                          \\
			& 600                                  & 350                                 & 22.19 (43.37)                & 0.71                     & 0.955 (2.16)                                            & 0.966 (2.03)                                            & 0.890                                                                                                          \\
			& 600                                  & 500                                 & 23.29 (44.33)                & 0.72                     & 0.925 (2.16)                                            & 0.944 (2.04)                                            & 0.891                                                                                                          \\ \cmidrule{2-8}
			& 750                                  & 50                                  & 28.59 (55.66)                & 0.67                     & 0.958 (2.19)                                            & 0.964 (2.04)                                            & 0.898                                                                                                          \\
			& 750                                  & 200                                 & 21.88 (46.98)                & 0.73                     & 0.964 (2.17)                                            & 0.973 (2.03)                                            & 0.887                                                                                                          \\
			& 750                                  & 350                                 & 22.16 (46.92)                & 0.74                     & 0.954 (2.18)                                            & 0.959 (2.04)                                            & 0.861                                                                                                          \\
			& 750                                  & 500                                 & 20.78 (48.78)                & 0.79                     & 0.952 (2.17)                                            & 0.960 (2.04)                                            & 0.866                                                                                                          \\ \bottomrule
	\end{tabular}}
	\caption{Results of Scenario B with $N=2$ based on 500 monte-carlo replications. Coverage metrics rounded to three decimals, all other metrics rounded to two decimals.}
	\label{tab:simB.N2}
\end{table}

Results of the simulation studies are in strong agreement with the theoretical developments. An expected deterioration in estimation from Gaussian to subexponential errors is observed. The component-wise coverage for $\tau^0_1$ in nearly all examined cases in all Scenarios and under both regimes is at the nominal level. Simultaneous coverage in Scenario A and A$'$ for both $N=2,4$ provides fairly precise control at nominal coverage levels ($(1-\al)^2=0.902$ and $(1-\al)^4=0.814$). Larger deviations of simultaneous coverage from nominal levels are observed in Scenario B and B$'$, especially for $N=4.$ This is despite the safeguard of evaluating conditional coverage on replicates that satisfy $\h N=N.$ Based on a close examination of the results for individual replicates, the main reason for observed deviations is related to the discussion in Remark \ref{rem:subseq.change}. Specifically, despite theoretical guarantees of $\text{Pr}(\h N=N)\to 1,$ the preliminary estimates do not necessarily obey this consistency effectively. Even for replicates with $\h N= N,$ there may be spurious change points and missed true changes. The simultaneous coverage metric being evaluated loses its meaning w.r.t the underlying theoretical results when change point indices are misidentified. Although the results of this article may still remain valid on some unknown subset of estimated change points, there is no clear observable metric to illustrate this numerically. We consider an additional Scenario C with larger values of $T,$ where per expectation simultaneous coverage appears to move closer to the nominal level (Table \ref{tab:simC} in Appendix \ref{app:numerical.supplement}).

\section{Application: Smartphone Based Human Activity Recognition}\label{sec:application}

Modern cellphones integrate a host of sensors, including accelerometers, gyroscopes and magnetometers that complement traditional telephony. These sensors obtain measurements of their respective users daily activities. Human Activity Recognition (HAR) is a research field that aims to identify activities of persons based on  on-body and environmental sensor information; for e.g., \cite{allen2006classification} illustrate how accelerometry can be used to retrieve body motion information.

We consider a data set obtained from smartphone embedded accelerometer and gyroscope measurements, made available by \cite{anguita2013public}, which is available at the repository \url{https://archive.ics.uci.edu/ml/datasets/Human+Activity+Recognition+Using+Smartphones}. The author described account of the data collection process is given next. Controlled experiments were carried out with a group of 30 volunteers within an age bracket of 19-48 years. Each person performed six activities (walking, walking upstairs, walking downstairs, sitting, standing, laying) wearing a smartphone ({\it Samsung Galaxy SII}) on the waist. Using the embedded sensors, measurements were obtained on the 3-axial linear acceleration and 3-axial angular velocity at a constant rate of 50Hz. A video of the experiment including an example of the six recorded activities with one of the participants can be seen at \url{http://www.youtube.com/watch?v=XOEN9W05_4A}. The sensor signals were pre-processed by applying noise filters and then sampled in fixed-width sliding windows of 2.56 sec and 50\% overlap (128 readings/window). The sensor acceleration signal, which has gravitational and body motion components, was separated using a Butterworth low-pass filter into body acceleration and gravity. The gravitational force is assumed to have only low frequency components, therefore a filter with 0.3 Hz cutoff frequency was used.  From each window, a vector of features was obtained by calculating variables from the time and frequency domain. 

The data used in our analysis comprise of $T=7352$ vectors, dimension $p=561.$ A detailed description of each collected feature can be found in the repository provided above. Each observed vector is labelled with the activity (one of six) that the subject performed at the time. Our objective is to perform an  unsupervised partitioning of observed vectors over the sampling period, via the change point model (\ref{model:rvmcp}) and the proposed methodology, in order to gauge the predictive power of such measurements in predicting the associated activity; namely, assess how well estimated change points match  segments of different known activities. For this purpose, we sort the data set by the associated activity labels, so that model (\ref{model:rvmcp}) becomes applicable, with the true change points (activity transitions) located at $\tau^0=(1226,2299,3285,4471,5945)^T.$ All observations are then randomized within each activity label in order to eliminate any local temporal artifacts that may have seeped in the data collection. The data are centered and scaled column-wise and method KFJS+BS+LR is used for estimating the locations of the change points and the corresponding confidence intervals. All implementation details are as described in Section \ref{sec:numerical} and Appendix \ref{app:numerical.supplement} with the following two distinctions: the minimum separation criteria enforced in this example was increased to $50$ indices ($0.6\%$ of the sampling period), and the regularization grid for soft thresholding of the mean parameters of Step 2 of Algorithm 1 was enlarged to $(0,1).$ Confidence intervals were obtained with  component-wise coverage set to $(1-\al)=0.95.$ Table \ref{tab:application.cp.est} provides the number and locations of the estimated change points and corresponding confidence intervals obtained under both the vanishing and non-vanishing jump size results. Table \ref{tab:application.jump.var} provides the estimated jump sizes and asymptotic variances. 

\begin{table}[]
	\parbox{.45\linewidth}{
		\centering
		\resizebox{0.56\textwidth}{!}{
			\begin{tabular}{ccclcl}
				\toprule
				\multirow{2}{*}{\begin{tabular}[c]{@{}c@{}}Estimated \\ Number of\\ Change Points\end{tabular}} & \multirow{2}{*}{\begin{tabular}[c]{@{}c@{}}Estimated \\ Locations\\  ($\tilde\tau$)\end{tabular}} & \multicolumn{4}{c}{\begin{tabular}[c]{@{}c@{}}Confidence Intervals \\ $(1-\al)=0.95$\end{tabular}} \\ \cmidrule{3-6} 
				&                                                                                                   & \multicolumn{2}{c}{Vanishing}                    & \multicolumn{2}{c}{Non-Vanishing}               \\ \midrule
				\multirow{5}{*}{$\tilde N=5$}                                                                   & $\tilde\tau_1=1228$                                                                               & \multicolumn{2}{c}{$[1225.91,\,\,1230.83]$}      & \multicolumn{2}{c}{$[1226,\,\,1230]$}           \\
				& $\tilde\tau_2=2299$                                                                               & \multicolumn{2}{c}{$[2297.46,\,\,2300.53]$}      & \multicolumn{2}{c}{$[2298,\,\,2300]$}           \\
				& $\tilde\tau_3=3285$                                                                               & \multicolumn{2}{c}{$[3284.74,\,\,3285.25]$}      & \multicolumn{2}{c}{$[3285,\,\,3285]$}           \\
				& $\tilde\tau_4=4570$                                                                               & \multicolumn{2}{c}{$[4567.79,\,\,4572.21]$}      & \multicolumn{2}{c}{$[4568,\,\,4572]$}           \\
				& $\tilde\tau_5=5945$                                                                               & \multicolumn{2}{c}{$[5944.78,\,\,5945.21]$}      & \multicolumn{2}{c}{$[5945,\,\,5945]$}           \\ \bottomrule
		\end{tabular}}
		\caption{Locations of Estimated change points and associated confidence intervals}
		\label{tab:application.cp.est}
	}
	\hfill
	\parbox{.45\linewidth}{
		\centering
		\resizebox{0.395\textwidth}{!}{
			\begin{tabular}{cccccc}
				\toprule
				\begin{tabular}[c]{@{}c@{}}Change Point \\ Index ($j$)\end{tabular}                                     & 1     & 2     & 3     & 4    & 5     \\ \midrule
				\begin{tabular}[c]{@{}c@{}}Estimated\\ Jump Sizes \\ ($\tilde\xi_j$)\end{tabular}                       & 10.63 & 15.66 & 38.23 & 4.07 & 10.22 \\ \midrule
				\begin{tabular}[c]{@{}c@{}}Estimated \\ Asymptotic \\ Variances\\ $\tilde\si^2_{(\iny,j)}$\end{tabular} & 21.36 & 34.20 & 34.34 & 3.31 & 2.02  \\ \bottomrule
		\end{tabular}}
		\caption{Estimated jump sizes and asymptotic variances}
		\label{tab:application.jump.var}
	}
\end{table}	

The proposed methodology identifies $\tilde N=5$ change points and estimates their locations with high precision, thus, clearly distinguishing all six tasks undertaken in the experiment solely based on accelerometric and gyroscopic measurements from a fairly rudimentary smartphone. The second, third and the fifth change point are estimated at exactly the true values, and the length of their confidence intervals is shorter than the first and fourth ones (those of the third and fifth ones are sufficiently narrow to allow only the exact value under both vanishing and non-vanishing regimes). The results highlight the predictive power of data collected from smartphones in distinguishing ordinary tasks which may seem physically very similar (e.g. sitting, standing and laying). These observations have clear potential beneficial applications in fields such as health care and assisted living. On the flip side, it also raises issues of privacy, since routine access to such data (that equipment manufacturers and service providers usually have) aids in indirect monitoring of daily activities of the phone users.

\bibliographystyle{plainnat}
\bibliography{meanchange}

\clearpage
\setcounter{page}{1}
\section*{Supplementary materials: Inference for Change Points in High Dimensional Mean Shift Models}

\appendix
\numberwithin{equation}{section}
\numberwithin{theorem}{section}

\section{Proofs of results in Section \ref{sec:main}}

To present the arguments of this section we require some additional notation. In all to follow define $\h\eta_{(j)}=\h\theta_{(j)}-\h\theta_{(j+1)},$ $j=1,...,N.$ For any non-negative sequences $0\le v_T\le u_T\le 1$ define the following collection,
\benr\label{def:setG}
\cG_j(u_T,v_T)&=&\Big\{\tau_j\in\{1,2,...,(T-1)\};\,\,Tv_T\le |\tau_j-\tau^0_j|\le Tu_T \Big\}
\eenr
Finally, for any vectors $\theta_{(j)}\in\R^p$ and any $(\tau_j,\tau_{-j}^T)^T\in\{1,...,(T-1)\}^N,$ define,
\benr\label{def:cU}
\cU_j(\tau_j,\tau_{-j},\theta)=Q_j(\tau_j,\tau_{-j},\theta)-Q_j(\tau^0_j,\tau_{-j},\theta),\quad j=1,...,N.
\eenr
Recall that $Q_j(\cdotp,\cdotp,\cdotp)$ is the squared loss defined in (\ref{def:Q}). Finally we also recall from Remark \ref{rem:jump.equivalence} that both models (\ref{model:rvmcp}) and (\ref{model:center.rvmcp}) have the same jump parameters thus no notational distinctions are made between the two for these parameters. Broadly speaking, the basic structure of the arguments to follow was first developed in \cite{kaul2020inference}, this has been further generalized in \cite{kaul2021graphical} and \cite{kaul2021segmentation} is other dynamic contexts. All mentioned articles in a single change point setting. The arguments below are further developed with delicate refinements in order to accommodate the disturbance caused by the presence of additional change points, in accordance with the assumed model (\ref{model:rvmcp}).

\bc$\rule{3.5in}{0.1mm}$\ec

\begin{lemma}\label{lem:lower.b.optimal} Suppose conditions A, B and C hold and let $0\le v_T\le u_T\le 1,$ be any non-negative sequences. Then, for any $0<a<1,$ choosing $c_{a}\ge \surd{(1/a)}$ and for any given $j=1,...,N,$ we have the following uniform lower bound,
	\benr
	\inf_{\tau_j\in\cG_j(u_T,v_T)}\cU_j(\tau_j,\h\tau_{-j},\h\theta)&\ge&\frac{\xi_j^2}{2}\Big[v_T- \frac{c_uc_{a}\si}{\xi_j}\Big(\frac{u_T}{T}\Big)^{\frac{1}{2}}\Big],\nn
	\eenr
	with probability at least $1-2a-o(1)-\pi_T.$
\end{lemma}

\begin{proof}[Proof of Lemma \ref{lem:lower.b.optimal}]
	We begin with a few observations that shall be required to obtain the desired lower bound of this lemma. Define, $\cA=\{\textrm{event where Condition C holds}\},$
	then we have by assumption $pr(\cA)\ge 1-\pi_T.$ First, begin by noting that from Condition C(i) we have, $\max_{1\le j\le N}|\h\tau_j-\tau^0_j|\le c_{u1}T\lm,$ for a suitably small constant $c_{u1}>0.$ Since by Condition B(ii), $T\lm$ is the least separation between consecutive change points, consequently, $\tau_j^0$ must lie between $\h\tau_{j-1}$ and $\h\tau_{j+1}$ on the event under consideration. Moreover, the same assumption also provides that there can only be atmost the immediate neighboring change points $\tau^0_{j-1}$ and $\tau^0_{j+1}$ in the interval $\h\tau_{j-1}$ and $\h\tau_{j+1}$ on the event $\cA,$ and no further change points can be contained in this interval on this event. Second, from Condition C(ii) we have the following relations on the event $\cA,$ for all $j=1,...,N,$
	\benr\label{eq:4}
	\|\h\eta_{(j)}-\eta^0_{(j)}\|_2&\le& \|\h\theta_{(j)}-\theta_{(j)}^0\|_2+\|\h\theta_{(j+1)}-\theta_{(j+1)}^0\|_2\le \frac{2c_{u1}\xim}{(Ns)^{1/2} \log(p\vee T)}\quad{\rm and},\nn\\
	\|\h\eta_{(j)}-\eta^0_{(j)}\|_1&\le& 4\surd{Ns}\|\h\theta_1-\theta_1^0\|_2+4\surd{Ns}\|\h\theta_2-\theta_2^0\|_2\le \frac{8c_{u1}\xim}{\log(p\vee T)}
	\eenr
	The third inequality follows from assumption $\|(\h\theta_{(j)})_{S_j^c}\|_1\le 3\|(\h\theta_{(j)}-\theta_{(j)}^0)_{S_j}\|_1$ of Condition C(iia) which implies that $\|\h\theta_{(j)}-\theta_{(j)}^0\|_1\le 4\surd{(Ns)}\|\h\theta_{(j)}-\theta_{(j)}^0\|_2.$ Next consider,
	\benr\label{eq:5}
	\|\h\eta_{(j)}\|_2&\le& \|\eta_{(j)}^0\|_2+\|\h\eta_{(j)}-\eta_{(j)}^0\|_2\le \xi_j\big\{1+\frac{2c_{u1}}{(Ns)^{1/2} \log(p\vee T)}\big\}\le \frac{3}{2}\xi_j,\quad{\rm and},\nn\\
	\|\h\eta_{(j)}\|_1&\le&\|\eta^0\|_1+\|\h\eta-\eta^0\|_1\nn\\
	&\le& \surd{(Ns)}\xi_j\big\{1+\frac{2c_{u1}}{(Ns)^{1/2} \log(p\vee T)}\big\}\le \frac{3}{2}\surd{(Ns)}\xi_j,
	\eenr
	The second inequality follows from (\ref{eq:4}). The $\ell_1$ bound follows analogously. In the following consider any $\tau_j\in\cG_j(u_T,v_T),$ and assume the ordering $\tau_j\ge \tau^0_{j+1}> \tau^0_j.$ The remaining permutations of the ordering of $\tau_j$ with respect to $\tau^0_{j-1},\tau_j^0$ and $\tau^0_{j+1},$ possible on the set $\cA$ are $\tau^0_{j+1}>\tau_j\ge \tau^0_j,$  $\tau_j\le \tau^0_{j-1}< \tau^0_j,$ and  $\tau_{j-1}^0\le \tau_{j}< \tau^0_j.$ These ordering permutations can be handled with analogous arguments as below, all yielding the same uniform lower bounds stated in the lemma. Another observation here is that under the assumed ordering, for any $\tau_j\in\cG_j(u_T,v_T)$ and on the set $\cA$ and applicable in the construction of the squared loss $Q_j,$  we have the following relation,
	\benr\label{eq:1a}
	\frac{(\tau_j-\tau^0_{j+1})}{(\tau^0_{j+1}-\tau^0_{j})}\le	\frac{(\h\tau_{j+1}-\tau^0_{j+1})}{(\tau^0_{j+1}-\tau^0_{j})}\le \frac{c_{u1}T\lm}{T\ell_j}\le c_{u1},
	\eenr
	for a suitably small constant $c_{u1}>0.$ Here the first inequality follows from the construction of the refitted squared loss $Q_j,$ i.e, the search space is restricted to $\{\h\tau_{j-1},...,\h\tau_{j+1}-1\}.$ The second inequality follows from Condition C(i) (on numerator) and the definition of $\ell_j$ (on denominator). The final inequality follows from definition of $\lm.$ The relation (\ref{eq:1a}) directly implies the following,
	\benr\label{eq:3}
	\frac{(\tau_j-\tau^0_{j+1})}{(\tau_{j}-\tau^0_{j})}=\frac{(\tau_j-\tau^0_{j+1})}{(\tau_j-\tau^0_{j})+(\tau^0_{j+1}-\tau^0_{j})}\le c_{u1},
	\eenr
	on the event $\cA$ and for all $\tau_j\in\cG_j(u_T,v_T)$ under consideration in context of the squared loss $Q_j.$ As a final observation, consider the expression,
	\benr
	\|\h\eta_{(j)}\|_2^2+2(\h\theta_{(j+1)}-\theta_{(j+1)}^0)^T\h\eta_{(j)}-2\frac{(\tau_j-\tau_{j+1}^0)}{(\tau_j-\tau_j^0)}(\theta_{(j+2)}^0-\theta_{(j+1)}^0)^T\h\eta_{(j)}\hspace{-3in}\label{eq:10}\\
	&=&\|\eta^0_{(j)}+(\h\eta_{(j)}-\eta_{(j)}^0)\|_2^2+2(\h\theta_{(j+1)}-\theta_{(j+1)}^0)^T\h\eta_{(j)}\nn\\
	-2\frac{(\tau_j-\tau_{j+1}^0)}{(\tau_j-\tau_j^0)}(\theta_{(j+2)}^0-\theta_{(j+1)}^0)^T\h\eta_{(j)}\hspace{-2.1in}\nn\\
	&\ge& \|\eta^0_{(j)}\|_2^2+2(\h\eta_{(j)}-\eta_{(j)}^0)^T\eta^{0}_{(j)}+2(\h\theta_{(j+1)}-\theta_{(j+1)}^0)^T\h\eta_{(j)}\nn\\
	-2\frac{(\tau_j-\tau_{j+1}^0)}{(\tau_j-\tau_j^0)}(\theta_{(j+2)}^0-\theta_{(j+1)}^0)^T\h\eta_{(j)}\hspace{-2.1in}\nn\\
	&\ge& \|\eta^0_{(j)}\|_2^2-2\|\h\eta_{(j)}-\eta^0_{(j)}\|_2\|\eta^{0}_{(j)}\|_2-2\|\h\theta_{(j+1)}-\theta_{(j+1)}^0\|_2\|\h\eta_{(j)}\|_2\hspace{-1cm}\nn\\
	-2\frac{(\tau_j-\tau_{j+1}^0)}{(\tau_j-\tau_j^0)}\|\theta_{(j+2)}^0-\theta_{(j+1)}^0\|_2\|\h\eta_{(j)}\|_2\hspace{-2.3in}\nn\\
	&\ge&\xi^2_j\Big(1-\frac{4c_{u1}}{(Ns)^{1/2} \log(p\vee T)}-\frac{3c_{u1}}{(Ns)^{1/2} \log(p\vee T)}-3c_{u1}c_u\Big)\ge \frac{\xi_j^2}{2},\nn
	\eenr
	which holds on the set $\cA$ which holds with probability at least $1-\pi_T.$ Here the first inequality is simply an algebraic manipulation. The second follows from the Cauchy-Schwartz inequality. The third follows from Condition C.1, (\ref{eq:4}) and (\ref{eq:5}) together with Condition B(iv) and (\ref{eq:3}).
	
	We can now proceed to the main proof of the uniform bound of this lemma. Consider the following decomposition under the assumed ordering $\tau_j\ge \tau^0_{j+1}> \tau^0_j,$
	\benr\label{eq:9}
	\cU_j(\tau_j,\h\tau,\h\theta)&=&Q_j(\tau_j,\h\tau_{-j},\h\theta)-Q_j(\tau_j^0,\h\tau_{-j},\h\theta)\nn\\
	&=&\sum_{t=\h\tau_{j-1}}^{\tau_j}\|x_t-\h\theta_{(j)}\|^2+\sum_{t=\tau_j+1}^{\h\tau_{j+1}}\|x_t-\h\theta_{(j+1)}\|^2\nn\\
	&&-\sum_{t=\h\tau_{j-1}}^{\tau_j^0}\|x_t-\h\theta_{(j)}\|^2+\sum_{t=\tau_j^0+1}^{\h\tau_{j+1}}\|x_t-\h\theta_{(j+1)}\|^2\nn\\
	&=&\sum_{t=\tau_j^0+1}^{\tau_j}\|x_t-\h\theta_{(j)}\|^2-\sum_{t=\tau_j^0+1}^{\tau_j}\|x_t-\h\theta_{(j+1)}\|^2\nn\\
	&=&(\tau_j-\tau_j^0)\Big\{\|\h\eta_{(j)}\|_2^2 +2(\h\theta_{(j+1)}-\theta_{(j+1)}^0)^T\h\eta_{(j)}\nn\\	&&-2\frac{(\tau_j-\tau_{j+1}^0)}{(\tau_j-\tau_j^0)}(\theta_{(j+2)}^0-\theta_{(j+1)}^0)^T\h\eta_{(j)}\Big\}	-2\sum_{t=\tau_j^0+1}^{\tau_j}\vep_{t}^{*T}\h\eta_{(j)}\nn\\
	&\ge &\frac{Tv_T\xi_j^2}{2}-2\sum_{t=\tau_j^0+1}^{\tau_j}\vep_t^{*T}\h\eta_{(j)}\nn\\
	&=&\frac{Tv_T\xi_j^2}{2}- 2\sum_{t=\tau_j^0+1}^{\tau_j}\vep_t^{*T}\eta^0_{(j)}-2\sum_{t=\tau_j^0+1}^{\tau_j}\vep_t^{*T}(\h\eta_{(j)}-\eta_{(j)}^0),
	\eenr
	on the event $\cA$ which holds with probability at least $1-\pi_T.$ Here the first inequality follows from (\ref{eq:10}) and construction of the set $\cG_j(u_T,v_T).$ The noise variables $\vep_t^*$ arise from the reparametrized model (\ref{model:center.rvmcp}). We now consider uniform upper bounds for each of the stochastic terms in the expression (\ref{eq:9}). First, applying Lemma \ref{lem:optimalcross} for any $0<a<1,$ with $c_{a}\ge \surd(1/a),$ we have,
	\benr\label{eq:13}
	\sup_{\substack{\tau_j\in\cG_j(u_T,v_T);\\ \tau_j\ge\tau_j^0}}2\Big|\sum_{t=\tau^0+1}^{\tau}\vep_t^{*T}\eta^0_{(j)}\Big|\le c_uc_{a}\si\xi_j\big(Tu_T\big)^{\frac{1}{2}}
	\eenr
	w.p. at least $1-2a.$ The second stochastic term in (\ref{eq:9}) can be bounded above as,
	\benr\label{eq:14a}
	2\sum_{t=\tau_j^0+1}^{\tau_j}\vep_t^{*T}(\h\eta_{(j)}-\eta_{(j)}^0)\le 2\Big\|\sum_{t=\tau_j^0+1}^{\tau_j}\vep_t^*\Big\|_{\iny}\big\|\h\eta_{(j)}-\eta_{(j)}^0\big\|_1
	\le c_u\si\xim \big(Tu_T\big)^{\frac{1}{2}},
	\eenr
	w.p. at least $1-o(1)-\pi_T.$ Here the second inequality follows using the deviation bounds in Lemma \ref{lem:nearoptimalcross} together with the $\ell_1$ error bound of (\ref{eq:4}). Substituting (\ref{eq:13}) and (\ref{eq:14a}) in (\ref{eq:9}), we obtain,
	\benr
	\inf_{\substack{\tau_j\in\cG_j(u_T,v_T);\\\tau_j\ge\tau_j^0}}\cU_j(\tau_j,\h\tau_{-j},\h\theta)&\ge& \frac{Tv_T\xi_j^2}{2}-c_uc_a\si\xi_j\big(Tu_T\big)^{\frac{1}{2}}-c_u\si\xim\big(Tu_T\big)^{\frac{1}{2}}\nn\\
	&\ge&\frac{T\xi_j^2}{2}\Big[v_T- \frac{c_uc_a\si}{\xi_j}\Big(\frac{u_T}{T}\Big)^{\frac{1}{2}}\Big]\nn
	\eenr
	w.p. at least $1-2a-o(1)-\pi_T.$ The remaining permutations of $\tau^0_{j+1}>\tau_j\ge \tau^0_j,$  $\tau_j\le \tau^0_{j-1}< \tau^0_j,$ and  $\tau_{j-1}^0\le \tau_{j}< \tau^0_j,$ can be handled with analogous arguments
	This completes the proof of this lemma.
\end{proof}

\bc$\rule{3.5in}{0.1mm}$\ec

\begin{lemma}\label{lem:lower.b.optimal.simul} Suppose conditions A, B and C hold  and let $0\le v_T\le u_T\le 1,$ be any non-negative sequences. Then, we have the following uniform lower bound,
	\benr
	\min_{1\le j\le N}\inf_{\tau_j\in\cG_j(u_T,v_T)}\cU_j(\tau_j,\h\tau_{-j},\h\theta)&\ge&\frac{\xim^2}{2}\Big[v_T- \frac{c_u\si}{\xim}\Big(\frac{u_T\log^2 T}{T}\Big)^{\frac{1}{2}}\Big],\nn
	\eenr
	with probability at least $1-o(1)-\pi_T.$
\end{lemma}

\begin{proof}[Proof of Lemma \ref{lem:lower.b.optimal.simul}] The structure of this proof is similar to that of Lemma \ref{lem:lower.b.optimal} with the distinction in the additional uniformity required over $j=1,...,N,$ which in turn requires utilizing stochastic bounds with this additional uniformity (Lemma \ref{lem:nearoptimalcross.maximums}). Proceeding identically as before in Lemma \ref{lem:lower.b.optimal}, under the ordering $\tau_j\ge \tau^0_{j+1}> \tau^0_j,$ we have (\ref{eq:9}), i.e.,
	\benr\label{eq:11}
	\cU_j(\tau_j,\h\tau,\h\theta)
	\ge \frac{Tv_T\xi_j^2}{2}- 2\sum_{t=\tau_j^0+1}^{\tau_j}\vep_t^{*T}\eta^0_{(j)}-2\sum_{t=\tau_j^0+1}^{\tau_j}\vep_t^{*T}(\h\eta_{(j)}-\eta_{(j)}^0),
	\eenr	
	on the event $\cA$ which holds with probability at least $1-\pi_T.$ Now consider each of the stochastic terms in (\ref{eq:11}) and apply the bounds of Lemma \ref{lem:nearoptimalcross.maximums} which possess the required additional uniformity over $j=1,...,N.$ First from Part (ii) of Lemma \ref{lem:nearoptimalcross.maximums},
	\benr\label{eq:12}
	\max_{1\le j\le N}\sup_{\substack{\tau_j\in\cG_j(u_T,v_T);\\ \tau_j\ge\tau_j^0}}\Big|\sum_{t=\tau^0+1}^{\tau}\vep_t^{*T}\eta^0_{(j)}\Big|\le c_u \xma\si\{Tu_T \log^2 T\}^{\frac{1}{2}}
	\eenr
	w.p. at least $1-o(1).$ Second,
	\benr\label{eq:14}
	\max_{1\le j\le N}\sup_{\substack{\tau_j\in\cG_j(u_T,v_T);\\ \tau_j\ge\tau_j^0}}\sum_{t=\tau_j^0+1}^{\tau_j}\vep_t^{*T}(\h\eta_{(j)}-\eta_{(j)}^0)\hspace{1.75in}\nn\\
	\le \max_{1\le j\le N}\sup_{\substack{\tau_j\in\cG_j(u_T,v_T);\\ \tau_j\ge\tau_j^0}}\Big\|\sum_{t=\tau_j^0+1}^{\tau_j}\vep_t^*\Big\|_{\iny}\big\|\h\eta_{(j)}-\eta_{(j)}^0\big\|_1\le c_u\si\xim \big(Tu_T\big)^{\frac{1}{2}},
	\eenr
	w.p. at least $1-o(1)-\pi_T,$ where the second inequality follows from Part (i) of Lemma \ref{lem:nearoptimalcross.maximums} together with the $\ell_1$ error bound of (\ref{eq:4}). Substituting (\ref{eq:12}) and (\ref{eq:14}) in (\ref{eq:11}) yields,
	\benr
	\min_{1\le j\le N}\inf_{\substack{\tau_j\in\cG_j(u_T,v_T);\\\tau_j\ge\tau_j^0}}\cU_j(\tau_j,\h\tau,\h\theta)&\ge&
	\frac{Tv_T\xim^2}{2}-c_u\si\xma\big(Tu_T\log^2 T\big)^{\frac{1}{2}}-c_u\si\xim\big(Tu_T\big)^{\frac{1}{2}}\nn\\
	&\ge&\frac{T\xim^2}{2}\Big[v_T- \frac{c_u\si}{\xim}\Big(\frac{u_T\log ^2 T}{T}\Big)^{\frac{1}{2}}\Big],\nn
	\eenr
	w.p. at least $1-o(1)-\pi_T.$ Here we have also utilized $\xma\le c_u\xim$ of Condition B(iv).
	The same bound can be obtained for all other permutations of the ordering of $\tau_j$s w.r.t $\tau^0_j$s via analogous arguments. This completes the proof of this lemma.
\end{proof}

\bc$\rule{3.5in}{0.1mm}$\ec

\begin{proof}[Proof of Theorem \ref{thm:cpoptimal}] The proof of this result relies on a recursive argument on Lemma \ref{lem:lower.b.optimal}, where the desired rate of convergence is obtained by a series of recursions, with this rate being sharpened at each step. We begin by considering any given $j=1,...,N,$ and any $v_T>0$ and applying Lemma \ref{lem:lower.b.optimal} on the set $\cG_j(1,v_T)$ to obtain,
	\benrr
	\inf_{\tau_j\in\cG_j(1,v_T)}\cU_j(\tau_j,\h\tau_{-j},\h\theta)\ge \frac{\xi_j^2}{2}\Big[v_T-\frac{c_uc_a\si}{\xi_j}\Big(\frac{1}{T}\Big)^{\frac{1}{2}}\Big].
	\eenrr
	with probability at least $1-2a-o(1)-\pi_T.$  Now upon choosing any,
	\benr
	v_T>v_T^*=\frac{c_uc_a\si}{\xi_j}\Big(\frac{1}{T}\Big)^{\frac{1}{2}},\nn
	\eenr
	we obtain $\inf_{\tau_j\in\cG_j(1,v_T)}\cU_j(\tau_j,\h\tau_{-j},\h\theta)>0,$ thus implying that $\tilde\tau_j\notin\cG_j(1,v_T^*),$ i.e., $|\tilde\tau_j-\tau^0_j|\le  Tv_T^*,$ with probability at least $1-2a-o(1)-\pi_T.$\footnote{Since by construction of $\tilde\tau_j,$ we have $\cU_j(\tau_j,\h\tau_{-j},\h\theta)\le 0.$} Now reset $u_T=v_T^*$ and reapply Lemma \ref{lem:lower.b.optimal} for any $v_T>0$ to obtain,
	\benrr
	\inf_{\tau_j\in\cG_j(u_T,v_T)}\cU_j(\tau_j,\h\tau_{-j},\h\theta)\ge \frac{\xi^2_j}{2}\Big[v_T-\Big(\frac{c_uc_a\si}{\xi_j}\Big)^{1+\frac{1}{2}}\Big(\frac{1}{T}\Big)^{\frac{1}{2}+\frac{1}{4}}\Big],\nn
	\eenrr
	with probability at least $1-2a-o(1)-\pi_T.$. Again choosing any,
	\benr
	v_T>v_T^*= \Big(\frac{c_uc_a\si}{\xi_j}\Big)^{1+\frac{1}{2}}\Big(\frac{1}{T}\Big)^{\frac{1}{2}+\frac{1}{4}},\nn
	\eenr
	we obtain $ \inf_{\tau_j\in\cG_j(u_T,v_T)}\cU_j(\tau_j,\h\tau_{-j},\h\theta)>0,$ thus yielding $\tilde\tau_j\notin\cG_j(u_T,v_T^*),$ i.e.,
	\benr
	|\tilde\tau_j-\tau^0_j|\le T\Big(\frac{c_uc_a\si}{\xi_j}\Big)^{a_2}\Big(\frac{1}{T}\Big)^{b_2},
	\eenr
	with probability at least $1-2a-o(1)-\pi_T.$ Where,
	\benr
	a_2=1+\frac{1}{2}=\sum_{j=0}^1\frac{1}{2^j},\quad{\rm and}\quad b_2=\frac{1}{2}+\frac{1}{4}=\sum_{j=1}^2\frac{1}{2^j}.\nn
	\eenr
	Note that the rate of convergence of $\tilde\tau$ has been sharpened at the second recursion in comparison to the first. Continuing these recursions by resetting $u_T$ to the bound of the previous recursion, and applying Lemma \ref{lem:lower.b.optimal}, we obtain for the $m^{th}$ recursion,
	\benr
	|\tilde\tau_j-\tau^0_j|\le T\Big(\frac{c_uc_a\si}{\xi_j}\Big)^{a_m}\Big(\frac{1}{T}\Big)^{b_m},
	\eenr
	with probability at least $1-2a-o(1)-\pi_T.$ Repeating these recursions an infinite number of times and noting that $a_\iny=\sum_{j=0}^{\iny}(1/2^j)=2,$ and $b_{\iny}=\sum_{j=1}^{\iny}(1/2^j)=1$ we obtain,
	\benr
	|\tilde\tau_j-\tau^0_j|\le T\Big(\frac{c_uc_a\si}{\xi_j}\Big)^2\frac{1}{T}=c_u^2c_a^2\si^2\xi_j^{-2}\nn
	\eenr
	with probability at least $1-2a-o(1)-\pi_T.$ Finally, note that despite the recursions in the above argument, the probability of the bound after every recursion is maintained to be at least $1-2a-o(1)-\pi_T.$ This follows since the probability statement of Lemma \ref{lem:lower.b.optimal} arises from stochastic upper bounds of Lemma \ref{lem:nearoptimalcross} and \ref{lem:optimalcross} applied recursively with a tighter bound at each recursion. This yields a sequence of events such that the event at each recursion is a proper subset of the event at the previous recursion. This completes the proof of this theorem.
\end{proof}

\bc$\rule{3.5in}{0.1mm}$\ec

\begin{proof}[Proof of Theorem \ref{thm:cpoptimal.simul}] The argument to follow is largely similar to that of the proof of Theorem \ref{thm:cpoptimal}. It is a recursive argument applied on Lemma \ref{lem:lower.b.optimal.simul} which possess uniformity over $j=1,...,N,$ in comparison to Lemma \ref{lem:lower.b.optimal} which does not. Recall that this uniformity is gained in exchange for a weaker bound in comparison to Lemma \ref{lem:lower.b.optimal}. Consider any $v_T>0$ and apply Lemma \ref{lem:lower.b.optimal.simul} on the sets $\cG_j(1,v_T),$ $j=1,...,N,$ to obtain,
	\benrr
	\min_{1\le j\le N}\inf_{\tau_j\in\cG_j(1,v_T)}\cU_j(\tau_j,\h\tau_{-j},\h\theta)\ge \frac{\xim^2}{2}\Big[v_T-\frac{c_u\si}{\xim}\Big(\frac{\log^2 T}{T}\Big)^{\frac{1}{2}}\Big].
	\eenrr
	with probability at least $1-o(1)-\pi_T.$  Upon choosing any,
	\benr
	v_T>v_T^*=\frac{c_u\si}{\xi_j}\Big(\frac{\log^2T}{T}\Big)^{\frac{1}{2}},\nn
	\eenr
	we obtain $\min_{1\le j\le N}\inf_{\tau_j\in\cG_j(1,v_T)}\cU_j(\tau_j,\h\tau_{-j},\h\theta)>0,$  thus implying that $\tilde\tau_j\notin\cG_j(1,v_T^*),$ $\forall j=1,...,N,$ with probability at least $1-o(1)-\pi_T.$  Note that this implies $\max_{1\le j\le N}|\tilde\tau_j-\tau^0_j|\le  Tv_T^*,$ with the same probability. Now reset $u_T=v_T^*$ and reapply Lemma \ref{lem:lower.b.optimal.simul} for any $v_T>0$ to obtain,
	\benrr
	\min_{1\le j\le N}\inf_{\tau_j\in\cG_j(u_T,v_T)}\cU_j(\tau_j,\h\tau_{-j},\h\theta)\ge \frac{\xim_j}{2}\Big[v_T-\Big(\frac{c_u\si}{\xim}\Big)^{1+\frac{1}{2}}\Big(\frac{\log^2T}{T}\Big)^{\frac{1}{2}+\frac{1}{4}}\Big],\nn
	\eenrr
	with probability at least $1-o(1)-\pi_T.$. Again choosing any,
	\benr
	v_T>v_T^*= \Big(\frac{c_u\si}{\xim}\Big)^{1+\frac{1}{2}}\Big(\frac{\log^2T}{T}\Big)^{\frac{1}{2}+\frac{1}{4}},\nn
	\eenr
	we obtain $\min_{1\le j\le N}\inf_{\tau_j\in\cG_j(u_T,v_T)}\cU_j(\tau_j,\h\tau_{-j},\h\theta)>0,$ thus yielding $\tilde\tau_j\notin\cG_j(u_T,v_T^*),$ $\forall j,$ i.e.,
	\benr
	\max_{1\le j\le N}|\tilde\tau_j-\tau^0_j|\le T\Big(\frac{c_u\si}{\xim}\Big)^{a_2}\Big(\frac{\log^2T}{T}\Big)^{b_2},
	\eenr
	with probability at least $1-o(1)-\pi_T.$ Where,
	\benr
	a_2=1+\frac{1}{2}=\sum_{j=0}^1\frac{1}{2^j},\quad{\rm and}\quad b_2=\frac{1}{2}+\frac{1}{4}=\sum_{j=1}^2\frac{1}{2^j}.\nn
	\eenr
	Continuing these recursions by resetting $u_T$ to the bound of the previous recursion, and applying Lemma \ref{lem:lower.b.optimal.simul}, we obtain for the $m^{th}$ recursion,
	\benr
	\max_{1\le j\le N}|\tilde\tau_j-\tau^0_j|\le T\Big(\frac{c_u\si}{\xim}\Big)^{a_m}\Big(\frac{\log^2T}{T}\Big)^{b_m},
	\eenr
	with probability at least $1-o(1)-\pi_T.$ Repeating these recursions an infinite number of times and noting that $a_\iny=\sum_{j=0}^{\iny}(1/2^j)=2,$ and $b_{\iny}=\sum_{j=1}^{\iny}(1/2^j)=1$ we obtain,
	\benr
	\max_{1\le j\le N}|\tilde\tau_j-\tau^0_j|\le T\Big(\frac{c_u\si}{\xim}\Big)^2\frac{\log^2T}{T}=c_u^2\si^2\xim^{-2}\log^2T\nn
	\eenr
	with probability at least $1-o(1)-\pi_T.$ As in the proof of Theorem \ref{thm:cpoptimal}, despite the recursions in the above argument, the probability of the bound after every recursion is maintained to be at least $1-o(1)-\pi_T,$ for the same reason as described there. Thus completing the proof of this theorem.
\end{proof}

\bc$\rule{3.5in}{0.1mm}$\ec

As the reader may have observed, a change of notation has been carried out for the results of Theorem \ref{thm:wc.vanishing} and Theorem \ref{thm:wc.non.vanishing}. These results are presented in more conventional {\it argmax} notation instead of the {\it argmin} notation of the problem setup in Section \ref{sec:intro}. This is purely a notational change and all results can equivalently be stated in the {\it argmin} language. Accordingly we define the following versions. Let $\cU_j(\tau_j,\tau_{-j},\theta)$ $j=1,...,N,$ be as defined in (\ref{def:cU}) and consider,
\benr\label{def:cC}
\cC_j(\tau,\tau_{-j},\theta)=-\cU_j(\tau,\tau_{-j},\theta),\quad j=1,...,N.
\eenr
Then, we can re-express the change point estimators $\tilde\tau_j(\h\tau_{-j},\h\theta)$ as,
\benr
\tilde\tau_j(\h\tau_{-j},\h\theta)=\argmax_{\h\tau_{j-1}\le \tau_j<\h\tau_{j+1}}\cC_j(\tau_j,\h\tau_{-j},\h\theta),\quad j=1,...,N.\nn
\eenr

The proofs of Theorem \ref{thm:wc.vanishing}, Theorem \ref{thm:wc.non.vanishing} and Theorem \ref{thm:wc.joint} below are applications of the Argmax Theorem (reproduced as Theorem \ref{thm:argmax} in Appendix \ref{app:auxiliary}). The arguments here are largely an exercise in verification of requirements of this theorem.

\bc$\rule{3.5in}{0.1mm}$\ec

\begin{proof}[Proof of Theorem \ref{thm:wc.vanishing}]
	Consider any fixed $j=1,...,N,$ then we begin by noting that although $(\tilde\tau_j-\tau^0_j)$ is discrete, however the sequence whose limiting distribution is being examined is $\xi^2_j(\tilde\tau_j-\tau^0_j),$ consequently, the underlying indexing metric space here is $\R.$ Now consider the two cases of known and unknown plug-in parameters.
	
	\vspace{1.5mm}	
	\noi{\bf Case I \big($\tau^0_{-j}$ and $\theta^0$ known\big):} Following is list of requirement of the Argmax theorem that require verification for this case (see, page 288 of \cite{vaart1996weak}).
	\begin{enumerate}
		\item The sequence $\xi_j^{2}(\tilde\tau_j^*-\tau^0_j)$ is uniformly tight (see, Definition \ref{def:utight} in Appendix \ref{app:auxiliary}).
		\item $\big\{2\si_{(\iny,j)}W_j(\z)-|\z|\}$ satisfies suitable regularity conditions\footnotemark.
		\item For any $\z\in [-c_u,c_u]$ we have
		\benr
		\cC_j(\tau_j^0+\z \xi^{-2}_j,\tau_{-j}^0,\theta^0)
		\Rightarrow \big\{2\si_{(\iny,j)}W_j(\z)-|\z|\}.\nn
		\eenr
	\end{enumerate}
	\footnotetext{Almost all sample paths $\z\to \big\{2\si_{(\iny,j)}W_j(\z)-|\z|\}$ are upper semicontinuous and posses a unique maximum at a (random) point $\argmax_{\z\in\R}\big\{2\si_{(\iny,j)}W_j(\z)-|\z|\},$ which as a random map in the indexing metric space is tight.}
	
	Note that by setting $\h\theta=\theta^0,$ and $\h\tau_{-j}=\tau_{-j}^0,$ the requirements of  Condition C are trivially satisfied. Now using Theorem \ref{thm:cpoptimal} we have that $\xi^{2}_j(\tilde\tau^*_j-\tau^0_j)=O_p(1).$ This yields requirement (1). The second requirement follows from well known properties of Brownian motion's. The only remaining requirement is (3), which is provided below.
	
	\benr
	\z \leftarrow (\z-\xi^{2}_j)\le \xi^{2}_j\lfloor \z\xi^{-2}_j\rfloor\le (\z+\xi^2_j)\to \z\nn
	\eenr
	Hence, w.l.o.g. we may assume $\z\xi^{-2}_j$ is integer valued. Now for any $\z\in(0,c_u],$ consider
	
	\benr\label{eq:16}
	\cC_j(\tau^0_j+\z\xi^{-2}_j,\tau_{-j}^0,\theta^0)&=&-\sum_{t=(\tau^0_j+1)}^{\tau^0_j+\z\xi^{-2}_j}\big\{\|x_t-\theta_{(j)}^0\|_2^2-\|x_t-\theta_{(j+1)}^0\|_2^2\big\}\nn\\
	&=&2\sum_{t=(\tau^0_j+1)}^{\tau^0_j+\z\xi^{-2}_j}\vep_t^{*T}\eta^0_{(j)}-\z\nn\\
	&=&2\sum_{t=(\tau^0_j+1)}^{\tau^0_j+\z\xi^{-2}_j}\vep_t^{T}\eta^0_{(j)}-\z-2\z\xi^{-2}_j\bar\vep^T\eta^0_{(j)}\nn\\
	&=&2\sum_{t=(\tau^0_j+1)}^{\tau^0_j+\z\xi^{-2}_j}\vep_t^{T}\eta^0_{(j)}-\z-o_p(1)
	\Rightarrow 2\si_{(\iny,j)}W_{1j}(\z)-\z,
	\eenr
	where the final equality follows from (\ref{eq:15}) together with Condition B(iv) and Condition E. The weak convergence follows from the functional central limit theorem. Repeating the same argument with $\z\in[-c_u,0),$ yields  	$\cC(\tau^0+\z\xi^{-2}_j,\tau_{-j}^0,\theta^0)\Rightarrow 2\si_{(\iny,j)}W_{2j}(-\z)-|\z|.$ his completes the proof of requirement (3) for the Argmax theorem and consequently an application of its results yields  $\xi^{2}_j(\tilde\tau^*_j-\tau^0_j)\Rightarrow \argmax_{\z\in\R}\big\{2\si_{(\iny,j)}W_j(\z)-|\z|\},$ which completes the proof of this case.
	
	\vspace{1.5mm}		
	\noi{\bf Case II \big($\tau^0_{-j}$ and $\theta^0$ unknown\big):} In this case the applicability of the argmax theorem requires verification of the following conditions.
	\begin{enumerate}
		\item The sequence $\xi_j^{2}(\tilde\tau_j-\tau^0_j)$ is uniformly tight.
		\item $\big\{2\si_{(\iny,j)}W_j(\z)-|\z|\}$ satisfies suitable regularity conditions.
		\item For any $\z\in [-c_u,c_u]$ we have
		\benr
		\cC_j(\tau_j^0+\z \xi^{-2}_j,\h\tau_{-j},\h\theta)
		\Rightarrow \big\{2\si_{(\iny,j)}W_j(\z)-|\z|\}.\nn
		\eenr
	\end{enumerate}
	Part (i) again follows from the result of Theorem \ref{thm:cpoptimal} under the assumed Condition C on the nuisance estimates $\h\tau_{-j}$ and $\h\theta.$  Part (2) is identical to the corresponding requirement of Case I. Finally to prove part (3) note that from Lemma \ref{lem:Capprox} we have that,
	\benr\label{eq:36a}	\sup_{\tau_j\in\cG_j\big(c_uT^{-1}\xi^{-2}_j,0\big)}|\cC_j(\tau_j,\h\tau_{-j},\h\theta)-\cC_j(\tau_j,\tau_{-j}^0,\theta^0)|=o_p(1).
	\eenr
	The approximation (\ref{eq:36a}) and Part (3) of Case I together imply Part (3) for this case. This completes the verification of all requirements for this case. The stated limiting distribution now follows by an application of the Argmax theorem.
\end{proof}

\bc$\rule{3.5in}{0.1mm}$\ec

\begin{proof}[Proof of Theorem \ref{thm:wc.non.vanishing}] The proof of this theorem is similar to Theorem \ref{thm:wc.vanishing} in that it is also an application of the Argmax theorem. The distinction here is in the limiting distribution that is induced by the change of regime of the jump size. Consider any given $j=1,...,N$ and the discrete sequence $(\tilde\tau_w-\tau^0_w),$ consequently the underlying indexing metric space here is $\Z.$ Now consider the two cases of known and unknown plug-in parameters.
	
	\vspace{1.5mm}	
	\noi{\bf Case I \big($\tau_{-j}^0$ and $\theta^0$ known\big):} The requirements to be verified here are as follows.
	\begin{enumerate}
		\item The sequence $(\tilde\tau_j^*-\tau^0_j)$ is uniformly tight.
		\item $\cC_{(\iny,j)}(\z)$ satisfies suitable regularity conditions.
		\item For any $\z\in \{-c_u,-c_u+1,...,-1,0,1,...c_u\},$ we have
		\benr
		\cC_j(\tau_j^0+\z,\tau^0_{-j},\theta^0)
		\Rightarrow \cC_{(\iny,j)}(\z).\nn
		\eenr
	\end{enumerate}
	As in the proof of Theorem \ref{thm:wc.vanishing}, requirement (1) follows directly from the result of Theorem \ref{thm:cpoptimal}.  Requirement (2) of regularity of the {\it argmax} of two sided negative drift random walk $\cC_{(\iny,j)}(\z)$ has been proved earlier in Lemma A.3 of the supplement of \cite{kaul2020inference}. The requirement (3) is verified in the following. For any $\z\in\{1,2,...,c_u\},$ consider,
	\benr\label{eq:24}
	\cC_j(\tau^0_j+\z,\tau^0_{-j},\theta^0)&=&-\sum_{t=(\tau^0_j+1)}^{\tau^0_j+\z}\big\{\|x_t-\theta_{(j)}^0\|_2^2-\|x_t-\theta_{(j+1)}^0\|_2^2\big\}\nn\\
	&=&\sum_{t=(\tau^0_j+1)}^{\tau^0_j+\z}\Big(2\vep_t^{*T}\eta^0_{(j)}-\xi^2_j\Big)\nn\\
	&=& \sum_{t=(\tau^0_j+1)}^{\tau^0_j+\z}\Big(2\vep_t^T\eta^0_{(j)}-\xi^2_j-o_p(1)\Big)\nn\\
	&\Rightarrow& \sum_{t=1}^{\z}\cP\big(-\xi_{(\iny,j)}^2,4\xi_{(\iny,j)}^2\si^2_{(\iny,j)}\big),
	\eenr
	The final equality follows similarly to the final equality of (\ref{eq:16}). The convergence in distribution follows from Condition A$',$ Condition D together with Slutsky's theorem. Repeating the same argument with $\z\in\{-c_u,-c_u+1,...,-1\},$ yields  	$\cC_j(\tau^0_j+\z,\tau^0_{-j},\theta^0)\Rightarrow \sum_{t=1}^{-\z}\cP\big(-\xi_{(\iny,j)}^2,4\xi_{(\iny,j)}^2\si^2_{(\iny,j)}\big).$ An application the Argmax theorem now yields $(\tilde\tau^*_j-\tau^0_j)\Rightarrow \argmax_{\z\in\Z}\cC_{(\iny,j)}(\z),$ which completes the proof of this case.
	
	\vspace{1.5mm}		
	\noi{\bf Case II \big($\tau^0_{-j}$ and $\theta^0$ unknown\big):} In this case, the applicability of the Argmax Theorem requires verification of the following.
	\benr
	&(i)& {\rm The\, sequence}\,\, (\tilde\tau_j-\tau^0_j)\,\, {\rm is\, uniformly\, tight}.\nn\\
	&(ii)&	\cC_{(\iny,j)}(\z)\,\, {\rm satisfies\, suitable\, regularity\, conditions}.\nn\\
	&(iii)& {\rm For\, any}\,\, \z\in \{-c_u,-c_u+1,...,-1,0,1,...c_u\}\,\, {\rm we\, have,}\,\, \cC(\tau^0_j+\z,\h\tau_{-j},\h\theta)\Rightarrow \cC_{(\iny,j)}(\z).\nn	\eenr
	Part (i) follows from Theorem \ref{thm:cpoptimal} under the assumed Condition C on the nuisance estimates $\h\tau_{-j}$ and $\h\theta.$  Part (ii) is identical to the corresponding requirement of Case I. Finally to prove part (iii) note that from Lemma \ref{lem:Capprox} we have that,
	\benr\label{eq:38}
	\sup_{\tau_j\in\cG_j(c_uT^{-1}\xi^{-2}_j,0)}|\cC_j(\tau_j,\h\tau_{-j},\h\theta)-\cC_j(\tau_j,\tau^0_{-j},\theta^0)|=o_p(1).
	\eenr
	The approximation (\ref{eq:38}) and Part (iii) of Case I together imply Part (iii) for this case. This completes the verification of all requirements for this case. The statement of the theorem now follows by an application of the Argmax theorem.
\end{proof}

\bc$\rule{3.5in}{0.1mm}$\ec

\begin{lemma}\label{lem:Capprox} Suppose Conditions A, B, C and E hold and let $\cC_j(\tau_j,\tau_{-j},\h\theta)$ be as in (\ref{def:cC}). Further, assume that $r_T$ in Condition C satisfies $r_T=\{o(1)\xim\}\big/\{{(Ns)^{1/2}\log (p\vee T)}\}.$ Then, for any given $j=1,...,N$ and any $c_u>0,$ we obtain
	\benr
	\sup_{\tau_j\in\cG_j\big(c_uT^{-1}\xi^{-2}_j,0\big)}\big|\cC_j(\tau_j,\h\tau_{-j},\h\theta)-\cC_j(\tau_j,\tau^0_{-j},\theta^0)\big|=o_p(1).\nn
	\eenr
\end{lemma}

\begin{proof}[Proof of Lemma \ref{lem:Capprox}] Recall that
	\benr
	\cC_j(\tau_j,\h\tau_{-j},\h\theta)&=&\sum_{t=\h\tau_{j-1}+1}^{\tau_j}\|x_t-\h\theta_{(j)}\|_2^2+\sum_{t=\tau_{j}+1}^{\h\tau_{j+1}}\|x_t-\h\theta_{(j+1)}\|_2^2\nn\\
	\cC_j(\tau_j,\tau_{-j}^0,\theta^0)&=&\sum_{t=\tau_{j-1}^0+1}^{\tau_j}\|x_t-\theta_{(j)}^0\|_2^2+\sum_{t=\tau_{j}+1}^{\tau_{j+1}^0}\|x_t-\theta_{(j+1)}^0\|_2^2,\nn
	\eenr
	Further, note that $\cC_j(\tau_j,\h\tau_{-j},\h\theta)$ and $\cC_j(\tau_j,\tau_{-j}^0,\theta^0)$ are sums over indices whose start and end points may differ. Despite this incoherence, the desired supremum over collection $\cG_j\big(c_uT^{-1}\xi^{-2}_j,0\big)$ is well defined. This is enforced by Condition C together with the rate assumption of Condition F. These ensure that the collection of $\tau_j'$s over which the supremum is evaluated remains between $\max\{\h\tau_{j-1},\tau_{j-1}^0\}$ and $\min\{\h\tau_{j+1},\tau_{j+1}^0\}$ with probability $1-\pi_T.$ Specifically, from Condition B(ii) we have that $(\tau^0_{j}-\tau^0_{j-1})\ge T\lm,$ $\forall j$ and from Condition C that $\max_{j}|\h\tau_{j}-\tau^0_j|\le c_{u1}T\lm,$ w.p. $1-\pi_T.$ In addition, the left and right end points of the set  $\cG_j\big(c_uT^{-1}\xi^{-2}_j,0\big),$ i.e., $\tau^0_{j}-c_u\xi_j^{-2}$ and $\tau^0_{j}+c_u\xi_j^{-2},$ respectively. Consequently, the rate assumption of Condition F forces these end points to be in a sufficiently small neighborhood of $\tau_{j}^0$, such that all values of $\tau_j$ in the collection $\cG_j\big(c_uT^{-1}\xi^{-2}_j,0\big),$ remain away from $\h\tau_{j-1},\tau_{j-1}^0$ as well as $\h\tau_{j+1},\tau_{j+1}^0,$ w.p. $1-\pi_T,$ thus allowing the desired supremum of interest to be well defined.
	
	The second observation is that by proceeding as in (\ref{eq:4}), under Condition C with $r_T=\{o(1)\xim\}\big/\{{(Ns)^{1/2}\log (p\vee T)}\},$ we get
	\benr\label{eq:33}
	\|\h\eta_{(j)}-\eta^0_{(j)}\|_1\le c_{u1}\surd{Ns}\|\h\eta_{(j)}-\eta^0_{(j)}\|_2= \frac{o(1)\xim}{\log (p\vee T)}
	\eenr
	w.p. at least $1-\pi_T.$ Next, consider any $\tau_j\ge \tau^0_j$ and define the following:
	\benr
	R_1=2\sum_{t=\tau^0_j+1}^{\tau_j}\vep_t^{*T}(\h\eta_{(j)}-\eta^0_{(j)}),\quad{\rm and}\quad R_2=(\tau_j-\tau^0_j)\big(\|\h\eta_{(j)}\|_2^2-\|\eta^0_{(j)}\|_2^2).\nn
	\eenr	
	Then, under the orientation $\min\{\h\tau_{j+1},\tau_{j+1}^0\}>\tau_j\ge\tau_j^0,$ we have the following algebraic expansion,	
	\benr\label{eq:34}
	\cC_j(\tau_j,\h\tau_{-j},\h\theta)-\cC_j(\tau_j,\tau_{-j}^0,\theta^0)\hspace{-1in}\nn\\
	&=& 2\sum_{t=\tau^0_j+1}^{\tau_j}\vep_t^{*T}(\h\eta_{(j)}-\eta^0_{(j)})- (\tau_j-\tau^0_j)\big(\|\h\eta_{(j)}\|_2^2-\|\eta^0_{(j)}\|_2^2)\nn\\
	&=&R_1-R_2
	\eenr
	Next, we provide uniform bounds for the terms $R_1$ and $R_2$ of (\ref{eq:34}). Consider
	\benr\label{eq:36}
	\sup_{\substack{\tau_j\in \cG_j(c_u T^{-1}\xi^{-2}_j,0);\\\tau_j\ge\tau^0_j}}|R_1|&\le& 2\sup_{\substack{\tau_j\in \cG_j(c_uT^{-1}\xi^{-2}_j,0);\\\tau_j\ge\tau^0_j}}\big\|\sum_{t=\tau^0_j+1}^{\tau_j}\vep_t^*\big\|_{\iny}\|\h\eta_{(j)}-\eta^0_{(j)}\|_1\nn\\
	&\le&  c_u\si\xi_j^{-1}\log(p\vee T)\|\h\eta_{(j)}-\eta^0_{(j)}\|_1= o(1),
	\eenr
	w.p. at least $1-o(1)-\pi_T.$ The second inequality follows from Lemma \ref{lem:nearoptimalcross}, while the final equality follows from an application of (\ref{eq:33}). Next, consider term $R_2$ of (\ref{eq:34})
	\benr\label{eq:37}
	\sup_{\substack{\tau_j\in \cG(c_u T^{-1}\xi^{-2}_j,0);\\\tau_j\ge\tau^0_j}}|R_2|&\le& c_u\xi^{-2}_j\big|\|\h\eta_{(j)}\|_2^2-\|\eta^0_{(j)}\|_2^2\big|= c_u\xi^{-2}_j\big|\|\h\eta_{(j)}-\eta^0_{(j)}\|_2^2+2(\h\eta_{(j)}-\eta^0_{(j)})^T\eta^0_{(j)}\big|\nn\\
	&\le& c_u\xi^{-2}_j\|\h\eta_{(j)}-\eta^0_{(j)}\|_2^2+2c_u\xi^{-1}_j\|\h\eta_{(j)}-\eta^0_{(j)}\|_2=o_p(1),
	\eenr
	wherein the second inequality follows as an application of the Cauchy-Schwarz inequality and the final equality follows from (\ref{eq:33}). Applying (\ref{eq:36}) and (\ref{eq:37}) in the expression (\ref{eq:34}) yields
	\benr
	\sup_{\substack{\tau_j\in \cG_j(c_u T^{-1}\xi^{-2}_j,0);\\\tau_j\ge\tau^0_j}}\big|\cC_j(\tau_j,\h\tau_{-j},\h\theta)-\cC_j(\tau_j,\tau_{-j}^0,\theta^0)\big|\hspace{1in}\nn\\
	\le \sup_{\substack{\tau_j\in \cG_j(c_u T^{-1}\xi^{-2}_j,0);\\\tau_j\ge\tau^0_j}}|R_1|
	+ \sup_{\substack{\tau_j\in \cG_j(c_u T^{-1}\xi^{-2}_j,0);\\\tau_j\ge\tau^0_j}}|R_2|
	=o_p(1)\nn
	\eenr
	Per the discussion in the first paragraph, the only other orientation allowed for any $\tau_j$ in the set $\cG_j(c_u T^{-1}\xi^{-2}_j,0)$ is $\max\{\h\tau_{j-1},\tau_{j-1}^0\}\le\tau_j<\tau_j^0,$ w.p. $1-\pi_T$. Hence, the same bound for this mirroring orientation can be obtained via symmetrical arguments. This completes the proof of the lemma.
\end{proof}

\bc$\rule{3.5in}{0.1mm}$\ec

The proof of Theorem \ref{thm:wc.joint} is also an application of the Argmax Theorem. However, we require preliminary work in order to establish a framework for this problem that can fit into the setup of the theorem. To that end, introduce some additional notation. Let $H\subseteq\{1,...,N\}$ be any finite subset, and let $\h\tau,\h\theta$ be the preliminary estimates as discussed in the main article. Define a new estimator
\benr\label{est:new.simul}
\breve\tau_H=\argmax_{\substack{\tau_H\in\Z^{|H|};\\\h\tau_{j-1}<\tau_j<\h\tau_{j+1};\\ \forall j\in H}}\sum_{j\in H} \cC_j(\tau_j,\h\tau_{-j},\h\theta),
\eenr
with $\cC_j(\tau_j,\h\tau_{-j},\h\theta)$ defined in (\ref{def:cC}). Then, all but the $j^{th}$ summand in (\ref{est:new.simul}) are constants in the $j^{th}$ component of the maximizing argument $\tau_H,$ and thus this estimator (\ref{est:new.simul}) is the same as the component-wise refitted estimates of (\ref{est:optimalcp}), i.e.,
\benr\label{eq:equality.simul.comp}
\breve\tau_j=\tilde\tau_j\quad \forall j\in H.
\eenr

\bc$\rule{3.5in}{0.1mm}$\ec

\begin{proof}[Proof of Theorem \ref{thm:wc.joint}] Let $H\subseteq\{1,...,N\}$ be any finite subset. Recall that by the non-vanishing jump size regime assumption, we have, $\xi_j\to \xi_{(\iny,j)},$ $0<\xi_{(\iny,j)}<\iny,$ $\forall j\in H.$ The proceeding argument shall apply the Argmax theorem in context of the $H$ dimensional sequence $(\breve\tau_H-\tau^0_H)$ of (\ref{est:new.simul}), the limiting result of which shall pass over to the proposed $\tilde\tau_H$ due to the equality (\ref{eq:equality.simul.comp}). Clearly, the underlying indexing metric space is $\Z^{|H|}.$ The reequirements to be verified for the Argmax theorem are:
	\begin{enumerate}
		\item The sequence $(\breve\tau_H-\tau^0_H)$ is uniformly tight in $\Z^{|H|}.$
		\item The $\Z^{|H|}\to \R$ random field $\sum_{j\in H}\cC_{(\iny,j)}(\z_j)$ satisfies suitable regularity conditions.
		\item For any $\z=(\z_1,...,\z_{|H|})^T\in \big\{{-\bf c_u},...,{\bf c_u}\big\}_{|H|\times 1},$ with ${\bf c_u}\in\Z^{+|H|},$ we have
		\benr\label{eq:17}
		\sum_{j\in H}\cC_j(\tau_j^0+\z_j,\h\tau_{-j},\h\theta)\Rightarrow \sum_{j\in H}\cC_{(\iny,j)}(\z_j).
		\eenr
	\end{enumerate}
	
	From Theorem \ref{thm:cpoptimal} we have for each fixed $j\in H,$ the sequence $(\tilde\tau_j-\tau^0_j)$ is uniformly tight in $\Z.$ Then, the equality (\ref{eq:equality.simul.comp}) together with the assumption that $|H|$ is finite, implies Requirement (1). The second requirement is verified in Lemma \ref{lem:mult.regularity} below. To prove requirement (3), first note that from Lemma \ref{lem:Capprox} and the finiteness of $|H|$, we obtain
	\benr
	\sup_{\tau_H\in\cG_H\big({\bf c_u}T^{-1},0\big)}\Big|\sum_{j\in H}\cC_j(\tau_j,\h\tau_{-j},\h\theta)-\sum_{j\in H}\cC_j(\tau_j,\tau^0_{-j},\theta^0)\Big|=o_p(1).\footnotemark\nn
	\eenr
	\footnotetext{Here $\cG_H\big({\bf c_u}T^{-1},0\big)=\cG_{j_1}\big( c_uT^{-1},0\big)\times....\times\cG_{j_{|H|}}\big( c_uT^{-1},0\big),$ for $H=\{j_1,....,j_{|H|}\}.$}
	Thus, to complete the proof of requirement (3) it only remains to show that
	\benr
	\sum_{j\in H}\cC_j(\tau_j^0+\z_j,\tau_{-j}^0,\theta^0)\Rightarrow \sum_{j\in H}\cC_{(\iny,j)}(\z_j),
	\eenr
	where the increments of $\cC_{(\iny,j)}(\z_j)$ are additionally independent over all $j$'s. In all arguments to follow, we assume w.l.o.g. $|H|=2,$ where $H=\{1,2\}.$  Let $\z_1>0,$ and $\z_2<0.$ Proceeding analogously as in (\ref{eq:24}), we get
	\benr\label{eq:25}
	\cC_1(\tau^0_1+\z_1,\tau^0_{-1},\theta^0)+\cC_2(\tau^0_2+\z_2,\tau^0_{-2},\theta^0)\hspace{-2.25in}\nn\\
	&=&\sum_{t=(\tau^0_1+1)}^{\tau^0_1+\z_1}\Big(2\vep_t^{*T}\eta^0_{(1)}-\xi^2_1\Big)+\sum_{t=(\tau^0_2+\z_2+1)}^{\tau^0_2}\Big(2\vep_t^{*T}\eta^0_{(2)}-\xi^2_2\Big)\nn\\
	&=& \sum_{t=(\tau^0_1+1)}^{\tau^0_1+\z_1}\Big(2\vep_t^T\eta^0_{(1)}-\xi^2_1\Big) + \sum_{t=(\tau^0_2+\z_2+1)}^{\tau^0_2}\Big(2\vep_t^T\eta^0_{(2)}-\xi^2_2\Big)-o_p(1)
	\Rightarrow  \sum_{j=1}^2\sum_{t=1}^{|\z_j|} z_{tj},
	\eenr
	where $z_{tj}\sim\cP\big(-\xi_{(\iny,j)}^2,4\xi_{(\iny,j)}^2\si^2_{(\iny,j)}\big),$ for each $t$ and each $j,$ which are independent over all $t$ and $j.$ Independence over $j$'s follows since by Condition B(ii), we have $(\tau_2^0-\tau^0_1)\ge T\lm\to \iny;$ consequently, for $T$ sufficiently large, the two sums of interest are over non-overlapping indices, i.e., $\tau^0_1+\z_1< \tau^0_2+\z_2.$ The second equality of (\ref{eq:25}) follows analogously to (\ref{eq:16}). The weak convergence follows from Condition A$'$. The remaining permutations of the signs of $\z_1,\z_2$ can be handled symmetrically to yield the same result. This completes the verification of requirement (3). An application of the Argmax theorem together with the equality (\ref{eq:equality.simul.comp}) now yields
	\benr
	(\tilde\tau_H-\tau^0_H)=(\breve\tau_H-\tau^0_H)\Rightarrow \argmax_{\z\in\Z^{|H|}}\sum_{j\in H}\cC_{(\iny,j)}(\z_j)\nn
	\eenr
	thereby establishing the first claim of the theorem. Next, note that
	\benr\label{eq:26}
	\argmax_{\z\in\Z^{|H|}}\sum_{j\in H}\cC_{(\iny,j)}(\z_j)= \Big(\argmax_{\z_1\in\Z}\cC_{(\iny,1)}(\z_1),\,\,\argmax_{\z_2\in\Z}\cC_{(\iny,2)}(\z_2)\Big)^T.
	\eenr
	This equality follows along the same lines as (\ref{eq:equality.simul.comp}).Also note that (\ref{eq:26}) is an exact equality and not just equality in distribution. The independence of $\cC_{(\iny,j)}(\z_j)$ over $j$ as discussed earlier implies
	\benr
	\Big(\argmax_{\z_1\in\Z}\cC_{(\iny,1)}(\z_1),\,\,\argmax_{\z_2\in\Z}\cC_{(\iny,2)}(\z_2)\Big)^T=^d\Pi_{j\in H}\argmax_{\z_j\in\Z}\cC_{(\iny,j)}(\z_j),\nn
	\eenr
	which proves the second claim of this theorem. The final claim of asymptotic independence of $\tilde\tau_j,$ over $j\in H$ now follows by comparing (\ref{eq:28}) to the marginal distributions obtained in Theorem \ref{thm:wc.non.vanishing}. This completes the proof of the theorem.
\end{proof}

\bc$\rule{3.5in}{0.1mm}$\ec

\begin{lemma}\label{lem:mult.regularity} Suppose Conditions A$'$, B and D hold, and let $H\subseteq\{1,2,...,N\}$ be any finite subset and assume the non-vanishing jump size regime of $\xi_j\to\xi_{(\iny,j)},$ $0<\xi_{(\iny,j)}<\iny,$ $\forall j\in H.$ Let $\cC_{(\iny,j)}(\z_j)$ be as defined in (\ref{eq:42}). Then, the $\Z^{|H|}\to \R$ map
	$\sum_{j\in H}\cC_{(\iny,j)}(\z_j)$ is continuous with respect to the domain space. Additionally, $\argmax_{\z\in\Z^{|H|}}\sum_{j\in H}\cC_{(\iny,j)}(\z_j)$ possesses an almost sure unique maximum at $\omega_{\iny}$ which as a random map in $\Z^{|H|}$ is tight.
\end{lemma}

\begin{proof}[Proof of Lemma \ref{lem:mult.regularity}] This proof has been adapted from Lemma A.3 of \cite{kaul2020inference} for the process under consideration. Continuity of sample paths of the random field $\sum_{j\in H}\cC_{(\iny,j)}(\z_j)$ follows trivially since the domain space $\Z^{|H|}$ is discrete. Next, from Condition A$'$ we have that incremental distributions are continuous, thus, if $\max_{\z\in\Z^{|H|}}\sum_{j\in H}\cC_{(\iny,j)}(\z_j)<\iny$ a.s. then $\omega_{\iny}$ must be unique and tight. Consequently, the only thing that remains to show is that $\max_{\z\in\Z^{|H|}}\sum_{j\in H}\cC_{(\iny,j)}(\z_j)<\iny$ a.s., for this purpose, w.l.o.g. let $|H|=2$ with $H=\{1,2\}.$  Now consider any fixed $\z_2=\z_2^0,$ and note that $\sum_{j\in H}\cC_{(\iny,j)}(\z_j)$ is a two sided random walk over $\z_1.$ Moreover, under the assumed non-vanishing jump size this two sided random walk is negative drift, from Condition D the incremental variances are finite, and from the assumed underlying subexponential distribution, all moments of incremental distributions exist.  Consequently, we have $\sum_{j\in H}\cC_{(\iny,j)}(\z_j)\to-\iny,$ as $\z_1\to \iny$ or $\z_1\to \iny,$ a.s. (strong law of large numbers). This implies that $\max_{\z_1\in\Z}\sum_{j\in H}\cC_{(\iny,j)}(\z_j)<\iny,$ a.s. (follows from the Hewitt-Savage $0$-$1$ law, see, e.g. (1.1) and (1.2) on Page 172, 173 of \cite{durrett2010probability}). Now applying union bounds over the countable collection of $\z_2\in\Z,$ yields $\max_{\z_2\in\Z}\max_{\z_1\in\Z}\sum_{j\in H}\cC_{(\iny,j)}(\z_j)<\iny,$ a.s. (countable intersection of a.s. events is a.s.), thereby completing the proof of the lemma.
\end{proof}

\bc$\rule{3.5in}{0.1mm}$\ec

The proof of Theorem \ref{thm:mean.rate} requires some preliminary work. For any non-negative sequence $u_T\le 1,$ define the collection,
\benr\label{def:set.cG.max}
\overline\cG(u_T)&=&\Big\{\tau\in\{1,...,T-1\}^N;\,\,\max_{1\le j\le N}|\tau_{j}-\tau^0_{j}|\le Tu_T\Big\}\nn
\eenr
We begin by first examining the behavior of the estimates $\tilde\theta_{(j)}(\tau),$ $j=1,...,N+1,$  uniformly over the collection $\overline\cG(u_T).$ This is provided in the following theorem.

\bc$\rule{3.5in}{0.1mm}$\ec

\begin{theorem}\label{thm:unifmean} Suppose Conditions A and B(i, ii) hold and let $\overline\cG$ be as defined in (\ref{def:set.cG.max}). Let $0\le u_{T}\le c_{u1}\lm,$ be any sequence with a suitably small constant $c_{u1}>0,$ and let $\psi=\max_j\|\eta^0_{(j)}\|_{\iny}.$ Further, assume $T\lm\ge \log(p\vee T)$ and for any constants $c_{u},c_{u2},>0,$ and $j=1,...,N,$ let
	\benr\label{eq:la}
	\la_j=\la=16\max\Big[\si\Big\{\frac{2c_{u2}\log(p\vee T)}{c_uT\lm}\Big\}^{\frac{1}{2}},\,\,\frac{u_T\psi}{c_u\lm}\Big].
	\eenr	
	Then, $\h\theta_{(j)}(\tau),$ $j=1,...,N+1$ of (\ref{est:softthresh}) satisfy the following two results\\~
	(i) For any $\tau\in \overline\cG_w(u_T),$ and any $j=1,...,N+1,$ such that $(\tau_j-\tau_{j-1})\ge c_uT\lm,$  we have $\big\|\big(\h\theta_{(j)}(\tau)\big)_{S_j^c}\big\|_1\le 3\big\|\big(\h\theta_{(j)}(\tau)-\theta_{(j)}^0\big)_{S_j}\big\|_1,$ for sets $S_j$ as defined in (\ref{def:setS}).\\~
	(ii) The following bound is satisfied
	\benr
	\max_{1\le j\le N+1}\,\,\sup_{\substack{\tau\in\overline\cG(u_{T});\\ \min_j(\tau_j-\tau_{j-1})\ge c_uT\lm}} \|\h\theta_{(j)}(\tau)-\theta_{(j)}^0\|_2\le 6\surd{(Ns)}\la,\nn
	\eenr
	where both parts (i) and (ii) hold with probability at least $1-4\exp\big\{-(c_{u3}-4)\log(p\vee T)\big\},$ $c_{u3}=c_{u2}\wedge \surd(c_{u2}c_{u}/2).$.
\end{theorem}

\begin{proof}[Proof of Theorem \ref{thm:unifmean}] We begin with an observation that proves useful for the ensuing argument. The assumption $u_T\le c_{u1}\lm,$ yields that for any $\tau=(\tau_1,...,\tau_N)^T\in\overline\cG(u_T),$ we obtain $\max_j|\tau_j-\tau_j^0|\le c_{u1}T\lm.$ recall from Condition B(ii), THAT all change points are separated by at least $T\lm,$ i.e., $\min_j(\tau_j-\tau_{j-1})\ge T\lm.$ Consequently, any $\tau\in\overline\cG(u_T)$ must satisfy any one of the four orientations $\tau^0_{j-1}\le\tau_{j-1}<\tau_j^0\le\tau_{j},$ $\tau_{j-1}\le\tau_{j-1}^0<\tau_j^0\le\tau_{j},$ $\tau^0_{j-1}\le\tau_{j-1}<\tau_j\le\tau_{j}^0$ or $\tau_{j-1}\le\tau_{j-1}^0<\tau_j\le\tau_{j}^0,$  for any $j=1,...,N.$ No other orientations are feasible under these assumed conditions. In view of this observation, w.l.o.g. we assume one of the first of these four possible orientations, $\tau_{j-1}^0\le\tau_{j-1}<\tau_{j}^0\le \tau_{j}$ in the argument to follow. The remaining three permutations of the ordering of $\tau_{j-1},\tau_{j}$ w.r.t. $\tau^0_{j-1},\tau^0_{j},$ can be proved using symmetrical arguments.
	
	Let $\tau\in\overline\cG(u_T)$ additionally satisfy the relation $\min_j(\tau_j-\tau_{j-1})\ge c_uT\lm,$ then an algebraic rearrangement of the elementary inequality $\big\|\bar x_{(j)}(\tau)-\h\theta_{(j)}(\tau)\big\|^2_2+\la_j\|\h\theta_{(j)}(\tau)\|_1\le \big\|\bar x_{(j)}(\tau)-\theta_{(j)}^0\big\|^2_2+\la_j\|\theta_{(j)}^0\|_1$ yields,
	
	\benr\label{eq:20}
	\big\|\h\theta_{(j)}(\tau)-\theta_{(j)}^0\big\|_2^2+\la_j\big\|\h\theta_{(j)}(\tau)\big\|_1&\le& \la_j\big\|\theta^0_{(j)}\big\|_1+ 2\sum_{t=\tau_{j-1}+1}^{\tau_j}\h\vep_t^{*T}(\h\theta_{(j)}(\tau)-\theta_{(j)}^0),\nn\\
	&=& \la_j\big\|\theta^0_{(j)}\big\|_1+\frac{2}{(\tau_j-\tau_{j-1})}\sum_{t=\tau_{j-1}+1}^{\tau_{j}}\vep_t^{*T}(\h\theta_{(j)}(\tau)-\theta_{(j)}^0)\nn\\
	&&-2\frac{(\tau_j-\tau^0_j)}{(\tau_j-\tau_{j-1})}(\theta_{(j)}^0-\theta_{(j+1)}^0)^T(\h\theta_{(j)}(\tau)-\theta_{(j)}^0)\nn\\
	&\le &\la_j\big\|\theta^0_{(j)}\big\|_1+\frac{2}{(\tau_j-\tau_{j-1})}\Big\|\sum_{t=\tau_{j-1}}^{\tau_j}\vep_t^*\Big\|_{\iny}\big\|\h\theta_{(j)}(\tau)-\theta_{(j)}^0\big\|_1\nn\\
	&&+\frac{2u_T}{c_u\lm}\psi\big\|\h\theta_{(j)}(\tau)-\theta_{(j)}^0\big\|_1,
	\eenr
	where in the first inequality we have $\h\vep_{t}^*=\big(x_t-\theta_{(j)}^0\big).$ The last inequality follows since $\tau\in\overline\cG(u_T),$ and by definition $\|\theta^0_{(j)}-\theta^0_{(j+1)}\|_{\iny}\le \psi.$  Now using the bound of Lemma \ref{lem:1.to.tau.bound} we have that,
	\benr\label{eq:30}
	\frac{2}{(\tau_j-\tau_{j-1})}\Big\|\sum_{t=\tau_{j-1}}^{\tau_j}\vep_t^*\Big\|_{\iny}\le 4\surd(2c_{u2}/c_{u})\si\Big\{\frac{\log(p\vee T)}{T\lm}\Big\}^{\frac{1}{2}}
	\eenr	
	with probability at least  $1-4\exp\big\{-(c_{u3}-4)\log(p\vee T)\big\},$ $c_{u3}=c_{u2}\wedge \surd(c_{u2}c_{u}/2).$ Consequently, upon choosing,
	\benr
	\la^*=8\max\Big\{\surd(2c_{u2}/c_{u})\si\Big\{\frac{\log(p\vee T)}{Tl_T}\Big\}^{\frac{1}{2}},\,\,\frac{u_T\psi}{c_ul_T}\Big\},\nn
	\eenr
	and substituting in (\ref{eq:20}), we obtain
	\benr\label{eq:21}
	\big\|\h\theta_{(j)}(\tau)-\theta_{(j)}^0\big\|_2^2+\la_j\big\|\h\theta_{(j)}(\tau)\big\|_1\le\la_j\big\|\theta^0_{(j)}\big\|_1+\la^*\big\|\h\theta_{(j)}(\tau)-\theta_{(j)}^0\big\|_1,
	\eenr
	with probability at least  $1-4\exp\big\{-(c_{u2}-4)\log(p\vee T)\big\}.$ choosing $\la_j= 2\la^*,$ leads to $\|\big(\h\theta_{(j)}(\tau)\big)_{S_j^c}\|_1\le 3\|\big(\h\theta_{(j)}(\tau)-\theta_{(j)}^0\big)_{S_j}\|_1,$ which upon noting that the bound (\ref{eq:30}) arises from Lemma \ref{lem:1.to.tau.bound} which holds uniformly over $j$ as well as over all considered values of $\tau,$ proves part (i) of this theorem.

	Next, from inequality (\ref{eq:21}) we also have that,
	\benr
	\|\h\theta_{(j)}(\tau)-\theta_{(j)}^0\|_2^2\le \frac{3}{2}\la_j\|\h\theta_{(j)}(\tau)-\theta_{(j)}^0\|_1\le 6\la_j\surd{(Ns)}\|\h\theta_{(j)}(\tau)-\theta_{(j)}^0\|_2
	\eenr
	This directly implies that  $\|\h\theta_{(j)}(\tau)-\theta_{(j)}^0\|_2\le 6\la_j\surd{(Ns)},$ where we have used $\|\h\theta_{(j)}(\tau)-\theta_{(j)}^0\|_1\le 4\sqrt{(Ns)}\|\h\theta_{(j)}(\tau)-\theta_{(j)}^0\|_2,$ which follows in turn Part (i). To complete the proof of this part recall that the only stochastic bound used here is the uniform bound of Lemma \ref{lem:1.to.tau.bound}; consequently, the final bound also holds uniformly over the same collection. This result can alternatively be proved using the properties of the soft-thresholding operator $k_{\la}(\cdotp),$ by building uniform versions of arguments such as those in \cite{rothman2009generalized}, or \cite{kaul2017structural}.
\end{proof}

\bc$\rule{3.5in}{0.1mm}$\ec

\begin{proof}[Proof of Theorem \ref{thm:mean.rate}] To prove the first claim, note that by Condition E$'$(i) we have,
	\benr\label{eq:31}
	c_u\si^2\xim^{-2}Ns\log^2(p\vee T)\le c_{u1}T\lm
	\eenr
	This relation together with assumed properties (\ref{eq:29}) of the preliminary change point estimates imply that Condition C(i) is satisfied. The remaining claims of this theorem are largely an application of Theorem \ref{thm:unifmean}.  Note that relations (\ref{eq:29}) and (\ref{eq:31}) imply that $\h\tau$ lies in the collection over which the uniform results of Theorem \ref{thm:mean.rate} are established, i.e., $\h\tau\in\overline\cG(u_T),$ $u_T=c_{u}T^{-1}\si^2\xi^{-2}Ns\log^2(p\vee T)\le\lm,$ and $\min_j(\h\tau_{j-1}-\h\tau_j)\ge c_uT\lm,$  w.p. $1-\pi_T.$  Now consider $\la$ as defined in (\ref{eq:la}) with this choice of $u_T,$
	\benr
	\la&=&c_u\max\Big[\si\Big\{\frac{\log(p\vee T)}{T\lm}\Big\}^{\frac{1}{2}},\,\,\frac{\psi \si^2\xim^{-2}Ns\log^2(p\vee T)}{T\lm}\Big]\nn\\
	&=& c_u\si\Big\{\frac{\log(p\vee T)}{T\lm}\Big\}^{\frac{1}{2}}\max\Big[1,\,\, \Big(\frac{\psi}{\xim}\Big)\Big(\frac{\si}{\xim}\Big)\Big\{\frac{Ns\log^{3/2}(p\vee T)} {\surd{(T\lm)}}\Big\}\Big]\nn\\
	&\le & c_u\si\Big\{\frac{\log(p\vee T)}{T\lm}\Big\}^{\frac{1}{2}},\nn
	\eenr
	wherein the inequality follows by using the assumption $\psi\big/\xim=O(1),$ together with Condition E$'$(ii). The second claim and the bound (\ref{eq:32}) now follows from the corresponding results of Theorem \ref{thm:unifmean}. To establish the final claim, note that from Condition E$'$(ii) we also have that,
	\benr\label{eq:35}
	c_u\si\Big\{\frac{Ns\log(p\vee T)}{T\lm}\Big\}^{\frac{1}{2}}\le \frac{c_{u1}\xim}{(Ns)^{1/2}\log(p\vee T)}
	\eenr
	Thus, (\ref{eq:32}) together with (\ref{eq:35}) imply that $\h\theta_{(j)}(\h\tau),$ $j=1,....,N+1$ satisfy all requirements of Condition C(ii). This completes the proof of the theorem.
\end{proof}

\bc$\rule{3.5in}{0.1mm}$\ec

\begin{proof}[Proof of Corollary \ref{cor:alg1.validity}] This result is a direct consequence of Theorem \ref{thm:mean.rate} and the results of Section \ref{sec:main}. In particular, under the assumed conditions, Theorem \ref{thm:mean.rate} yields that the preliminary estimates $\h\tau$ and $\h\theta$ satisfy all requirements of Condition C. All claims of this now follow from corresponding results of Section \ref{sec:main}.
\end{proof}

\bc$\rule{3.5in}{0.1mm}$\ec

\section{Deviation bounds}\label{app:deviation}	

\begin{lemma}\label{lem:near.opt.unif.comp} Assume Condition A holds and let $\bar\vep=\sum_{t=1}^T\vep_t\big/T.$ Then, for any $c_u>0,$ we get
	\benr
	\|\bar\vep\|_{\iny}\le
	\begin{cases}
		\si\surd\{2c_u\log (p\vee T)/T\}, & {\rm when}\,\, T\ge 2c_u\log (p\vee T)\\
		2c_u\si\log (p\vee T)/\surd T, & {\rm when}\,\, T\ge 1,\\ 	
	\end{cases}
	\eenr
	with probability at least $1-2\exp\{-(c_u-1)\log(p\vee T)\}.$ Further, for any non-random $\delta\in\R^p,$ $\|\delta\|_2=1,$ we have,  $\surd{T}\delta^T\bar\vep=O_p(1).$ More precisely, for any $0<a<1,$ choosing $c_a=\surd{(1/a)},$ we have, $pr\big(|\surd{T}\delta^T\bar\vep|>4c_a\si\big)\le a.$  	
\end{lemma}

\begin{proof}[Proof of Lemma \ref{lem:near.opt.unif.comp}] Applying Bernstein's inequality  (Lemma \ref{lem:bernstein}) for each $k=1,...,p,$, we obtain
	\benr\label{eq:2}
	pr\Big(\big|\sum_{t=1}^{T}\vep_{tk}\big|>dT\Big)\le 2\exp\Big\{-\frac{T}{2}\Big(\frac{d^2}{\si^2}\wedge\frac{d}{\si}\Big)\Big\}.
	\eenr
	In the case where $T\ge 2c_u\log (p\vee T),$ select $d=\si\{2c_u\log(p\vee T)/T\}^{1/2}$ to get $(d^2/\si^2)\wedge (d/\si)=d^2/\si^2.$ Substituting $d$ in (\ref{eq:2}) and applying union bounds over $k=1,...p$ yields the desired bound for this case. In the case where $T\ge 1,$ select $d=2c_u\si\{\log^2(p\vee T)/T\}^{1/2},$ and note that
	\benr
	\frac{T}{2}\Big(\frac{d^2}{\si^2}\Big)= 2c_u\log^2(p\vee T),\quad{\rm and}\quad
	\frac{T}{2}\Big(\frac{d}{\si}\Big)\ge c_u \log(p\vee T).\nn
	\eenr
	Since in this case the latter expression is smaller,  substituting this choice of $d$ in (\ref{eq:2}) and applying union bounds over $k=1,...p$ yields the desired bound. The second claim follows from the Markov inequality upon noting that $\surd{T}\delta^T\bar\vep\sim {\rm subE}(\si^2)$ (Lemma \ref{lem:lcsubE}) together with a second moment bound for subexponential distributions (Lemma \ref{lem:momentprop}).
\end{proof}

\bc$\rule{3.5in}{0.1mm}$\ec

\begin{lemma}\label{lem:nearoptimalcross} Assume that Conditions A and B(i) hold and $\vep_t^*,$ $t=1,...,T$ be as defined in (\ref{model:center.rvmcp}). Let	
	$0\le v_T\le u_T\le 1,$ be any non-negative sequences. Then, for any $c_u\ge 1,$ we get
	\benr
	\sup_{\substack{\tau_j\in\cG_j(u_T,v_T);\\\tau_j\ge\tau^0_j}}\Big\|\sum_{t=\tau^0_j+1}^{\tau_j}\vep_{t}^*\Big\|_{\iny}\le 4c_u\si\log (p\vee T)\surd \big(Tu_T\big),\quad\textrm{for any given}\,\, j=1,...,N.
	\eenr
	with probability at least $1-4\exp\{-(c_u-2)\log(p\vee T)\}.$ 	
\end{lemma}

\begin{proof}[Proof of Lemma \ref{lem:nearoptimalcross}]
	Without loss of generality assume $v_T\ge (1/T)$ (else, the sum of interest is over an empty set of indices and trivially zero). Consider any $k\in\{1,2,...,p\}$ and any $\tau_j>\tau^0_j,$ and apply Bernstein's inequality (Theorem \ref{lem:bernstein}) for any $d>0$ to obtain
	\benr\label{eq:2b}
	pr\Big(\big|\sum_{t=\tau^0_j+1}^{\tau_j}\vep_{tk}\big|>d(\tau_j-\tau^0_j)\Big)\le 2\exp\Big\{-\frac{(\tau_j-\tau^0_j)}{2}\Big(\frac{d^2}{\si^2}\wedge\frac{d}{\si}\Big)\Big\}.
	\eenr
	Select $d=2c_u\si\{\log^2 (p\vee T)/(\tau_j-\tau^0_j)\}^{1/2},$ and note that
	\benr
	(\tau_j-\tau^0_j)\frac{d^2}{2\si^2}&=&2c_u^2\log^2(p\vee T),\quad {\rm and},\nn\\
	(\tau_j-\tau^0_j)\frac{d}{2\si}&\ge& c_u\log(p\vee T),\nn
	\eenr
	where we have used $(\tau_j-\tau^0_j)\ge Tv_T\ge 1$ to obtain the inequality. Substituting this choice of $d$ in (\ref{eq:2b}), we obtain
	\benr
	\big|\sum_{t=\tau^0_j+1}^{\tau_j}\vep_{tk}\big|\le 2c_u\si(\tau_j-\tau^0_j)^{1/2}\{\log^2(p\vee T)\}^{1/2}\le 2c_u\si\{Tu_T\log^2 (p\vee T)\}^{1/2},\nn
	\eenr
	w.p. at least $1-2\exp\{-c_u\log (p\vee T)\}.$ Applying union bounds over $k=1,...,p$ and $T$ possible distinct values of $\tau_j,$ yields,
	\benr
	\sup_{\substack{\tau_j\in\cG_j(u_T,v_T);\\\tau_j\ge\tau^0_j}}\big\|\sum_{t=\tau^0_j+1}^{\tau_j}\vep_{t}\big\|_{\iny}\le 2c_u\si\{Tu_T\log^2 (p\vee T)\}^{1/2},\nn
	\eenr
	w.p. at least $1-2\exp\{-(c_u-2)\log (p\vee T)\}.$ Finally, recall from (\ref{model:center.rvmcp}) that $\vep_t^*=\vep_t-\bar\vep,$ consequently,
	\benr\label{eq:7}
	\sup_{\substack{\tau_j\in\cG_j(u_T,v_T);\\ \tau_j\ge\tau^0_j}}\Big\|\sum_{t=\tau^0_j+1}^{\tau_j}\vep_t^*\Big\|_{\iny}&\le& \sup_{\substack{\tau_j\in\cG_j(u_T,v_T);\\\tau_j\ge\tau^0_j}}\Big\|\sum_{t=\tau^0_j+1}^{\tau_j}\vep_t\Big\|_{\iny} + Tu_T\|\bar\vep\|_{\iny}\nn\\
	&\le& 2c_u\si\{Tu_T\log^2(p\vee T)\}^{\frac{1}{2}}+ 2c_u\si u_T\{T\log^2(p\vee T)\}^{\frac{1}{2}}\nn\\
	&\le& 2c_u\si\{Tu_T\log^2(p\vee T)\}^{\frac{1}{2}}\big[1+\surd{u_T}\big]\nn\\
	&\le& 4c_u\si\{Tu_T\log^2(p\vee T)\}^{\frac{1}{2}},
	\eenr
	w.p. at least $1-2\exp\{-(c_u-2)\log (p\vee T)\}-2\exp\{-(c_u-1)\log(p\vee T)\}\ge 1-4\exp\{-(c_u-2)\log(p\vee T)\}.$ The second inequality follows from Lemma \ref{lem:near.opt.unif.comp} and the final inequality follows from $u_T\le 1.$ This completes the proof of the lemma.
\end{proof}

\bc$\rule{3.5in}{0.1mm}$\ec

\begin{lemma}\label{lem:nearoptimalcross.maximums} Assume that Conditions A and B(i) hold and $\vep_t^*,$ $t=1,...,T$ be as defined in (\ref{model:center.rvmcp}). Let	
	$0\le v_T\le u_T\le 1,$ be any non-negative sequences. Then, for any $c_u\ge 1,$ we have
	\benr
	(i)\,\,\max_{1\le j\le N}\sup_{\substack{\tau_j\in\cG_j(u_T,v_T);\\\tau_j\ge\tau^0_j}}\Big\|\sum_{t=\tau^0_j+1}^{\tau_j}\vep_{t}^*\Big\|_{\iny}\le 4c_u\si\log (p\vee T)\surd\big(Tu_T\big),
	\eenr
	with probability at least $1-4\exp\{-(c_u-3)\log(p\vee T)\}.$ Additionally, let $\xma$ be as in (\ref{def:jumpsize}), then,
	\benr
	(ii)\,\, \max_{1\le j\le N} \sup_{\substack{\tau_j\in\cG_j(u_T,v_T);\\\tau_j\ge\tau^0_j}}\Big|\sum_{t=\tau^0_j+1}^{\tau_j}\vep_{t}^{*T}\eta^0_{(j)}\Big|\le 2c_u\xma\si\{Tu_T\log^2 T\}^{1/2},
	\eenr
	with probability at least $1-4\exp\{-(c_u-2)\log T\}.$
\end{lemma}

\begin{proof}[Proof of Lemma \ref{lem:nearoptimalcross.maximums}] The first part of this lemma is a direct application of Lemma \ref{lem:nearoptimalcross} and is obtained by supplying an additional union bound over $j=1,..,N,$ and noting that $N\le T.$ To establish Part (ii), we have $\vep_t^T\eta^0_{(j)}\sim {\rm subE}(\xi_j^2\si^2),$ for each $t=1,...,T.$ Now proceed analogously to Lemma \ref{lem:nearoptimalcross} by applying Bernstein's inequality to obtain for any given $j=1,...,N$ and $d>0,$
	\benr\label{eq:6}
	pr\Big(\big|\sum_{t=\tau^0_j+1}^{\tau_j}\vep_{t}^{T}\eta^0_{(j)}\big|>d(\tau_j-\tau^0_j)\Big)\le 2\exp\Big\{-\frac{(\tau_j-\tau^0_j)}{2}\Big(\frac{d^2}{\xi_j^2\si^2}\wedge\frac{d}{\xi_j\si}\Big)\Big\}.
	\eenr	
	Selecting $d=2c_u\xi_j\si\{\log^2 T/(\tau_j-\tau^0_j)\}^{1/2}$ and substituting in (\ref{eq:6}), we obtain	
	\benr
	\big|\sum_{t=\tau^0_j+1}^{\tau_j}\vep_{t}^T\eta^0_{(j)}\big|\le 2c_u\xi_j\si\{Tu_T\log^2  T\}^{1/2},\nn
	\eenr
	w.p. at least $1-2\exp\{-c_u\log T\}.$ Supplying union bounds over $T$ possible distinct values of $\tau_j$ and over $j=1,...,N,$ and that by definition $\xi_j\le \xma,$ we obtain
	\benr
	\max_{1\le j\le N}\sup_{\substack{\tau_j\in\cG_j(u_T,v_T);\\\tau_j\ge\tau^0_j}}\big|\sum_{t=\tau^0_j+1}^{\tau_j}\vep_{t}^T\eta^0_{(j)}\big|\le 2c_u\xma\si\{Tu_T\log^2  T\}^{1/2},\nn
	\eenr
	w.p. at least $1-2\exp\{-(c_u-2)\log T\}.$ In order to obtain the analogous bound w.r.t. $\vep_t^*,$ note that $\surd{T}\bar\vep^T\eta^{0}_{(j)}\sim {\rm subE}(\xi_j^2\si^2).$ Again employing Bernstein's inequality together with union bounds over $j=1,...,N,$ we get
	\benr\label{eq:15}
	\max_{1\le j\le N}\big|\bar\vep^T\eta^{0}_{(j)}\big|\le 2c_u\xma\si \{\log^2 T/T\}^{1/2},
	\eenr
	w.p. at least $1-2\exp\{-(c_u-1)\log T\}.$ Next, proceeding as in (\ref{eq:7}), we obtain
	\benr
	\max_{1\le j\le N}\sup_{\substack{\tau_j\in\cG_j(u_T,v_T);\\\tau_j\ge\tau^0_j}}\big|\sum_{t=\tau^0_j+1}^{\tau_j}\vep_{t}^{*T}\eta^0_{(j)}\big| &\le& \max_{1\le j\le N} \sup_{\substack{\tau_j\in\cG_j(u_T,v_T);\\\tau_j\ge\tau^0_j}}\big|\sum_{t=\tau^0_j+1}^{\tau_j}\vep_{t}^T\eta^0_{(j)}\big|+ Tu_T\max_{1\le j\le N}\big|\bar\vep^T\eta^{0}_{(j)}\big|\nn\\
	&\le& 2c_u\xma\si\{Tu_T\log^2 T\}^{1/2}+ 2c_u\xma\si \{T\log^2 T\}^{1/2}\nn\\
	&\le& 4c_u\xma\si\{Tu_T\log^2 T\}^{1/2},\nn
	\eenr
	w.p. at least $1-4\exp\{-(c_u-2)\log T\},$ which completes the proof of the lemma.
\end{proof}

\bc$\rule{3.5in}{0.1mm}$\ec

\begin{lemma}\label{lem:optimalcross} Assume that Conditions A and B(i) hold and let $u_T,v_T$ be any non-negative sequences satisfying $0\le v_T\le u_T\le 1.$ Then, for any $0<a<1,$ with $c_a\ge \surd(1/a)$ and for any given $j=1,...,N,$ we get
	\benr
	\sup_{\substack{\tau_j\in\cG_j(u_T,v_T);\\\tau_j\ge\tau^0_j}}\Big|\sum_{t=\tau^0_j+1}^{\tau_j}\vep_{t}^{*T}\eta^0_{(j)}\Big|\le 8c_{a}\si\xi_j\surd (T u_T),\nn
	\eenr
	with probability at least $1-2a.$	
\end{lemma}

\begin{proof}[Proof of Lemma \ref{lem:optimalcross}]
	This result is largely an application of Kolmogorov's inequality (Theorem \ref{thm:kolmogorov}). For any given $j=1,...,N,$ we have
	\benr
	{\rm var}(\vep_t^T\eta^0_{(j)})\le 16\xi_j^2\si^2\nn
	\eenr
	where the inequality follows from Lemma \ref{lem:momentprop}. Next, note that there are at most $Tu_T$ distinct values of $\tau_j$ in the set $\cG_j(u_T,v_T).$ Now apply Kolmogorov's inequality (Theorem \ref{thm:kolmogorov}) for any $d>0$ to obtain
	\benr
	pr\Big(\sup_{\substack{\tau_j\in\cG_j(u_T,v_T);\\\tau_j\ge\tau^0_j}}\Big|\sum_{t=\tau^0_j+1}^{\tau_j}\vep_{t}^T\eta^0_{(j)}\Big|>d\Big)\le \frac{Tu_T}{d^2}16\si^2\xi_j^2.\nn
	\eenr
	Selecting $d=4c_{a}\si\xi_j\surd (Tu_T)$ with $c_{a}\ge \surd (1/a)$ yields
	\benr\label{eq:8}
	\sup_{\substack{\tau_j\in\cG_j(u_T,v_T);\\\tau_j\ge\tau^0_j}}\Big|\sum_{t=\tau^0_j+1}^{\tau_j}\vep_{t}^{T}\eta^0_{(j)}\Big|\le 4c_{a}\si\xi_j\surd (T u_T),\nn
	\eenr
	w.p. at least $1-a.$ The analogous bound  w.r.t $\vep_t^*$ can be obtained as
	\benr
	\sup_{\substack{\tau_j\in\cG_j(u_T,v_T);\\\tau_j\ge\tau^0_j}}\Big|\sum_{t=\tau^0_j+1}^{\tau_j}\vep_{t}^{*T}\eta^0_{(j)}\Big|&\le& 	\sup_{\substack{\tau_j\in\cG_j(u_T,v_T);\\\tau_j\ge\tau^0_j}}\Big|\sum_{t=\tau^0_j+1}^{\tau_j}\vep_{t}^{T}\eta^0_{(j)}\Big| + Tu_T|\bar\vep^T\eta^0_{(j)}|\nn\\
	&\le& 4c_{a}\si\xi_j\surd (T u_T)+ 4c_{a}\si\xi_j u_T\surd{T}\nn\\
	&\le& 8c_{a}\si\xi_j\surd (T u_T),\nn
	\eenr
	w.p. at least $1-2a.$ Here the second inequality follows from (\ref{eq:8}) together with the second claim of Lemma \ref{lem:near.opt.unif.comp}. This completes the proof of the lemma.
\end{proof}

\bc$\rule{3.5in}{0.1mm}$\ec

\begin{lemma}\label{lem:1.to.tau.bound} Assume Conditions A and B(i) hold and that $T\lm\ge\log (p\vee T).$ Then, for any $c_u,c_{u1}>0,$ we have
	\benr\label{eq:1}\label{eq:22}
	\max_{1\le j\le N+1}\sup_{\substack{\tau_{j-1},\tau_{j}\in\{1,.....,T-1\};\\ (\tau_{j}-\tau_{j-1})\ge c_uT\lm}}\frac{1}{(\tau_{(j)}-\tau_{(j-1)})}\Big\|\sum_{t=\tau_{j-1}+1}^{\tau_{j}}\vep_{t}^*\Big\|_{\iny}\hspace{0.75in}\nn\\
	\le  2\si\Big\{\frac{2c_{u1}\log(p\vee T)}{c_uT\lm}\Big\}^{\frac{1}{2}}\nn
	\eenr
	with probability at least $1-4\exp\big\{-(c_{u2}-4)\log (p\vee T)\big\},$ where $c_{u2}=c_{u1}\wedge\surd(c_uc_{u1}/2).$
\end{lemma}

\begin{proof}[Proof of Lemma \ref{lem:1.to.tau.bound}]
	For any given $j=1,...,N$ consider any  $\tau_{j-1}<\tau_{j}\in\{1,...,T\}$ satisfying $(\tau_j-\tau_{j-1})\ge c_uT\lm,$ and any $k\in\{1,...,p\}.$ Then applying the Bernstein's inequality (Lemma \ref{lem:bernstein}) for any $d>0,$ we obtain,
	\benr\label{eq:23}
	pr\Big(\Big|\sum_{t=\tau_{j-1}+1}^{\tau_{j}}\vep_{tk}\Big|>d(\tau_{j}-\tau_{j-1})\Big)\le 2\exp\Big\{-\frac{(\tau_{j}-\tau_{(j-1)})}{2}\Big(\frac{d^2}{\si^2}\wedge\frac{d}{\si}\Big)\Big\}.
	\eenr	
	Choose $d=\si\Big\{2c_{u1}\log (p\vee T)\big/\big(\tau_{(j)}-\tau_{(j-1)}\big)\Big\}^{1/2},$ then, we have,
	\benr
	\big(\tau_{j}-\tau_{j-1}\big)\frac{d^2}{2\si^2}&=&c_{u1}\log(p\vee T),\quad {\rm and},\nn\\
	\big(\tau_{j}-\tau_{j-1}\big)\frac{d}{2\si}&\ge&\surd(c_{u1}/2)(c_uT\lm)^{1/2}\{\log(p\vee T)\}^{1/2}\nn\\
	&\ge& \surd(c_uc_{u1}/2)\log(p\vee T).\nn
	\eenr	
	The first inequality follows since by choice $\big(\tau_{j+1}-\tau_{j-1}\big)\ge c_u T\lm,$ and the second inequality follows by assumption $T\lm\ge\log (p\vee T).$ Substituting this choice of $d$ in (\ref{eq:23}),  we obtain,	
	\benr
	\frac{1}{\big(\tau_{j}-\tau_{j-1}\big)}\Big|\sum_{t=\tau_{j-1}+1}^{\tau_{j}}\vep_{tk}\Big|&\le& \si\Big\{2c_{u1}\log (p\vee T)\big/\big(\tau_{j}-\tau_{j-1}\big)\Big\}^{1/2}\nn\\
	&\le&\si\Big\{\frac{2c_{u1}\log(p\vee T)}{c_uT\lm}\Big\}^{1/2}\nn
	\eenr
	with probability at least $1-2\exp\{-c_{u2}\log (p\vee T)\},$ where $c_{u2}=c_{u1}\wedge\surd(c_uc_{u1}/2).$ Applying union bounds over $k=1,...,p,$ the upper bound $T^2$ of at most distinct combinations of $\tau_{j-1}$ and $\tau_{j+1},$ and then over $j=1,...,N+1,$ ($N\le T$) yields,
	\benr
	\max_{1\le j\le N+1}\sup_{\substack{\tau_{j-1},\tau_{j}\in\{1,.....,T-1\};\\ (\tau_{j}-\tau_{j-1})\ge c_uT\lm}}\frac{1}{\big(\tau_{j}-\tau_{j-1}\big)}\Big|\sum_{t=\tau_{j-1}+1}^{\tau_{j+1}}\vep_{tk}\Big|\le\si\Big\{\frac{2c_{u1}\log(p\vee T)}{c_uT\lm}\Big\}^{1/2},\nn
	\eenr
	w.p. at least $1-2\exp\big\{-(c_{u2}-4)\log (p\vee T)\big\}.$ Finally utilizing the form $\vep_t^*=\vep_t-\bar\vep,$ $t=1,...,T,$ together with the first bound for $\|\bar\vep\|_{\iny}$ of Lemma \ref{lem:near.opt.unif.comp} by an argument analogous to that in (\ref{eq:7}) yields the statement of the lemma.
\end{proof}

\bc$\rule{3.5in}{0.1mm}$\ec

\section{Definitions and auxiliary results}\label{app:auxiliary}

The following definitions and results provide basic properties of subexponential distributions. These are largely reproduced from \cite{vershynin2019high} and \cite{rigollet201518}. Theorem \ref{thm:kolmogorov} and \ref{thm:argmax} below reproduce Kolmogorov's inequality and the Argmax Theorem. We also refer to Appendix B and Appendix F of \cite{kaul2020inference} and \cite{kaul2021graphical}, respectively, where these results and some additional proofs have been compiled.

\begin{definition}\label{def:sube}[Subexponential r.v.] A random variable $X\in\R$ is said to be sub-exponential with parameter $\si^2>0$ \big(denoted by $X\sim{\rm subE(\si^2)}$\big) if $E(X)=0$ and its moment generating function
	\benr
	E(\e^{tX})\le \e^{t^2\si^2/2},\qquad \forall\,\, |t|\le \frac{1}{\si}\nn
	\eenr
\end{definition}

\begin{definition}\label{def:submult} A random vector $X\in\R^p$ is subexponential with parameter $\si^2,$ if the inner product $\langle X, v\rangle\sim {\rm subE}(\si^2),$ respectively, for any $v\in\R^p$ with $\|v\|_2 = 1.$
\end{definition}

Following is the elementary definition of uniform tightness of a sequence of random variables reproduced from Page 166, Chapter 2 of \cite{durrett2010probability}.
\begin{definition}\label{def:utight} A sequence of random variables $X_n$ is said to be uniformly tight if for every $\ep>0,$ there is a compact set $K$ such that $pr(X_n\in K)>1-\ep.$
\end{definition}

\bc$\rule{3.5in}{0.1mm}$\ec

\begin{lemma}\label{lem:tailb}[Tail bounds] If $X\sim {\rm subE}(\si^2),$ then
	\benr
	pr(|X|\ge \la)\le 2\exp\Big\{-\frac{1}{2}\Big(\frac{\la^2}{\si^2}\wedge\frac{\la}{\si}\Big) \Big\}.\nn
	\eenr
\end{lemma}

\bc$\rule{3.5in}{0.1mm}$\ec

\begin{lemma}[Moment bounds]\label{lem:momentprop} If $X\sim {\rm subE}(\si^2),$ then
	\benr
	E|X|^k\le 4\si^k k^k, \qquad k> 0.\nn
	\eenr	
\end{lemma}
\bc$\rule{3.5in}{0.1mm}$\ec

\begin{lemma}\label{lem:lcsubE} Assume that $X\sim{\rm subE(\si^2)},$ and that $\al\in\R,$ then $\alpha X\sim{\rm subE}(\alpha^2\si^2).$ Moreover, assume that $X_1\sim{\rm subE(\si_1^2)}$ and $X_2\sim{\rm subE(\si_2^2)},$ then $X_1+X_2\sim{\rm subE((\si_1+\si_2)^2)},$ additionally, if $X_1$ and $X_2$ are independent, then $X_1+X_2\sim{\rm subE(\si_1^2+\si_2^2)}.$
\end{lemma}

\bc$\rule{3.5in}{0.1mm}$\ec

\begin{lemma}[Bernstein's inequality]\label{lem:bernstein} Let $X_1,X_2,...,X_T$ be independent random variables such that $X_t\sim {\rm subE}(\la^2).$ Then for any $d>0$ we have,
	\benr
	pr(|\bar X|>d)\le 2\exp\Big\{-\frac{T}{2}\Big(\frac{d^2}{\la^2}\wedge \frac{d}{\la}\Big)\Big\}\nn
	\eenr
\end{lemma}

\bc$\rule{3.5in}{0.1mm}$\ec

The next result is Kolmogorov's inequality reproduced from \cite{hajek1955generalization}
\begin{theorem}[Kolmogorov's inequality]\label{thm:kolmogorov} If $\xi_1,\xi_2,...$ is a sequence of mutually independent random variables with mean values $E(\xi_k)=0$ and finite variance ${\rm var}(\xi_k)=D_k^2$ $(k=1,2,...),$ we have, for any $\vep>0,$
	\benr
	pr\Big(\max_{1\le k\le m}\big|\xi_1+\xi_2+...+\xi_k\big|>\vep\Big)\le \frac{1}{\vep^2}\sum_{k=1}^mD_k^2\nn
	\eenr	
\end{theorem}

\bc$\rule{3.5in}{0.1mm}$\ec

Next, we provide the Argmax Theorem reproduced from Theorem 3.2.2 of \cite{vaart1996weak}.
\begin{theorem}[Argmax Theorem]\label{thm:argmax} Let $\cM_n,\cM$ be stochastic processes indexed by a metric space $H$ such that $\cM_n\Rightarrow\cM$ in $\ell^{\iny}(K)$ for every compact set $K\subseteq H$. Suppose that almost all sample paths $h\to \cM(h)$ are upper semicontinuous and posses a unique maximum at a (random) point $\h h,$ which as a random map in $H$ is tight. If the sequence $\h h_n$ is uniformly tight and satisfies $\cM_n(\h h_n)\ge \sup_h \cM_n(h)-o_p(1),$ then $\h h_n\Rightarrow \h h$ in $H.$
\end{theorem}

\section{Additional details and numerical results}\label{app:numerical.supplement}

\subsection{Estimation of drifts, asymptotic variances and quantiles}\label{subsec:app:par.est}

Next, we provide a discussion on the estimation of $\xi_j,$ and $\si^2_{(\iny,j)},$ $j=1,...,N,$ employed to obtain confidence intervals for $\tau^0=(\tau^0_1,...,\tau^0_N)^T,$ using the results of Theorems \ref{thm:wc.vanishing}, \ref{thm:wc.non.vanishing} and \ref{thm:wc.joint}.

First, to alleviate finite sample regularization biases we employ refitted mean estimates computed as $\tilde\theta_{(j)}=\big[\bar x_{(j)}(\tilde\tau)\big]_{\h S_j}$ $j=1,...,N$ wherein $\tilde\tau$ is the change point estimate of Algorithm 1. Here $\h S_j=\{k\,\,\h\theta_{(j)k}\ne 0\},$ $j=1,...,N$ correspond to the estimated sparsity sets, where $\h\theta_{(j)},$ $j=1,...,N$ are the Step 2 mean estimates of Algorithm 1. All remaining indices of these mean estimates are set to zero. It is known that refitted mean estimates preserve the rate of convergence of the regularized version while reducing finite sample biases, e.g. \cite{belloni2011square}. The jump vectors and jump sizes $\tilde\eta_{(j)}$ and $\tilde\xi_j,$ $j=1,...,N,$ are then evaluated as plug-in estimates per the defining relations (\ref{def:jumpsize}).

Next, consider the asymptotic variances $\si^2_{(\iny,j)},$ $j=1,...,N$ of Condition D. Note the finite sample representation of this parameter, $\xi_{j}^{-2}\eta^{0T}_{(j)}\Si\eta^0_{(j)}.$ Plug-in versions $\tilde\si^2_{(\iny,j)},$ $j=1,...,N,$ are computed by employing the above described estimated parameters. The covariance matrix $\Si$ is estimated as the sample covariance $\tilde\Si$ computed by utilizing the entire data set centered with the estimated mean parameters $\tilde\theta_{(j)},$ $j=1,...,N$ over estimated partitions induced by $\tilde\tau$ of Algorithm 1. Note that since we are not interested in the estimation of $\Si$ itself, but instead the quadratic form described above, employing the sample covariance is effectively identical to employing the refitted covariance on the adjacency matrix estimated by the jump vectors $\tilde\eta_{(j)}'s,$ in turn making this shortcut valid despite potential high dimensionality.

Finally, for quantiles of the limiting distributions characterized in Theorems \ref{thm:wc.vanishing} and  \ref{thm:wc.non.vanishing} in the vanishing and non-vanishing regimes, respectively, we note the following: in the former case, we employ the cdf of this distribution which was first presented in \cite{yao1987approximating}. In the latter case, we assume in all calculations that the underlying distribution is Gaussian and consequently the distribution of the increments $\cP$ of Condition A$'$ is also Gaussian. The above estimated parameters are then used to produce realizations of the increments' distribution, and thus realizations of the two-sided random walk and in turn those of its {\it argmax}. The quantiles are then estimated by a Monte Carlo approximation.

\subsection{Additional numerical results of Section \ref{sec:numerical}}\label{subsec:app:numerical}


\vspace{1.5mm}
{\noi{\bf Results of Scenarios A an B (Gaussian errors)}}: Tables \ref{tab:simA.N4} and \ref{tab:simB.N4} below provide results for these scenarios for $N=4$ change points, respectively.

\begin{table}[H]
	\centering
	\resizebox{0.6\textwidth}{!}{
		\begin{tabular}{cccccc}
			\toprule
			\multicolumn{2}{c}{\begin{tabular}[c]{@{}c@{}}$N=4,$\\ $s=4$\end{tabular}} & \multirow{2}{*}{haus.d (sd)} & \multicolumn{2}{c}{\begin{tabular}[c]{@{}c@{}}Comp. coverage (av. ME)\\ $(1-\alpha)=0.95$\end{tabular}} & \multirow{2}{*}{\begin{tabular}[c]{@{}c@{}}Simul. \\ Coverage\\ $(1-\alpha)^N=0.814$\end{tabular}} \\ \cmidrule{1-2} \cmidrule{4-5}
			$T$                                  & $p$                                 &                              & Vanishing                                          & Non-Vanishing                                      &                                                                                                    \\ \midrule
			450                                  & 50                                  & 1.31 (1.28)                  & 0.964 (2.08)                                       & 0.978 (2.02)                                       & 0.794                                                                                              \\
			450                                  & 200                                 & 1.31 (1.15)                  & 0.936 (2.07)                                       & 0.958 (2)                                          & 0.79                                                                                               \\
			450                                  & 350                                 & 1.33 (1.42)                  & 0.946 (2.07)                                       & 0.978 (2.01)                                       & 0.798                                                                                              \\
			450                                  & 500                                 & 1.35 (1.37)                  & 0.954 (2.12)                                       & 0.976 (2.06)                                       & 0.79                                                                                               \\ \midrule
			600                                  & 50                                  & 1.41 (1.32)                  & 0.96 (2.12)                                        & 0.976 (2.03)                                       & 0.79                                                                                               \\
			600                                  & 200                                 & 1.31 (1.3)                   & 0.934 (2.08)                                       & 0.96 (2.02)                                        & 0.808                                                                                              \\
			600                                  & 350                                 & 1.38 (1.17)                  & 0.952 (2.05)                                       & 0.978 (1.99)                                       & 0.798                                                                                              \\
			600                                  & 500                                 & 1.36 (1.19)                  & 0.94 (2.05)                                        & 0.97 (1.98)                                        & 0.772                                                                                              \\ \midrule
			750                                  & 50                                  & 1.39 (1.22)                  & 0.958 (2.12)                                       & 0.972 (2.02)                                       & 0.786                                                                                              \\
			750                                  & 200                                 & 1.33 (1.28)                  & 0.946 (2.09)                                       & 0.962 (2.02)                                       & 0.788                                                                                              \\
			750                                  & 350                                 & 1.41 (1.41)                  & 0.95 (2.09)                                        & 0.968 (2)                                          & 0.768                                                                                              \\
			750                                  & 500                                 & 1.42 (1.38)                  & 0.954 (2.08)                                       & 0.972 (2.02)                                       & 0.774                                                                                              \\ \bottomrule
	\end{tabular}}
	\caption{Results of Scenario A with $N=4$ based on 500 replicates. Coverage metrics rounded to three decimals, all other metrics rounded to two decimals.}
	\label{tab:simA.N4}
\end{table}

\begin{table}[H]
	\centering
	\resizebox{0.8\textwidth}{!}{
		\begin{tabular}{ccclllll}
			\toprule
			\multirow{2}{*}{Method}                                                    & \multicolumn{2}{c}{\begin{tabular}[c]{@{}c@{}}$N=4,$\\ $s=4$\end{tabular}} & \multicolumn{1}{c}{\multirow{2}{*}{haus.d (sd)}} & \multicolumn{1}{c}{\multirow{2}{*}{N-match}} & \multicolumn{2}{c}{\begin{tabular}[c]{@{}c@{}}Comp. cov. (av. ME) $\big|\h N=N$\\ $(1-\alpha)=0.95$\end{tabular}} & \multicolumn{1}{c}{\multirow{2}{*}{\begin{tabular}[c]{@{}c@{}}Simul. \\ cov. $\big| \h N= N$\\ $(1-\alpha)^N=0.814$\end{tabular}}} \\ \cmidrule{2-3} \cmidrule{6-7}
			& $T$                                  & $p$                                 & \multicolumn{1}{c}{}                             & \multicolumn{1}{c}{}                         & \multicolumn{1}{c}{Vanishing}                         & \multicolumn{1}{c}{Non-Vanishing}                         & \multicolumn{1}{c}{}                                                                                                               \\ \midrule
			\multirow{12}{*}{\begin{tabular}[c]{@{}c@{}}KFJS+\\ BS+\\ LR\end{tabular}} & 450                                  & 50                                  & 2.81   (6.46)                                    & 0.95                                         & 0.951 (2.08)                                          & 0.975 (1.99)                                              & 0.753                                                                                                                              \\
			& 450                                  & 200                                 & 5.09 (10.6)                                      & 0.88                                         & 0.939 (2.08)                                          & 0.964 (2.02)                                              & 0.722                                                                                                                              \\
			& 450                                  & 350                                 & 4.15 (9.57)                                      & 0.90                                         & 0.924 (2.09)                                          & 0.955 (2.02)                                              & 0.753                                                                                                                              \\
			& 450                                  & 500                                 & 4.27 (9.33)                                      & 0.89                                         & 0.921 (2.11)                                          & 0.946 (2.05)                                              & 0.742                                                                                                                              \\ \cmidrule{2-8}
			& 600                                  & 50                                  & 3.79 (9.86)                                      & 0.94                                         & 0.932 (2.11)                                          & 0.951 (2.02)                                              & 0.756                                                                                                                              \\
			& 600                                  & 200                                 & 4.86 (11.71)                                     & 0.92                                         & 0.954 (2.07)                                          & 0.963 (2.00)                                              & 0.771                                                                                                                              \\
			& 600                                  & 350                                 & 5.19 (12.6)                                      & 0.92                                         & 0.948 (2.10)                                          & 0.969 (2.03)                                              & 0.734                                                                                                                              \\
			& 600                                  & 500                                 & 4.57 (11.4)                                      & 0.92                                         & 0.939 (2.07)                                          & 0.961 (2.02)                                              & 0.747                                                                                                                              \\ \cmidrule{2-8}
			& 750                                  & 50                                  & 6.18 (16.19)                                     & 0.91                                         & 0.956 (2.14)                                          & 0.978 (2.04)                                              & 0.776                                                                                                                              \\
			& 750                                  & 200                                 & 7.10 (17.78)                                     & 0.90                                         & 0.942 (2.10)                                          & 0.969 (2.02)                                              & 0.752                                                                                                                              \\
			& 750                                  & 350                                 & 6.78 (16.79)                                     & 0.90                                         & 0.951 (2.11)                                          & 0.976 (2.03)                                              & 0.734                                                                                                                              \\
			& 750                                  & 500                                 & 6.16 (16.14)                                     & 0.91                                         & 0.941 (2.10)                                          & 0.956 (2.03)                                              & 0.742                                                                                                                              \\ \midrule
			\multirow{12}{*}{\begin{tabular}[c]{@{}c@{}}WS\\ +LR\end{tabular}}         & 450                                  & 50                                  & 12.54 (17.97)                                    & 0.64                                         & 0.934 (2.06)                                          & 0.959 (1.99)                                              & 0.723                                                                                                                              \\
			& 450                                  & 200                                 & 14.16 (22.25)                                    & 0.64                                         & 0.938 (2.07)                                          & 0.959 (1.99)                                              & 0.716                                                                                                                              \\
			& 450                                  & 350                                 & 24.83 (53.64)                                    & 0.58                                         & 0.911 (2.08)                                          & 0.955 (2.01)                                              & 0.712                                                                                                                              \\
			& 450                                  & 500                                 & 84.14 (116.91)                                   & 0.49                                         & 0.918 (2.07)                                          & 0.951 (2.00)                                              & 0.694                                                                                                                              \\ \cmidrule{2-8}
			& 600                                  & 50                                  & 15.77 (23.71)                                    & 0.62                                         & 0.926 (2.10)                                          & 0.945 (2.00)                                              & 0.746                                                                                                                              \\
			& 600                                  & 200                                 & 18.55 (25.95)                                    & 0.62                                         & 0.942 (2.04)                                          & 0.971 (1.97)                                              & 0.759                                                                                                                              \\
			& 600                                  & 350                                 & 18.09 (29.49)                                    & 0.64                                         & 0.928 (2.09)                                          & 0.950 (2.00)                                              & 0.704                                                                                                                              \\
			& 600                                  & 500                                 & 30.79 (68.63)                                    & 0.60                                         & 0.957 (2.06)                                          & 0.973 (2.01)                                              & 0.739                                                                                                                              \\ \cmidrule{2-8}
			& 750                                  & 50                                  & 20.57 (30.58)                                    & 0.60                                         & 0.94 (2.15)                                           & 0.967 (2.05)                                              & 0.779                                                                                                                              \\
			& 750                                  & 200                                 & 20.14 (29.03)                                    & 0.62                                         & 0.935 (2.09)                                          & 0.955 (2.00)                                              & 0.718                                                                                                                              \\
			& 750                                  & 350                                 & 19.38 (29.51)                                    & 0.66                                         & 0.961 (2.10)                                          & 0.976 (2.02)                                              & 0.721                                                                                                                              \\
			& 750                                  & 500                                 & 23.25 (38.99)                                    & 0.65                                         & 0.923 (2.09)                                          & 0.944 (2.02)                                              & 0.728                                                                                                                              \\ \bottomrule
	\end{tabular}}
	\caption{Results of Scenario B with $N=4$ based on 500 replicates. Coverage metrics rounded to three decimals, all other metrics rounded to two decimals.}
	\label{tab:simB.N4}
\end{table}

\vspace{1.5mm}
{\noi\bf {Results of Scenarios A$'$ and B$'$ (subexponential errors)}}: Tables \ref{tab:simA'.N2} and \ref{tab:simB'.N2} below provide results for these scenarios with $N=2$ change points, and  Tables \ref{tab:simA'.N4} and  \ref{tab:simB'.N4} with $N=4.$. Under Scenario B$'$ with subexponential errors, the method WS for preliminary estimation was found to have a very low proportion of replicates wherein $\h N=N.$ Hence, it was rendered unsuitable for calculating the coverage metrics. Consequently in Scenario B$'$ we only report results obtained by the KFJS+BS+LR method.

\begin{table}[H]
	\centering
	\resizebox{0.6\textwidth}{!}{
		\begin{tabular}{cccccc}
			\toprule
			\multicolumn{2}{c}{\begin{tabular}[c]{@{}c@{}}$N=2,$\\ $s=4$\end{tabular}} & \multicolumn{1}{c}{\multirow{2}{*}{haus.d (sd)}} & \multicolumn{2}{c}{\begin{tabular}[c]{@{}c@{}}Comp. coverage (av. ME)\\ $(1-\alpha)=0.95$\end{tabular}} & \multicolumn{1}{c}{\multirow{2}{*}{\begin{tabular}[c]{@{}c@{}}Simul. \\ Coverage\\ $(1-\alpha)^N=0.902$\end{tabular}}} \\ \cmidrule{1-2} \cmidrule{4-5}
			$T$                                  & $p$                                 & \multicolumn{1}{c}{}                             & \multicolumn{1}{c}{Vanishing}                    & \multicolumn{1}{c}{Non-Vanishing}                    & \multicolumn{1}{c}{}                                                                                                   \\ \midrule
			450                                  & 50                                  & 0.79 (1.00)                                      & 0.962 (2.16)                                     & 0.982 (2.05)                                         & 0.888                                                                                                                  \\
			450                                  & 200                                 & 0.76 (1.10)                                      & 0.956 (2.15)                                     & 0.968 (2.05)                                         & 0.894                                                                                                                  \\
			450                                  & 350                                 & 0.75 (1.02)                                      & 0.950 (2.12)                                     & 0.964 (2.05)                                         & 0.882                                                                                                                  \\
			450                                  & 500                                 & 0.79 (1.06)                                      & 0.954 (2.11)                                     & 0.968 (2.02)                                         & 0.878                                                                                                                  \\ \midrule
			600                                  & 50                                  & 0.71 (1.11)                                      & 0.958 (2.18)                                     & 0.964 (2.05)                                         & 0.908                                                                                                                  \\
			600                                  & 200                                 & 0.81 (1.13)                                      & 0.940 (2.15)                                     & 0.952 (2.03)                                         & 0.886                                                                                                                  \\
			600                                  & 350                                 & 0.84 (1.03)                                      & 0.946 (2.16)                                     & 0.964 (2.04)                                         & 0.862                                                                                                                  \\
			600                                  & 500                                 & 0.70 (0.96)                                      & 0.948 (2.14)                                     & 0.972 (2.03)                                         & 0.898                                                                                                                  \\ \midrule
			750                                  & 50                                  & 0.74 (0.99)                                      & 0.958 (2.17)                                     & 0.966 (2.03)                                         & 0.888                                                                                                                  \\
			750                                  & 200                                 & 0.81 (1.07)                                      & 0.954 (2.17)                                     & 0.962 (2.03)                                         & 0.864                                                                                                                  \\
			750                                  & 350                                 & 0.70 (0.97)                                      & 0.956 (2.16)                                     & 0.968 (2.02)                                         & 0.888                                                                                                                  \\
			750                                  & 500                                 & 0.72 (0.92)                                      & 0.962 (2.17)                                     & 0.974 (2.03)                                         & 0.898                                                                                                                  \\ \bottomrule
	\end{tabular}}
	\caption{Results of Scenario A$'$ with $N=2$ based on 500 replicates. Coverage metrics rounded to three decimals, all other metrics rounded to two decimals.}
	\label{tab:simA'.N2}
\end{table}

\begin{table}[H]
	\centering
	\resizebox{0.6\textwidth}{!}{
		\begin{tabular}{cccccc}
			\toprule
			\multicolumn{2}{c}{\begin{tabular}[c]{@{}c@{}}$N=4,$\\ $s=4$\end{tabular}} & \multicolumn{1}{c}{\multirow{2}{*}{haus.d (sd)}} & \multicolumn{2}{c}{\begin{tabular}[c]{@{}c@{}}Comp. coverage (av. ME)\\ $(1-\alpha)=0.95$\end{tabular}} & \multicolumn{1}{c}{\multirow{2}{*}{\begin{tabular}[c]{@{}c@{}}Simul. \\ Coverage\\ $(1-\alpha)^N=0.814$\end{tabular}}} \\ \cmidrule{1-2} \cmidrule{4-5}
			$T$                                  & $p$                                 & \multicolumn{1}{c}{}                             & \multicolumn{1}{c}{Vanishing}                    & \multicolumn{1}{c}{Non-Vanishing}                    & \multicolumn{1}{c}{}                                                                                                   \\ \midrule
			450                                  & 50                                  & 1.28 (1.33)                                      & 0.962 (2.08)                                     & 0.972 (2.00)                                         & 0.808                                                                                                                  \\
			450                                  & 200                                 & 1.34 (1.23)                                      & 0.958 (2.06)                                     & 0.976 (1.98)                                         & 0.788                                                                                                                  \\
			450                                  & 350                                 & 1.31 (1.31)                                      & 0.958 (2.08)                                     & 0.968 (2.00)                                         & 0.796                                                                                                                  \\
			450                                  & 500                                 & 1.37 (1.15)                                      & 0.946 (2.09)                                     & 0.964 (2.02)                                         & 0.770                                                                                                                   \\ \midrule
			600                                  & 50                                  & 1.32 (1.24)                                      & 0.958 (2.11)                                     & 0.980 (2.04)                                         & 0.798                                                                                                                  \\
			600                                  & 200                                 & 1.26 (1.20)                                      & 0.940 (2.06)                                     & 0.958 (1.98)                                         & 0.798                                                                                                                  \\
			600                                  & 350                                 & 1.40 (1.30)                                      & 0.956 (2.07)                                     & 0.982 (2.00)                                         & 0.768                                                                                                                  \\
			600                                  & 500                                 & 1.44 (1.47)                                      & 0.956 (2.06)                                     & 0.974 (1.99)                                         & 0.770                                                                                                                   \\ \midrule
			750                                  & 50                                  & 1.29 (1.25)                                      & 0.940 (2.14)                                     & 0.952 (2.04)                                         & 0.814                                                                                                                  \\
			750                                  & 200                                 & 1.36 (1.31)                                      & 0.946 (2.09)                                     & 0.962 (2.02)                                         & 0.772                                                                                                                  \\
			750                                  & 350                                 & 1.29 (1.25)                                      & 0.956 (2.09)                                     & 0.970 (2.00)                                         & 0.806                                                                                                                  \\
			750                                  & 500                                 & 1.46 (1.35)                                      & 0.950 (2.10)                                     & 0.972 (2.03)                                         & 0.784                                                                                                                  \\ \bottomrule
	\end{tabular}}
	\caption{Results of Scenario A$'$ with $N=4$ based on 500 replicates. Coverage metrics rounded to three decimals, all other metrics rounded to two decimals.}
	\label{tab:simA'.N4}
\end{table}

\begin{table}[H]
	\centering
	\resizebox{0.8\textwidth}{!}{
		\begin{tabular}{cccccccc}
			\toprule
			\multirow{2}{*}{Method}                                                    & \multicolumn{2}{c}{\begin{tabular}[c]{@{}c@{}}$N=2,$\\ $s=4$\end{tabular}} & \multirow{2}{*}{haus.d (sd)} & \multirow{2}{*}{N-match} & \multicolumn{2}{c}{\begin{tabular}[c]{@{}c@{}}Comp. coverage (av. ME)\\ $(1-\alpha)=0.95$\end{tabular}} & \multirow{2}{*}{\begin{tabular}[c]{@{}c@{}}Simul. \\ Coverage\\ $(1-\alpha)^N=0.902$\end{tabular}} \\ \cmidrule{2-3} \cmidrule{6-7}
			& $T$                                  & $p$                                 &                              &                          & Vanishing                                          & Non-Vanishing                                      &                                                                                                    \\ \midrule
			\multirow{12}{*}{\begin{tabular}[c]{@{}c@{}}KFJS+\\ BS+\\ LR\end{tabular}} & 450                                  & 50                                  & 18.14 (25.7)                 & 0.65                     & 0.957 (2.16)                                       & 0.957 (2.12)                                       & 0.877                                                                                              \\
			& 450                                  & 200                                 & 18.96 (26.59)                & 0.66                     & 0.955 (2.16)                                       & 0.970 (2.12)                                       & 0.882                                                                                              \\
			& 450                                  & 350                                 & 17.16 (26.31)                & 0.70                     & 0.940 (2.18)                                       & 0.974 (2.11)                                       & 0.883                                                                                              \\
			& 450                                  & 500                                 & 16.66 (26.04)                & 0.71                     & 0.924 (2.21)                                       & 0.941 (2.18)                                       & 0.876                                                                                              \\ \cmidrule{2-8} 
			& 600                                  & 50                                  & 25.75 (34.03)                & 0.60                     & 0.940 (2.18)                                       & 0.957 (2.10)                                       & 0.881                                                                                              \\
			& 600                                  & 200                                 & 26.55 (35.15)                & 0.62                     & 0.913 (2.14)                                       & 0.926 (2.06)                                       & 0.859                                                                                              \\
			& 600                                  & 350                                 & 25.90 (35.85)                & 0.65                     & 0.938 (2.18)                                       & 0.969 (2.12)                                       & 0.864                                                                                              \\
			& 600                                  & 500                                 & 23.78 (34.47)                & 0.66                     & 0.952 (2.16)                                       & 0.967 (2.08)                                       & 0.891                                                                                              \\ \cmidrule{2-8} 
			& 750                                  & 50                                  & 24.18 (39.17)                & 0.70                     & 0.936 (2.21)                                       & 0.954 (2.09)                                       & 0.858                                                                                              \\
			& 750                                  & 200                                 & 31.50 (42.29)                & 0.62                     & 0.964 (2.18)                                       & 0.964 (2.09)                                       & 0.877                                                                                              \\
			& 750                                  & 350                                 & 33.38 (44.07)                & 0.61                     & 0.944 (2.17)                                       & 0.958 (2.06)                                       & 0.869                                                                                              \\
			& 750                                  & 500                                 & 33.99 (45.38)                & 0.62                     & 0.929 (2.18)                                       & 0.936 (2.09)                                       & 0.878                                                                                              \\ \bottomrule
	\end{tabular}}
	\caption{Results of Scenario B$'$ with $N=2$ based on 500 replicates. Coverage metrics rounded to three decimals, all other metrics rounded to two decimals.}
	\label{tab:simB'.N2}
\end{table}

\begin{table}[H]
	\centering
	\resizebox{0.8\textwidth}{!}{
		\begin{tabular}{cccccccc}
			\toprule
			\multirow{2}{*}{Method}                                                    & \multicolumn{2}{c}{\begin{tabular}[c]{@{}c@{}}$N=4,$\\ $s=4$\end{tabular}} & \multicolumn{1}{c}{\multirow{2}{*}{haus.d (sd)}} & \multicolumn{1}{c}{\multirow{2}{*}{N-match}} & \multicolumn{2}{c}{\begin{tabular}[c]{@{}c@{}}Comp. coverage (av. ME)\\ $(1-\alpha)=0.95$\end{tabular}} & \multicolumn{1}{c}{\multirow{2}{*}{\begin{tabular}[c]{@{}c@{}}Simul. \\ Coverage\\ $(1-\alpha)^N=0.814$\end{tabular}}} \\ \cmidrule{2-3} \cmidrule{6-7}
			& $T$                                  & $p$                                 & \multicolumn{1}{c}{}                             & \multicolumn{1}{c}{}                         & \multicolumn{1}{c}{Vanishing}                    & \multicolumn{1}{c}{Non-Vanishing}                    & \multicolumn{1}{c}{}                                                                                                   \\ \midrule
			\multirow{12}{*}{\begin{tabular}[c]{@{}c@{}}KFJS+\\ BS+\\ LR\end{tabular}} & 450                                  & 50                                  & 3.57 (8.71)                                      & 0.93                                         & 0.961 (2.09)                                     & 0.970 (2.08)                                         & 0.773                                                                                                                  \\
			& 450                                  & 200                                 & 4.32 (9.75)                                      & 0.91                                         & 0.928 (2.08)                                     & 0.952 (2.08)                                         & 0.740                                                                                                                  \\
			& 450                                  & 350                                 & 3.86 (8.63)                                      & 0.90                                         & 0.934 (2.11)                                     & 0.962 (2.09)                                         & 0.723                                                                                                                  \\
			& 450                                  & 500                                 & 4.01 (9.21)                                      & 0.91                                         & 0.940 (2.14)                                     & 0.962 (2.14)                                         & 0.753                                                                                                                  \\ \cmidrule{2-8} 
			& 600                                  & 50                                  & 4.42 (11.27)                                     & 0.93                                         & 0.950 (2.11)                                     & 0.974 (2.07)                                         & 0.767                                                                                                                  \\
			& 600                                  & 200                                 & 4.92 (12.32)                                     & 0.92                                         & 0.939 (2.08)                                     & 0.961 (2.05)                                         & 0.762                                                                                                                  \\
			& 600                                  & 350                                 & 5.51 (13.30)                                     & 0.91                                         & 0.945 (2.08)                                     & 0.967 (2.06)                                         & 0.750                                                                                                                  \\
			& 600                                  & 500                                 & 4.50 (11.71)                                     & 0.93                                         & 0.940 (2.09)                                     & 0.970 (2.08)                                         & 0.755                                                                                                                  \\ \cmidrule{2-8} 
			& 750                                  & 50                                  & 5.64 (14.81)                                     & 0.92                                         & 0.948 (2.14)                                     & 0.952 (2.08)                                         & 0.754                                                                                                                  \\
			& 750                                  & 200                                 & 5.55 (14.99)                                     & 0.93                                         & 0.942 (2.09)                                     & 0.952 (2.05)                                         & 0.769                                                                                                                  \\
			& 750                                  & 350                                 & 5.47 (14.83)                                     & 0.93                                         & 0.959 (2.09)                                     & 0.974 (2.05)                                         & 0.754                                                                                                                  \\
			& 750                                  & 500                                 & 6.26 (16.56)                                     & 0.91                                         & 0.949 (2.08)                                     & 0.967 (2.04)                                         & 0.786                                                                                                                  \\ \bottomrule
	\end{tabular}}
	\caption{Results of Scenario B$'$ with $N=4$ based on 500 replicates. Coverage metrics rounded to three decimals, all other metrics rounded to two decimals.}
	\label{tab:simB'.N4}
\end{table}

\vspace{1.5mm}
{\noi\bf Setup and results of Scenario C:} The design of this simulation is largely identical to that of Scenario B (Gaussian errors) described in Section \ref{sec:numerical}, with the only distinction being that we consider larger values of the sampling period $T\in\{1000,1500\}.$ Results are provided in Table \ref{tab:simC} below.

\begin{table}[H]
	\centering
	\resizebox{0.8\textwidth}{!}{
		\begin{tabular}{cccccccc}
			\toprule
			\multirow{2}{*}{Method}                                                   & \multicolumn{2}{c}{\begin{tabular}[c]{@{}c@{}}$N=4,$\\ $s=4$\end{tabular}} & \multirow{2}{*}{haus.d (sd)} & \multirow{2}{*}{N-match} & \multicolumn{2}{c}{\begin{tabular}[c]{@{}c@{}}Comp. coverage (av. ME)\\ $(1-\alpha)=0.95$\end{tabular}} & \multirow{2}{*}{\begin{tabular}[c]{@{}c@{}}Simul. \\ Coverage\\ $(1-\alpha)^N=0.814$\end{tabular}} \\ \cmidrule{2-3} \cmidrule{6-7}
			& $T$                                  & $p$                                 &                              &                          & Vanishing                                          & Non-Vanishing                                      &                                                                                                    \\ \midrule
			\multirow{6}{*}{\begin{tabular}[c]{@{}c@{}}KFJS+\\ BS+\\ LR\end{tabular}} & 1000                                 & 200                                 & 8.01 (21.17)                 & 0.90                    & 0.967 (2.12)                                       & 0.978 (2.01)                                       & 0.792                                                                                              \\
			& 1000                                 & 350                                 & 9.85 (23.37)                 & 0.87                     & 0.945 (2.12)                                       & 0.956 (2.02)                                       & 0.793                                                                                              \\
			& 1000                                 & 500                                 & 8.91 (22.70)                 & 0.89                    & 0.935 (2.11)                                       & 0.957 (2.02)                                       & 0.763                                                                                              \\ \cmidrule{2-8} 
			& 1500                                 & 200                                 & 12.85 (34.33)                & 0.89                     & 0.957 (2.18)                                       & 0.969 (2.02)                                       & 0.804                                                                                              \\
			& 1500                                 & 350                                 & 13.78 (35.74)                & 0.88                    & 0.957 (2.17)                                       & 0.973 (2.02)                                       & 0.758                                                                                              \\
			& 1500                                 & 500                                 & 11.77 (32.69)                & 0.89                   & 0.955 (2.16)                                       & 0.971 (2.01)                                       & 0.786                                                                                              \\ \midrule
			\multirow{6}{*}{\begin{tabular}[c]{@{}c@{}}WS+\\ LR\end{tabular}}         & 1000                                 & 200                                 & 24.70 (41.09)                & 0.66                    & 0.964 (2.12)                                       & 0.976 (2.00)                                       & 0.781                                                                                              \\
			& 1000                                 & 350                                 & 24.99 (38.80)                & 0.63                    & 0.937 (2.13)                                       & 0.947 (2.03)                                       & 0.767                                                                                              \\
			& 1000                                 & 500                                 & 21.60 (37.41)                & 0.71                   & 0.949 (2.11)                                       & 0.966 (2.02)                                       & 0.764                                                                                              \\ \cmidrule{2-8} 
			& 1500                                 & 200                                 & 34.72 (61.29)                & 0.67                    & 0.941 (2.17)                                       & 0.962 (2.01)                                       & 0.776                                                                                              \\
			& 1500                                 & 350                                 & 35.65 (61.65)                & 0.68                   & 0.959 (2.17)                                       & 0.974 (2.01)                                       & 0.752                                                                                              \\
			& 1500                                 & 500                                 & 37.09 (63.98)                & 0.68                   & 0.950 (2.16)                                       & 0.962 (2.01)                                       & 0.776                                                                                              \\ \bottomrule
	\end{tabular}}
	\caption{Results of Scenario C with $N=4$ based on 500  replicates. Coverage metrics rounded to three decimals, all other metrics rounded to two decimals.}
	\label{tab:simC}
\end{table}

\subsection{Description of second method (KFJS+BS) employed for preliminary estimation in Section \ref{sec:numerical}}

\cite{kaul2020inference} considers a mean shift model with a single change point under potential high dimensionality, i.e., model (\ref{model:rvmcp}) with  $N=1.$ They propose a two step algorithmic procedure which yields an estimate that is optimal is its rate of convergence (this estimate is the same to $\tilde\tau$ in Algorithm 1 in Section \ref{sec:main} for $N=1$). While not of direct interest, the paper also establishes that the first update in their algorithm is near optimal, i.e., obeys the bound (\ref{eq:29}) with $N=1,$ under identical assumptions as those assumed here, including the relaxation to subexponential distributions. Remark 4.2 of the paper provides an $\ell_0$ regularization that also enables boundary selection of the change point estimate, i.e., identifying that a change point \textit{is not present}. This estimator is compiled as Algorithm 2 below.

\begin{figure}[H]
	\noi\rule{\textwidth}{0.5pt}
	
	\vspace{-2mm}
	\flushleft {\bf Algorithm 2 (KFJS):} Near optimal estimation of $\tau^0$ with boundary selection (under $N=1$)
	
	\vspace{-1.25mm}
	\noi\rule{\textwidth}{0.5pt}
	
	\vspace{-1.25mm}
	\flushleft{\bf (Initialize):} Select a preliminary evenly spaced coarse grid $\cD\subset\{1,...,T\}$ of cardinality $\log T.$ Select an initializer $\check\tau\in\cD$ as the best fitting value to the data $\big\{x_t\big\}_{t=1}^T.$
	
	\vspace{-1.25mm}
	\flushleft{\bf Step 1:} Obtain mean estimates $\check\theta_{(j)}=\h\theta_{(j)}(\check\tau),$ $j=1,2,$ and update change point estimates as
	\benr\label{eq:44}
	\h\tau=\argmin_{\tau\in \{1,...,(T-1)\}}Q(\tau,\check\theta),\nn
	\eenr
	and perform an $\ell_0$ regularization as
	\benr
	\h\tau^*=\begin{cases} T ({\rm no\,\, change}) & {\rm if}\,\, \{Q(T,\check\theta)-Q(\h\tau,\check\theta)\}<\g\\
		\h\tau & {\rm else}.
	\end{cases}\nn
	\eenr
	
	\vspace{-3.25mm}
	\flushleft{\bf (Output):} $\h\tau^*$
	
	\vspace{-1.25mm}
	\noi\rule{\textwidth}{0.5pt}
\end{figure}

The mean estimates $\h\theta(\tau)$ of Algorithm 2 are the soft-thresholded sample means as defined in (\ref{est:softthresh}), and $Q(\tau,\theta)$ represents the squared loss under a single change point assumption defined as
\benr
Q(\tau,\theta)=\sum_{t=1}^{\tau} \|x_t-\theta_{(1)}\|_2^2 +\sum_{t=\tau+1}^{T} \|x_t-\theta_{(2)}\|_2^2.\nn
\eenr
It can be observed that the regularization carried out in Step 1 of Algorithm 2 is equivalent to
\benr
\h\tau^*=\argmin_{\tau\in \{1,...,(T-1)\}}\big\{Q(\tau,\check\theta)+\g {\bf 1}[\tau\ne T]\big\}.\nn
\eenr
Further, the BIC criterion to tune this regularization reduces to $\g=(|\h S_2|+1)\log T$. Note that at the boundary value $\tau=T,$ the model has $|S_2|$ fewer mean parameters and one less change point parameter. It can be shown that in addition to near optimal estimation of $\tau^0,$ Algorithm 2 also provides selection consistency, i.e., $pr(\h\tau^*=T)\to 1$ when $\tau^0=T.$ A natural extension of Algorithm 2 is employed in Section \ref{sec:numerical} by leveraging binary segmentation, i.e., recursive application of Algorithm 2 on estimated partitions, performed until no further change points are detected. This extension is summarized in Algorithm 3.

\begin{figure}[]
	\noi\rule{\textwidth}{0.5pt}
	
	\vspace{-2mm}
	\flushleft {\bf Algorithm 3 (KJFS+BS):} Extension of KJFS to multiple changes via binary segmentation

	\vspace{-1.25mm}
	\noi\rule{\textwidth}{0.5pt}
	
	\vspace{-1.25mm}
	\flushleft{\bf (Initialize):}  $\h\tau_{{\rm st}}=\phi$ collecting all change points to be estimated.

	Implement $\h\tau$= Alg. 2 $\big(\{1,...,T\}\big).$\\~
	{\bf If} {$\h\tau=T$ (no change)} then {\bf Stop}\\~	
	{\bf Else} \,\,$\h\tau_{{\rm up}}=(\tau_{{\rm st}},\h\tau)$ (updated vector of estimated change points)

	{\bf While} ${\rm length}(\h\tau_{up})> {\rm length}(\h\tau_{st})$ {\bf do}\\~
	$\h\tau_{st}=\h\tau_{up}$\\~
	{\bf for} $m \in 1: ({\rm length}{(\tau_{st})}+1)$ {\bf do}\\~
	${\rm partition}_m=\{\tau_{{st}(m-1)},...,\tau_{{st}(m)}\}$\\~
	$\h\tau= {\rm Alg.\,2}({\rm partition_m})$\\~
	{\bf If} $\h\tau$ is away from boundary of sampling period of partition {\bf then}\\~
	$\h\tau_{up}=(\h\tau_{st},\h\tau)$

	\flushleft{\bf (Output):} all estimated change points of vector $\h\tau_{up}$ sorted in ascending order.
	
	\vspace{-1.25mm}
	\noi\rule{\textwidth}{0.5pt}
\end{figure}

\end{document}